\documentclass[useAMS,usenatbib]{mn2e}

\usepackage{amsmath}
\usepackage{amssymb}
\usepackage{subfigure}
\usepackage{flafter}
\usepackage[pdfborder={0.1 0.1 0.7}]{hyperref}
\hypersetup{pdflinkmargin=0.1mm}
\usepackage{adjustbox}
\usepackage{blindtext}
\usepackage{color}
\usepackage{graphicx}
\usepackage[T1]{fontenc}
\usepackage{aecompl}

\DeclareMathOperator\erf{erf}

\newcommand{\hmpc}{h^{-1}\mathrm{Mpc}}

\newcommand{\Ncen}{\ensuremath{N_{\mathrm{cen}}}}
\newcommand{\Nsat}{\ensuremath{N_{\mathrm{sat}}}}
\newcommand{\Mmin}{M_{\mathrm{min}}}
\newcommand{\sigLM}{\ensuremath{\sigma_{\log M}}}
\newcommand{\Mz}{M_0}
\newcommand{\Mop}{M^{\prime}_1}

\def\simlt{\lower.5ex\hbox{$\; \buildrel < \over \sim \;$}}
\def\simgt{\lower.5ex\hbox{$\; \buildrel > \over \sim \;$}}

\title[Understanding BOSS/CMASS Galaxies]
{Building a Better Understanding of the High Redshift BOSS Galaxies as Tools for Cosmology} 

\author[Favole, et al. 2015]{
\parbox[t]{\textwidth}{
Ginevra Favole$^{1,2}$\thanks{E-mail: g.favole@csic.es}\thanks{MultiDark Fellow}, Cameron K. McBride$^{3}$, Daniel J. Eisenstein$^{3}$, Francisco Prada$^{1,2,4}$, Molly E. Swanson$^{3}$, Chia-Hsun Chuang$^{1,2}$, Donald P. Schneider$^{5,6}$}
\vspace*{15pt} \\ 
$^1$ Instituto de F\'{\i}sica Te\'orica, (UAM/CSIC), Universidad Aut\'onoma de
Madrid, Cantoblanco, E-28049 Madrid, Spain\\
$^2$ Campus of International Excellence UAM/CSIC, Cantoblanco, E-28049 Madrid, Spain\\
$^3$ Center for Astrophysics, Harvard University, 60 Garden Street, Cambridge, MA 02138, USA \\
$^4$ Instituto de Astrof\'isica de Andaluc\'ia (CSIC), Granada, E-18008, Spain \\
$^5$ Department of Astronomy and Astrophysics, The Pennsylvania State University, University Park, PA 16802 \\
$^6$ Institute for Gravitation and the Cosmos, The Pennsylvania State University, University Park, PA 16802
}
\date{ }
\begin{document}
\bibliographystyle{mnras}
\maketitle
\begin{abstract}

\noindent We explore the bluer star-forming population of the Sloan Digital Sky Survey (SDSS) III/BOSS CMASS DR11 galaxies at $z>0.55$ 
 to quantify their differences, in terms of redshift-space distortions and large-scale bias, with respect to the luminous 
red galaxy sample. We perform a qualitative analysis to understand the significance of these differences and whether we can model and reproduce them in mock catalogs. Specifically, we measure galaxy clustering in CMASS on small and intermediate scales 
($r\lesssim50 \; \hmpc$) by computing the two-point correlation function \textemdash\; both projected and 
redshift-space \textemdash\; of these galaxies, and a new statistic, $\Sigma(\pi)$, 
able to provide robust information about redshift-space distortions and large-scale bias. 
We interpret our clustering measurements by adopting a Halo Occupation Distribution (HOD) scheme that maps them onto 
high-resolution N-body cosmological simulations to produce suitable mock galaxy catalogs.
The traditional HOD prescription can be applied to the red and the blue samples, independently, but this approach is unphysical since it allows 
the same mock galaxies to be either red or blue. To overcome this failure, we modify the standard formulation and infer the red
 and the blue mock catalogs directly from the full one, so that they are complementary and non-overlapping. This separation is performed 
 by matching the observed CMASS red and blue galaxy fractions and produces reliable and accurate models.
\end{abstract}


\section{Introduction}
\label{sec:intro}
In the last decade, an enormous effort has been spent to explore the formation and evolution of the 
large scale structure of our Universe. The standard cold dark matter ($\Lambda$CDM) model with 
cosmological constant, together with the theory of cosmic inflation, has become the leading 
theoretical picture in which structures can form, providing a clear prediction for their initial 
conditions and hierarchical growth through gravitational instability \citep[e.g., ][]{Primack1997}. Testing this model requires 
one to combine large N-body simulations with measurements from last generation large-volume 
photometric and spectroscopic galaxy surveys, as the Sloan Digital Sky Survey (SDSS),
 \citep[]{York2000, Gunn2006, Smee2013} and the SDSS-III Baryon Oscillation Spectroscopic Survey \citep[BOSS;][]{Eisenstein2011, Dawson2013}.
   In particular, BOSS has been able to measure the Baryon Acoustic Oscillation
  (BAO) feature in the clustering of galaxies and Lyman-$\alpha$ forest with unprecedented accuracy, 
  by collecting spectra of 1.5 million galaxies up to z=0.7  \citep[]{Anderson2014}, over a 10,000 $\deg^2$ 
  area of sky, and about 160,000 Lyman-$\alpha$ forest spectra of quasars in the redshift range $2.2<z<3$ 
  \citep[]{Slosar2011}. 

The $\Lambda$CDM paradigm claims that galaxies form at the center of dark matter halos, 
thus estimating the clustering features of such complex structures, is currently one of the main targets 
of modern cosmology \citep[]{Kravtsov2012}. Despite the recent dramatic improvement in the observational data,  
what primarily prevents us from achieving this goal immediately is the theoretical uncertainty of galaxy bias 
i.e., the difference between the distribution of galaxies and that of the matter. Galaxies are treated 
as biased tracers of the underlying matter distribution, and observations of their clustering properties are 
used to infer those cosmological parameters that govern the matter content of the Universe. In this context, 
the Halo Occupation Distribution \citep[HOD; ][]{Berlind2002, 
Kravtsov2004, Zheng2005, Zheng2007} framework has emerged as a powerful tool to bridge the gap between 
galaxies and dark matter halos, providing a theoretical framework able to  characterize their mutual relation 
in terms of the probability, $P(N|M)$, that a halo of virial mass $M$ contains $N$ galaxies of a given type. 
At the same time, it provides a robust prediction of the relative spatial and velocity distributions of galaxies 
and dark matter within halos. In this approach, the use of large-volume N-body cosmological simulations 
is crucial to produce reliable maps of the dark matter sky distribution. 

In this work, we explore the red/blue color bimodality observed in the CMASS sample of BOSS DR11 \citep{Alam2015} galaxies. 
In order to quantify and model the differences between these two galaxy populations, we measure their 
clustering signal on small and intermediate scales, from $r\sim0.1 \; \hmpc$ up to $r\sim50 \; \hmpc$. 
We compute the two-point correlation function (2PCF) \textemdash\; both projected and in redshift-space \textemdash\; 
of the BOSS CMASS galaxies, and a new metric, 
$\Sigma(\pi)$, designed to extract information about the small-scale nonlinear redshift-space distortion effects. 
We then map our results to the MultiDark cosmological simulation \citep{Prada2011, Riebe2011} using an 
HOD approach \citep{Zheng2007, White2011}, to generate suitable mock galaxy catalogs. 
In this context, we investigate whether we can find an HOD parametrization able to model both the blue and red 
observed clustering amplitudes, with small variations in its parameters. 
As an alternative to HOD models, one can interpret clustering observations with an 
Halo Abundance Matching (HAM) prescription \citep[e.g., ][]{Trujillo2011, Nuza2013} 
with the advantage of avoiding free parameters, only assuming that more luminous 
galaxies are associated to more massive halos, monotonically, through their number 
densities. HAM is a straightforward technique that provides accurate predictions for clustering 
measurements; nevertheless, we choose to model our CMASS clustering measurements 
using a five-parameter HOD scheme because it is a general method, based on a halo mass parametrization, and 
does not require a specific luminosity (stellar mass) function \citep[]{MonteroDorta2014}
 to reproduce the observations.

Besides the traditional HOD approach, where each galaxy population has its own independent model defined 
by a different set of parameters, we test an alternative prescription, in which the red and the blue models are 
recovered by splitting the full mock catalog using suitable conditions to mimic the observed CMASS red and blue galaxy fractions, 
as a function of the central halo mass. In this way, the resulting mocks are no longer independent \textemdash\; they are based on 
the same HOD parameter set \textemdash\; and the total number of degrees of freedom is reduced from 15 (three independent models, 
with five parameters each) to 5 (full HOD) plus 2 (galaxy fraction constraint). The main motivation of this new 
approach is that the classical HOD parametrization reproduces well the full CMASS population, and it provides non-physical predictions 
when applied to the red and blue sub-samples, independently. In fact, in the process of populating a halo with
 central and satellite galaxies, this kind of modeling allows the same galaxy to be either red or blue i.e., to be placed in 
 halos with different masses. To overcome this problem, we adopt a new HOD formulation, in which the red/blue split observed 
 in our data sample is used as a discriminant condition to perform an univocal galaxy assignment. 

We investigate the impact of redshift-space distortions on the clustering signal, both on small ($1$-halo term) 
and intermediate ($2$-halo level) scales. Our new metrics, $\Sigma(\pi)$, allows us to separate and quantify both 
the nonlinear elongation seen in the two-point correlation function below $2 \; \hmpc$, and the Kaiser compression 
at scales beyond $10 \; \hmpc$. We model these effects in terms of two parameters, $A$ and $G$, respectively 
encoding the galaxy velocity dispersion with respect to the surrounding Hubble flow, and the linear large-scale bias. 
In agreement with several previous works \citep[see, for instance,][]{Wang2007, Zehavi2005b, Swanson2008}, we find that red galaxies 
are more clustered (i.e. higher peculiar velocity contribution) and biased, compared to their blue star-forming companions. 

The paper is organized as follows. In Section\;\ref{sec:Methods} we introduce the methodology used to measure and 
model galaxy clustering in the BOSS CMASS DR11 galaxy sample: we define the metrics we examine, the correlation function and the 
covariance estimators. We then provide an overview of the MultiDark simulation, we discuss 
the HOD formalism adopted to create mock galaxy catalogs, and introduce the analytic tools 
used to model both finger-of-god and Kaiser effects. In Section\;\ref{sec:Data}, we present the 
CMASS DR11 sample and the specific red/blue color selection used in our analysis, we illustrate how to weight 
the data to account for fiber collision and redshift failure effects, and outline the procedure adopted to generate randoms.
In Section\;\ref{sec:Modeling_full_CMASS_Sample}, we describe how we model our full CMASS clustering measurements 
building reliable mock galaxy catalogs that take into account the contribution of redshift-space distortions, and present 
the first results for the three metrics of interest: $\xi(s)$, $w_p(r_p)$, $\Sigma(\pi)$. In Section  \ref{sec:Modeling_Color}, 
we first apply the same procedure individually to the red and blue CMASS galaxy sub-samples to create their own independent mock 
catalogs; then, we propose an alternative 
method to separate the red and blue populations using, as a constraint, the observed CMASS red/blue galaxy fractions. 
Our data versus mock $\Sigma(\pi)$ results, compared to the $A$, $G$ analytic models are shown 
in Section \ref{sec:results}. Section\;\ref{sec:discussion} reports our main conclusions.

\section{Methods}
\label{sec:Methods}


\subsection{Clustering Measurements}
\label{sec:Clustering_Measurements}

We quantify the clustering of galaxies by computing the two-point correlation function i.e., the 
excess probability over random to find a pair of galaxies typically parameterized as a function of 
their co-moving separation \cite[see, e.g., ][]{Peebles1980}.  The galaxy correlation function is well 
known to approximate a power-law across a wide range of scales, 
\begin{equation} \label{eq:powerlaw}
  \xi(r)=\left(\frac{r}{r_0}\right)^{-\gamma} \; ,
\end{equation}
where $r_0$ is the correlation length, and $\gamma$ is the power-law slope or spectral index.  
However, improved models \citep[see review at][]{Cooray2002} have been shown to better match the 
data \citep{Zehavi2004}.

The redshift-space correlation function differs from the real-space one due to the distortion 
effects caused by our inability to separate the peculiar velocities of galaxies from their recession 
velocity when we estimate distances from the redshift. These distortions introduce anisotropies in 
the 2PCF in two different ways.  On large scales, where the linear regime 
holds, galaxies experience a slow infall toward an over-dense region, and the peculiar velocities 
make structures appear squashed in the line-of-sight direction, an effect commonly known as ``Kaiser 
compression'' \citep{Kaiser1987, Hamilton1998}.  At smaller scales, nonlinear gravitational collapse 
creates virialized systems and thereby relatively large velocity differences arise between close 
neighbors resulting in structures appearing significantly stretched along the line-of-sight 
\citep{Jackson1972}. This effect is commonly referred to as the ``finger-of-god''(FoG).  

We are interested in using three related two-point clustering metrics: the redshift-space monopole, 
$\xi(s)$, the projected correlation function, $w_p(r_p)$, and a new line-of-sight focused measurement to 
capture small-scale redshift-space distortion effects, $\Sigma(\pi)$, which we define below.
In our formalism, $s$ represents the redshift-space pair separation, while $r_p$ and $\pi$ are the 
perpendicular and parallel components with respect to the line-of-sight such that 
$s=\sqrt{r_p^2+\pi^2}$. We can parameterize the redshift-space correlation function as a function of 
redshift-space separation $s$ or, equivalently, in terms of $r_p$ and $\pi$. We can mitigate the impact of redshift-distortions 
by integrating along the line-of-sight to approximate real-space clustering \citep{Davis1983} in the 
projected correlation function, 
\begin{equation} \label{eq:projectedCF}
  w_p(r_p)=2\int_0^{\infty}\xi (r_p,\pi)\; d\pi \; .
\end{equation}
This integration is performed over a finite line-of-sight distance as a discrete sum,  
\begin{equation} \label{eq:discrete}
  w_p(r_p)=2\sum_{i}^{\pi_{max}} \xi (r_p,\pi) \; \Delta\pi_i \; , 
\end{equation}
where $\pi_i$ is the $i^{th}$ bin of the line-of-sight separation, and $\Delta \pi_i$ is the 
corresponding bin size. We use $\pi_{max}=80 \; \hmpc$ and $\Delta \pi = 10 \; \hmpc$. 

Since $w_p(r_p)$ is not affected by redshift-space distortions, the best fit power-law is equivalent 
to a real-space measurement. One can therefore quantify the deviation of the redshift-space 
$\xi(r_p,\pi)$ correlation function from the real-space behavior by measuring the ratio, 
\begin{equation} \label{eq:sigma_new_metrics}
  \Sigma(\pi)=\frac{\xi(\bar{r}_p,\pi)}{\xi(\pi)} \; , 
\end{equation}
where $\xi(\pi)$ is the best-fit power law to $w_p(r_p)$, evaluated at the $\pi$ scale, and 
$\bar{r}_p$ indicates that we perform a spherical average in the range 
0.5 $\leq r_p \leq$ 2 $\hmpc$. 
This statistic illuminates the nonlinear FoG effects by normalizing out the expected 
real-space clustering along the line-of-sight direction. It is therefore preferable to measuring the 
quadrupole-to-monopole ratio, $\xi_2(s)/\xi_0(s)$ \citep{Hamilton1992, Hamilton1998, Peacock2001},
in the attempt to interpret the small-scale nonlinear redshift-space clustering effects.


\subsection{Correlation Function Estimation}
\label{sec:CF_Estimation}

For our clustering statistics, we use the estimator of \cite{Landy1993}:
\begin{equation} \label{eq:ls}
  \xi(s)=\frac{DD(s)-2DR(s)+RR(s)}{RR(s)}
\end{equation}
where $DD$, $DR$ and $RR$ are the data-data, data-random and random-random weighted pair counts 
computed from a data sample of $N$ galaxies and a random catalog of $N_R$ points. These pair counts 
are normalized by the number of all possible pairs, typically by dividing by $N(N-1)/2$, $NN_R$ and 
$N_R(N_R-1)/2$, respectively, and weighted by \cite[]{Ross2012}
\begin{equation}
  DD(r_p,\pi)=\sum_i\sum_jw_{tot,i}w_{tot,j}\Theta_{ij}(r_p,\pi)
\end{equation}
with $w_{tot}$ given by Eq. \eqref{eq:totweight}, and $\Theta_{ij}(r_p,\pi)$ represents a 
step-function which is $1$ if $r_p$ belongs to the $i^{th}$ and $\pi$ to the $j^{th}$ bin, and $0$ 
otherwise.  These weights correct the galaxy densities to provide a more isotropic selection, 
therefore they should not be applied to the random catalog, which is based on an isotropic 
distribution. For randoms $w_{tot,i}=w_{tot,j}=1$, therefore
\begin{align}
  DR(r_p,\pi) &=\sum_i\sum_jw_{tot,i}\Theta_{ij}(r_p,\pi) \; ,\\
  RR(r_p,\pi) &=\sum_i\sum_j\Theta_{ij}(r_p,\pi) \; .
\end{align}

To evaluate the correlation function, we create a random catalog that has the same selection as the 
BOSS CMASS galaxy data matching both the redshift distribution and sky footprint \citep[see, e.g., ][]{Anderson2014}. 
The method of random catalog construction is almost identical to that described in \citet{Anderson2014}, 
but constructed to be ten times as dense as the galaxy data. We down-sample random points based on sky 
completeness, and ``shuffle'' the observed galaxy redshifts assigning them to random sky positions so 
as to exactly reproduce the observed redshift distribution.


\subsection{Covariance Estimation}
\label{sec:Covariance_Estimation}

To estimate the uncertainties in our clustering measurements, we utilize the jackknife re-sampling 
technique \citep[]{Quenouille1956, Turkey1958, Miller1974, Norberg2009, Norberg2011}. 
There are known limitations to this type of error estimation 
\citep[see, e.g., ][]{Norberg2009}, but they have proven sufficient in analyses on scales similar to 
our analysis \citep{Zehavi2002,Zehavi2005,Zehavi2011,Guo2012, Ross2012, Anderson2012}.
The jackknife covariance matrix for $N_{res}$ re-samplings is computed by 
\begin{equation} 
  C_{ij}=\frac{N_{res}-1}{N_{res}}\sum_{a=1}^{N_{res}} (\xi^a_{i}-\bar{\xi}_{i})(\xi^a_{j}-\bar{\xi}_{j}),
  \label{eq:covma}
\end{equation}
where $\bar{\xi}_{i}$ is the mean jackknife correlation function estimate in the specific $i^{th}$ bin,
\begin{equation} 
  \bar{\xi}_i=\sum_{a=1}^{N_{res}} \xi^a_{i}/N_{res}.
\end{equation}
The overall factor in Eq.\;\ref{eq:covma} takes into account the lack of independence
between the $N_{res}$ jackknife configurations: from one copy to the next, only two sub-volumes are different 
or, equivalently, $N_{res}-2$ sub-volumes are the same \citep{Norberg2011}.


\subsection{The MultiDark Simulation}
\label{sec:MD_Simulation}

MultiDark \citep{Prada2011} is a N-body cosmological simulation with $2048^3$ dark matter 
particles in  a periodic box of $L_{box}=1$ Gpc h$^{-1}$ on a side. The first run, MDR1, was 
performed in 2010, with an initial redshift of $z=65$, and a mass resolution of 
$8.721\times10^9$ $h^{-1}$M$_{\odot}$. 
It is based on the WMAP5 cosmology \citep[]{Komatsu2009}, with parameters: 
$\Omega_m=0.27$, $\Omega_b=0.0469$, $\Omega_{\Lambda}=0.73$, $n_s=0.95$ and $\sigma_8=0.82$. 
Here $\Omega$ is the present day contribution of each component to the matter-energy density of the 
Universe; $n_s$ is the spectral index of the primordial density fluctuations, and $\sigma_8$ is the 
linear RMS mass fluctuation in spheres of $8 \;\hmpc$ at $z=0$.

MultiDark includes both the Bound Density Maxima \citep[BDM; ][]{Klypin1997,Riebe2011}, 
and the Friends-of-Friends \citep[FOF; ][]{Davis1985} halo-finders. For the current analysis, 
we use only BMD halos that are identified as local density maxima truncated at some spherical 
cut-off radius, from which unbound particles (i.e., those particles whose velocity exceeds the escape 
velocity) are removed. According to the overdensity limit adopted, two different BDM halo catalogs 
are produced:  
(i) BDMV \textemdash\; halos extend up to $\Delta_{vir}\times\rho_{back}$, where $\Delta_{vir}=360$ is the 
virial overdensity threshold, $\rho_{back}=\Omega_m\times\rho_c$ is the background or average 
matter density, and $\rho_{c}$ is the critical density of the Universe. 
(ii) BDMW \textemdash\; the maximum halo density is $\Delta_{200}\times\rho_{c}$, where $\Delta_{200}=200$, which implies 
that BDMW halos are smaller than BDMV ones.
The bound density maxima algorithm treats halos and sub-halos (those sub-structures whose virial 
radius lies inside a larger halo) in the same way, with no distinction. In this work we use the BDMW 
halo catalogs, since they resolve better the distribution of sub-structures in distinct halos, leading to a clearer small-scale clustering signal.


\subsection{Halo Occupation Distribution Model using Subhalos}
\label{sec:HOD_general}
The halo model \citep[reviewed in][]{Cooray2002} is a powerful tool to understand the clustering of 
galaxies.  The Halo Occupation Distribution \citep[HOD; ][]{Berlind2002} is a commonly used method 
of mapping galaxies to dark matter halos, which characterizes the bias between galaxies and the 
underlying dark matter distribution.  The HOD is based on the conditional probability, $P(N|M)$, that 
a halo with mass $M$ contains $N$ galaxies of a given type.  In our analysis, we apply the 
five-parameter HOD formalism presented in \citet{Zheng2007} using the MDR1 simulation at $z=0.53$. 
First, we populate distinct halos with central galaxies whose mean is given by the function form of:
\begin{equation} \label{eq:Ncen}
  \langle \Ncen(M)\rangle = 
  \frac{1}{2}\left[1+\erf {\left(\frac{\log M-\log \Mmin}{\sigLM}\right)}\right] \; , 
\end{equation}
where $\erf{}$ is the error function, $\erf{(x)}=2\int^x_0 e^{-t^2} dt/\sqrt\pi$.

The free parameters are $\Mmin$, the minimum mass scale of halos that can host a central galaxy, and 
$\sigLM$, the width of the cutoff profile. At a halo mass of $\Mmin$, $50$\% of halos host a central 
galaxy, which in terms of probability means that $P(1)=1-P(0)$. If the relation between galaxy 
luminosity and halo mass had no scatter, $\langle \Ncen(M) \rangle$ would be modeled by a hard step 
function. In reality, this relation must possess some scatter, resulting in a gradual transition from 
$\Ncen \simeq 0$ to $\Ncen \simeq 1$. The width of this transition is $\sigLM$. In order to place the 
satellite galaxies, we assume their number in halos of a given mass follows a Poisson distribution, 
which is consistent with theoretical predictions \citep{Berlind2002, Kravtsov2004, Zheng2005}.
We approximate the mean number of satellite galaxies per halo with a power law truncated at a 
threshold mass of $\Mz$
\begin{equation} \label{eq:Nsat}
\langle \Nsat \rangle=\langle \Ncen(M) \rangle\left(\frac{M-\Mz}{\Mop}\right)^{\alpha'}.
\end{equation}

\begin{figure}
  \begin{center}
    \includegraphics[width=1.05\linewidth]{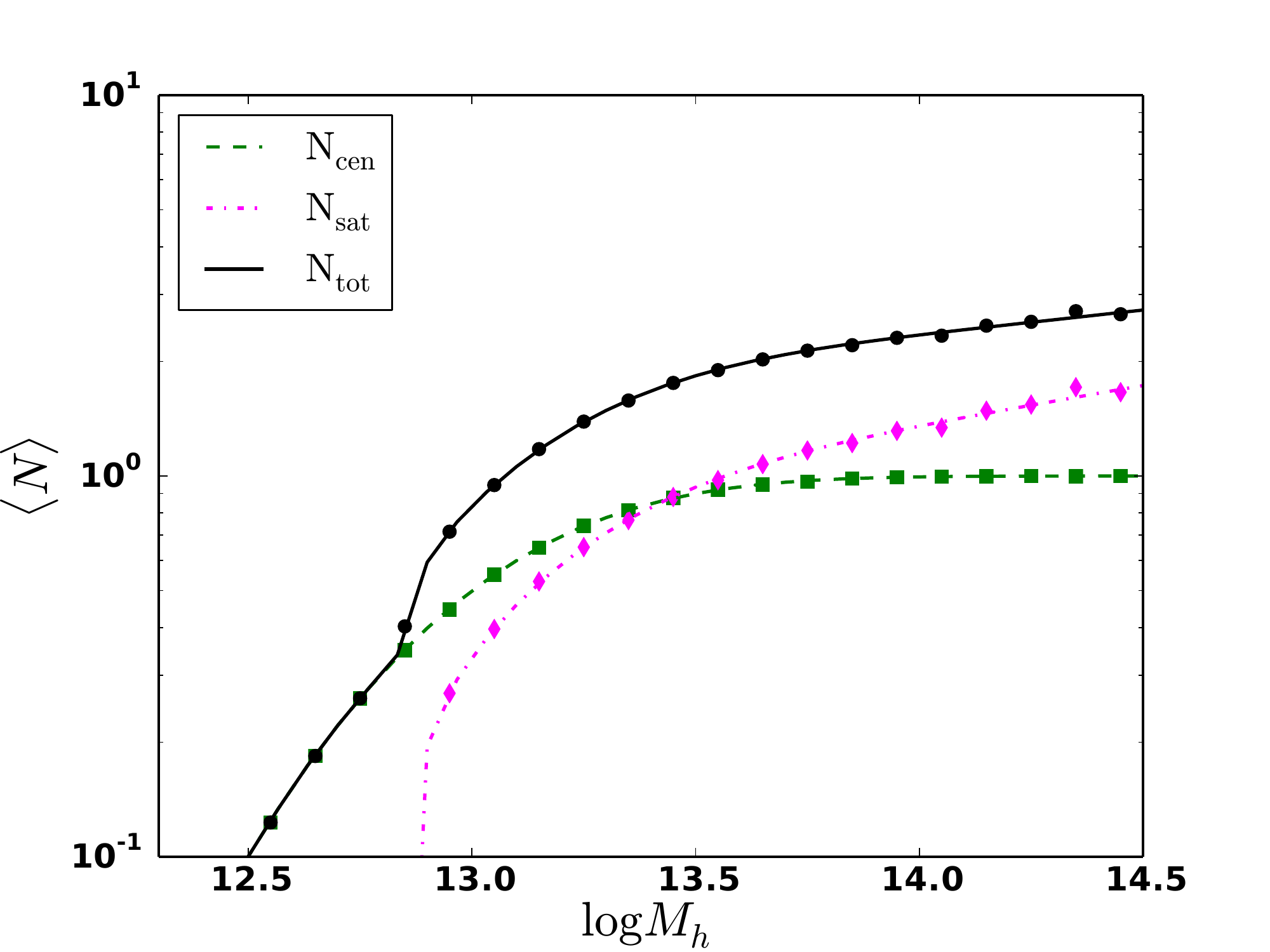}
    \caption{Five-parameter Halo Occupation Distribution model for MDR1, at $z=0.53$. The parametrization 
    is from \citeauthor{Zheng2007} (\citeyear{Zheng2007}), and the input values from 
    \citeauthor{White2011} (\citeyear{White2011}). The total (solid line) population of galaxies is 
    the sum of two contributions: central (dashed) and satellite (dot-dashed) galaxies.}
    \label{fig:HOD}
  \end{center}
\end{figure}

The parameter $\Mop$ corresponds to the halo mass where $\Nsat \simeq 1$, when (as in our case) 
$\Mop > \Mz$ and $\Mop > \Mmin$. When $\alpha'=1$ and $M > \Mz$, the mean number of satellites per 
halo is proportional to the halo mass. 
To populate with satellite galaxies, we randomly extract from each host halo a certain number of its 
sub-halos, following a Poisson distribution with mean given by Eq.~\ref{eq:Nsat}. The coordinates 
of these sub-halos become the locations for satellites. This approach, explored in previous 
works as \cite{Kravtsov2004}, \cite{White2011}, is intrinsically different from the more commonly used procedure, 
in which satellites are assigned by randomly assigning the positions of dark-matter particles \citep[see, e.g., ][]{Reid2009}.
In our case, satellite galaxies are assigned by reflecting the original halo structure 
made of one central halo plus none, one, or many sub-halos. 

Figure~\ref{fig:HOD} shows our HOD model built from MultiDark BDMW at $z=0.53$, for the full 
CMASS sample: central galaxies are represented by the dashed curve; satellites are the 
dot-dashed line and the total contribution is the solid curve. As input parameters, 
we adopt the values consistent with the BOSS CMASS HOD modeling in \citet{White2011}.
 

\subsection{Analytic models}
\label{sec:Analytic_Models}

\citet{Kaiser1987} demonstrated that on large scales, where the linear regime holds, the redshift-space correlation 
function can be factorized in terms of its real space version, $\xi(r)$, as
\begin{equation}
  \xi(s)=\xi(r)\left(1+\frac{2}{5}\beta+\frac{1}{5}\beta^2\right),
  \label{eq:kaiser_simple}
\end{equation}
where $\beta$ is the Kaiser factor encoding  the compression effect (Sec. \ref{sec:Clustering_Measurements}) seen in the clustering signal and $b$ is the linear bias between galaxies and the 
underlying matter distribution. These two quantities can be related \citep[e.g., ][]{Peebles1980} through the following approximation: 
\begin{equation}
\beta\simeq\Omega_m^{0.6}/b.
\label{eq:bias_approx}
\end{equation} 

In general, one can decompose the redshift-space separation $s$ into its 
parallel and transverse components to the line-of-sight and approximate $\xi(r)$ with the power law in Eq. 
\ref{eq:powerlaw} to produce \citep{Matsubara1996}: 
\begin{equation}
  \begin{aligned}
    \xi(r_p,\pi)=
    \xi(r)\left\{1+\frac{2(1-\gamma\mu^2)}{3-\gamma}\beta+\right.\hspace{2cm}\\
    \hspace{2cm}\left.\frac{3-6\gamma\mu^2+\gamma(2+\gamma)\mu^4}{(3-\gamma)(5-\gamma)}\beta^2\right\}.
    \label{eq:kaiser_fac}
  \end{aligned}
\end{equation} 

Here $\gamma$ is the power law spectral index and $\mu$ is the cosine of the angle between the separation 
and the line-of-sight direction.
We include the small-scale nonlinear FoG by convolving with a pairwise velocity distribution 
\citep{Fisher1994,Hamilton1998,Croom2005}, which can be modeled as an exponential, 
\begin{equation} \label{eq:vpec}
 f_{exp}(w)=\frac{1}{\sqrt{2}\alpha}\exp\left(-\sqrt{2}\frac{|w|}{\alpha}\right) \; , 
\end{equation}
or a Gaussian form, 
\begin{equation} \label{eq:vpec_norm}
 f_{norm}(w)=\frac{1}{\sqrt{2\pi}\alpha}\exp\left(-\frac{w^2}{2\alpha^2}\right) \;,
\end{equation}
\begin{figure}
\begin{center}
\includegraphics[width=\linewidth]{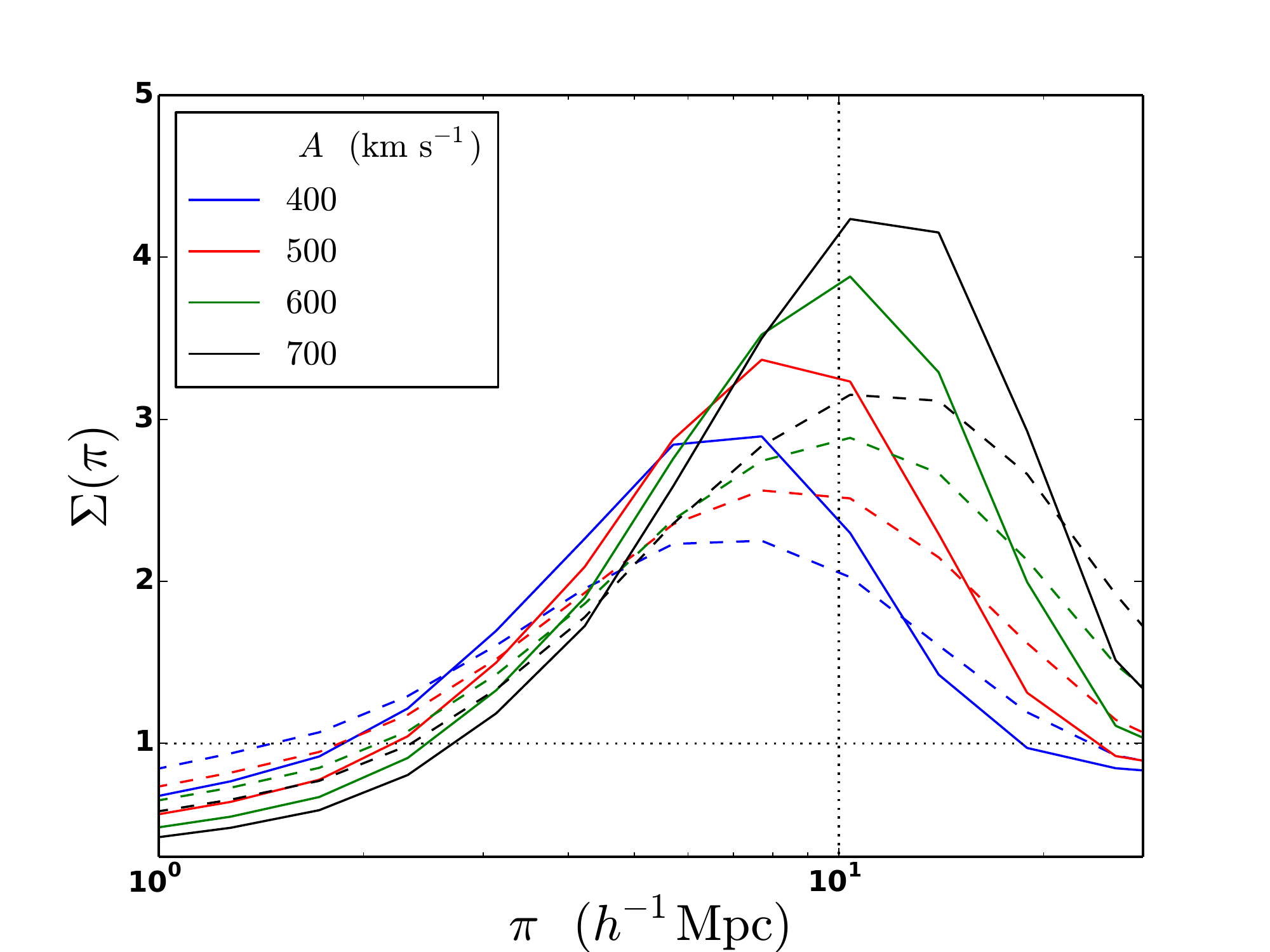}\hfill
\includegraphics[width=\linewidth]{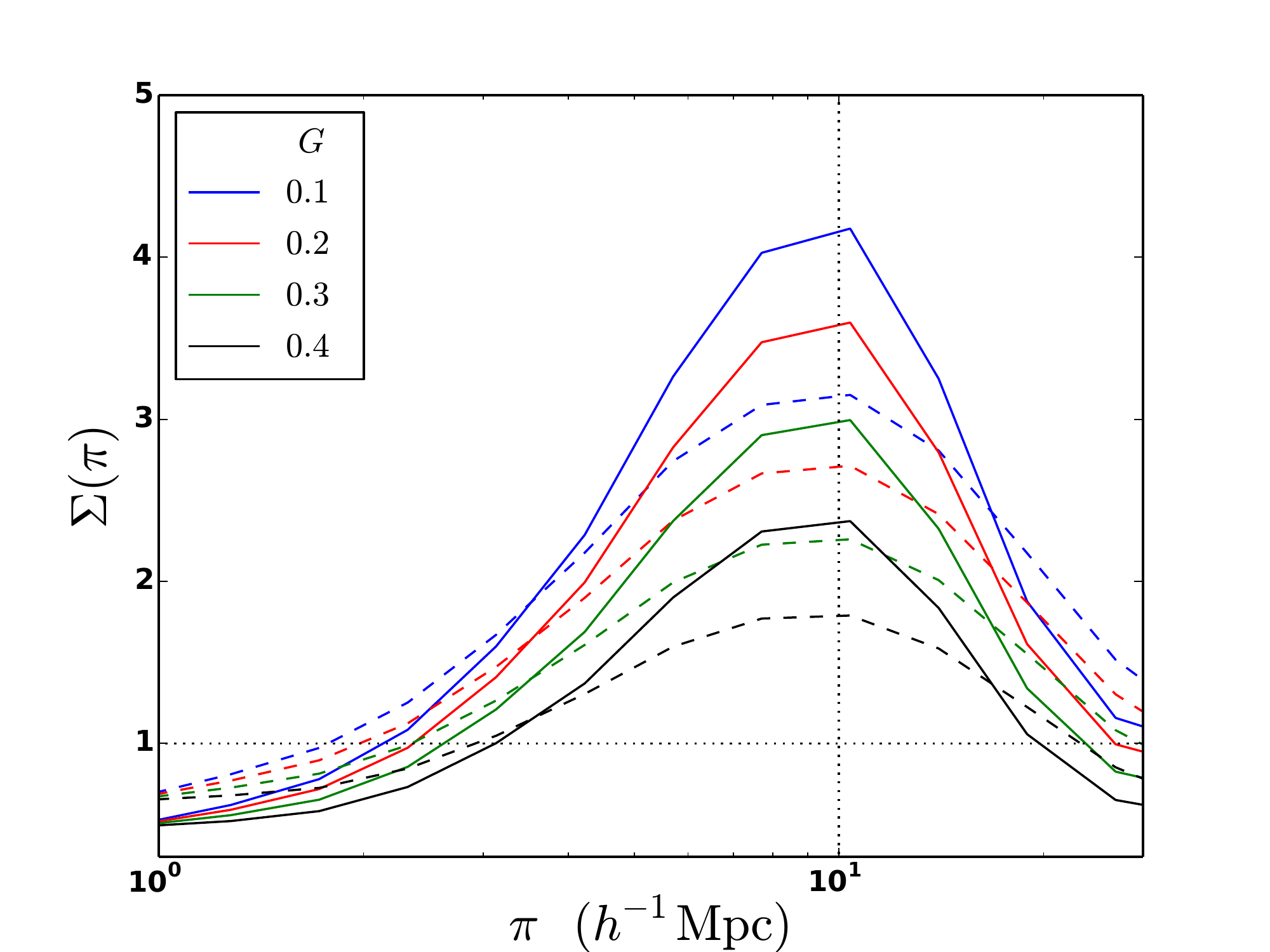}\hfill
\end{center}
 \caption{$\Sigma(\pi)$ analytic model as a function of the pairwise velocity dispersion, $A$, (top panel) and the 
 parameter $G$, encoding the Kaiser factor (bottom panel). Solid lines represent the Gaussian model given 
 in Eq.\;\ref{eq:vpec_norm}; dashed curves are the exponential functions in Eq.\;\ref{eq:vpec}. We 
choose to model our $\Sigma(\pi)$ measurements using the normal functional form only, since it reproduces more accurately 
the small-scale feature provoqued by the FoG distortions and peak at larger scales.}
\label{fig:sigma_reticolo}
\end{figure}
where $\alpha$ is the pairwise velocity dispersion. The full model then becomes
\begin{equation} \label{eq:modello_analitico}
  \xi(r_p,\pi)=\int_{-\infty}^{+\infty}\xi(r_p,r_z(w))f(w)dw \;, 
\end{equation} 
with $\xi(r_p,r_z(w))$ given by Equation\;\ref{eq:kaiser_fac}. The quantity 
$r_z(w)\equiv (\pi-w)/(aH(z))$ is the line-of-sight component of the real-space distance $r$, $a=(1+z)^{-1}$ is the scale factor, 
and $H(z)$ is the Hubble parameter evaluated at redshift $z$. The full $\Sigma(\pi)$ analytic model, as a function 
of $\alpha$ and $\beta$, is obtained by averaging Eq.\;\ref{eq:modello_analitico} in the range 
0.5 $\leq r_p \leq$ 2 $\hmpc$ and integrating the result in $\pi$ bins, as explained in Section 
\ref{sec:Clustering_Measurements}. 

Combining these definitions and matching the binning in $\Delta r_p $ and $\Delta \pi$, we have:
\begin{equation}
  \Sigma(\pi) = 
\frac{ 
    \int\frac{dZ}{\Delta \pi}\int \frac{dR}{\Delta r_p}\int\xi\left(R,\frac{Z-w}{aH(z)}\right) f(w)dw
  } {\large{ 
    \int\frac{dZ}{\Delta \pi}\int \frac{dR}{\Delta r_p} \left( \frac{r_0^2}{R^2 + Z^2} \right)^{\gamma/2}
  }}\; 
  \label{eq:modello_analitico_sigma}
\end{equation}
Finally, we rename the parameters $\alpha$ and $\beta$ respectively $A$ and $G$ to emphasize they are fitted 
parameters that might differ slightly from their theoretically motivated meaning. In this formalism, Eq. \ref{eq:bias_approx} 
simply becomes

\begin{equation}
G\simeq\Omega_m^{0.6}/b.
\label{eq:bias_approx_AG}
\end{equation}

 The FoG and Kaiser effects could be overlapping and, as fit parameters in a model, they are correlated.  The importance of our modeling 
is not to isolate their value, but to differentiate between models and data with sub-populations of galaxies.  
Figure\;\ref{fig:sigma_reticolo} shows how both effect contribute to modulate our $\Sigma(\pi)$ model. There is a
degeneracy between the parameter values, in the sense that both increasing $A$ or reducing $G$ produces an enhancement in the $\Sigma(\pi)$ peak. This dependence prevents us from interpreting the $G$ parameter as 
the only one responsible of the $\Sigma(\pi)$ amplitude.


\subsection{Fitting $w_p(r_p)$}
\label{sec:Fitting_wp}

\begin{figure}
  \begin{center}
    \includegraphics[width=\linewidth]{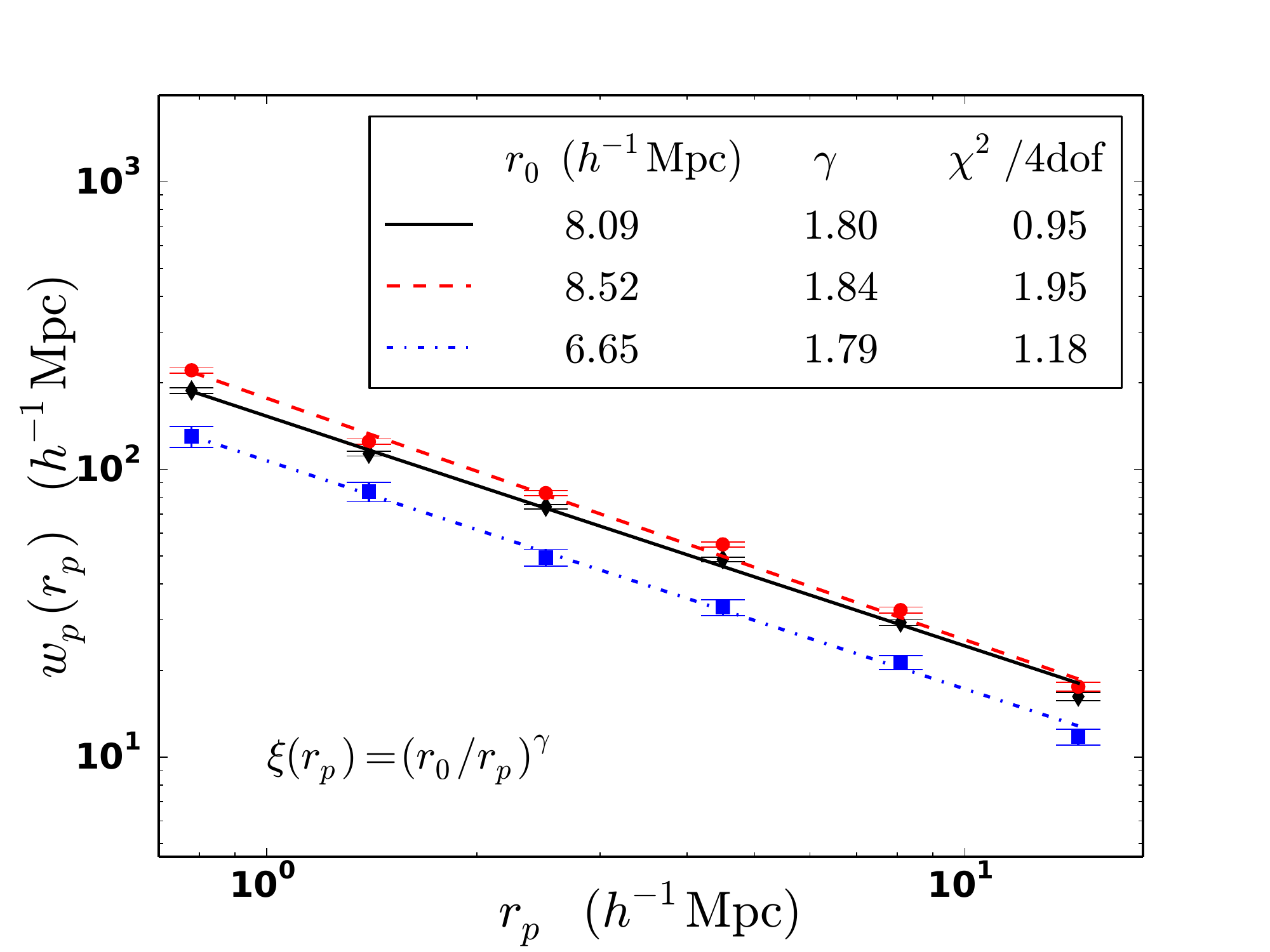}
    \caption{Power-law fits to the CMASS full, red and blue projected correlation functions, 
which define the denominator in Eq. \ref{eq:modello_analitico_sigma}. The $r_0$ and $\gamma$ values 
we find are consistent with \citeauthor{Zehavi2005} \citeyear{Zehavi2005}, and show that red galaxies cluster more than blue star-forming 
ones. The error bars correspond the $1\sigma$ uncertainties 
    estimated using 200 jackknife resamplings (Sec.\;\ref{sec:Covariance_Estimation}).}
    \label{fig:PL}   
  \end{center}
\end{figure}

To implement the integral in Eq.\;\ref{eq:projectedCF}, to estimate the projected correlation function $w_p(r_p)$, 
we need to truncate it at some upper value, $\pi_{max}$, 
above which the contribution to correlation function becomes negligible.
If one includes very large scales, the measurement will be affected by noise; inversely, if we consider only very 
small scales, the clustering amplitude will be underestimated. In our case,  CMASS results are not sensitive to $\pi\ge80$ $h^{-1}$Mpc, 
therefore we adopt this value as our $\pi_{max}$ limit. The projected auto-correlation function is related to the real-space one
 by (\citeauthor{Davis1983} \citeyear{Davis1983})
\begin{equation}
  w_p(r_p)=2\int^{\pi_{max}}_{r_p}\frac{r\xi(r)}{\sqrt{r^2-r_p^2}}dr \; .
\end{equation}
\citet{Zehavi2005b} demonstrates that for a generic power law, $\xi(r)=(r/r_0)^{\gamma}$, the equation above can be written
 in terms of the Euler's Gamma function as 
\begin{equation}
w_p(r_p)=r_p\left(\frac{r_p}{r_0}\right)^{\gamma}\Gamma\left(\frac{1}{2}\right)\Gamma\left(\frac{\gamma-1}{2}\right) /\Gamma\left(\frac{\gamma}{2}\right).
\label{eq:projection}
\end{equation}
allowing one to infer the best-fit power law for $\xi(r)$ from $w_p(r_p)$,
corresponding to the full CMASS galaxy sample, blue and red sub-samples. 
Figure \ref{fig:PL} presents the power-law fits to the full, red and blue CMASS projected correlation functions, 
and the resulting $(r_0,\gamma)$ optimal values.


\section{BOSS CMASS Data}
\label{sec:Data}

\indent BOSS target galaxies primarily lie within two main samples: CMASS, with $0.43<z<0.7$ and LOWZ, with $z<0.43$ \citep[]{Ross2012, Anderson2012, Bolton2012}. 
These samples are selected on the basis of photometric observations done with the dedicated 2.5-m Sloan Telescope \citep{Gunn2006}, located at Apache Point Observatory in New Mexico, using a drift-scanning mosaic CCD camera with five color-bands, $ugriz$ \citep[][]{Gunn1998, Fuku1996}. Spectra of the LOWZ and CMASS samples are obtained using the double-armed BOSS spectrographs, which are significantly upgraded from those used by SDSS-I/II, covering the wavelength range $3600-10000{\buildrel _{\circ} \over {\mathrm{A}}}$ with a resolving power of 1500 to 2600 \citep[]{Smee2013}. Spectroscopic redshifts are then measured using the minimum-$\chi^2$ template-fitting procedure described in \cite{Aihara2011}, with templates and methods updated for BOSS data as described in \cite{Bolton2012}.

We select galaxies from CMASS DR11 \citep[]{Alam2015} -- North plus South Galactic caps -- which is defined by a series of color cuts designed to obtain a galaxy sample with approximately constant stellar mass. Specifically, these cuts are:
\begin{equation}
         17.5<i_{cmod}<19.9
\end{equation} 
\begin{equation}
          r_{mod}-i_{mod}<2
\end{equation} 
\begin{equation}
         d_{\perp}>0.55
\end{equation}
\begin{equation}
         i_{fib2}<21.5
\end{equation} 
\begin{equation}
          i_{cmod}<19.86+1.6(d_{\perp}-0.8),
\end{equation}
where $i_{cmod}$ is the $i-$band cmodel magnitude. The quantities $i_{mod}$ and $r_{mod}$ are model magnitudes, $i_{fib2}$ is the $i-$band magnitude within a 2$"$ aperture and $d_{\perp}$ is defined as
\begin{equation}
d_{\perp}=r_{mod}-i_{mod}-(g_{mod}-r_{mod})/8.0.
\end{equation}
All the magnitudes are corrected for Galactic extinction using the dust maps from \cite{Schlegel1998}. In addition to the above color cuts, CMASS objects must also pass two star-galaxy separation constraints:
\begin{equation}
i_{psf}-i_{mod}>0.2+0.2(20.0-i_{mod})
\end{equation}
\begin{equation}
z_{psf}-z_{mod}>9.125-0.46z_{mod},
\end{equation}
unless the objects also pass the LOWZ criteria. Therefore, to distinguish CMASS from LOWZ candidates, it is necessary to select them by redshift. 


\subsection{Color Selection}
\label{sec:color_selection}
\begin{figure}
 \begin{center}
 \includegraphics[width=0.8\linewidth]{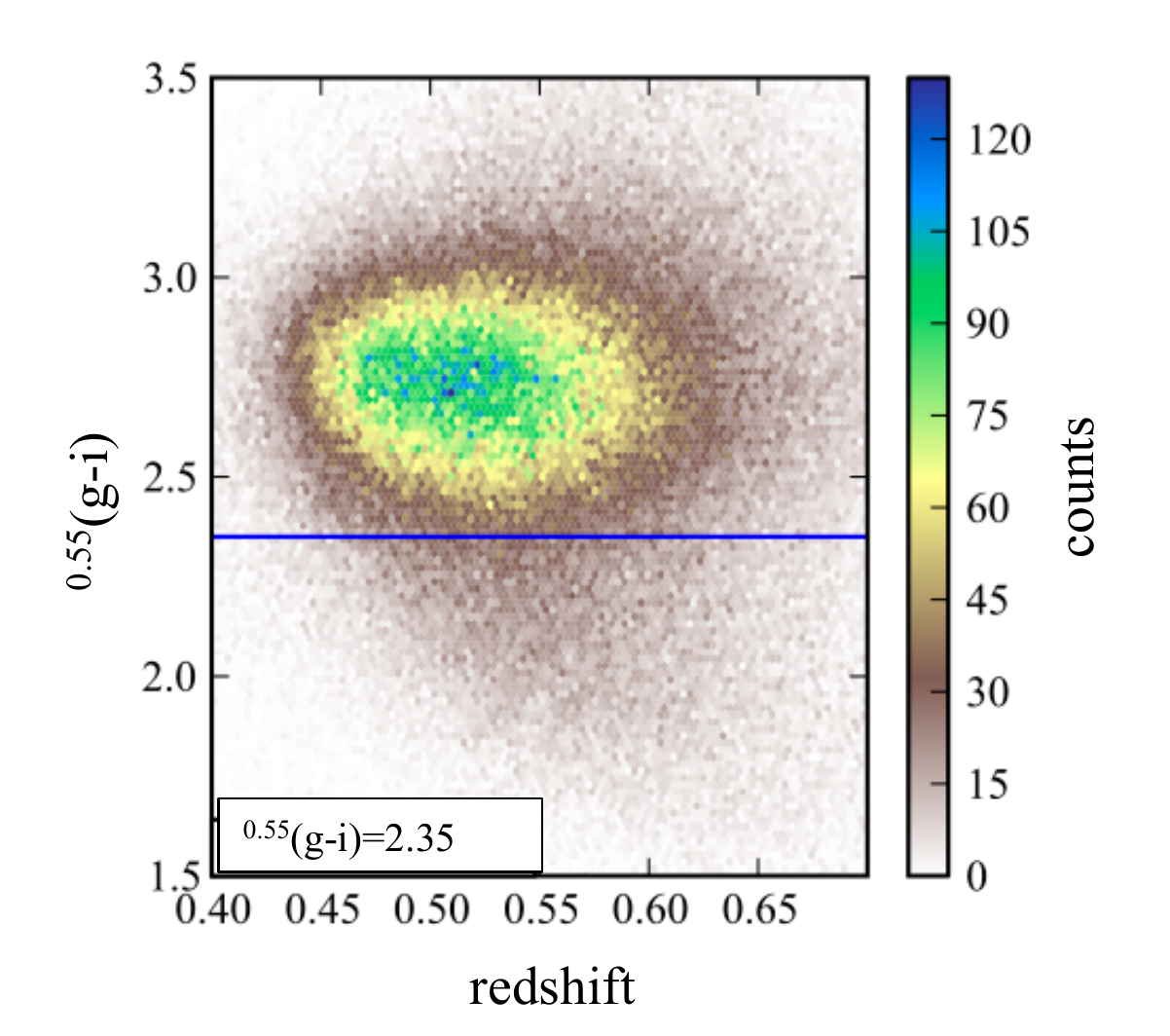} 
 \caption{BOSS CMASS DR11 color selection:  the $(g-i)$ color cut divides the full sample into a red 
dense population (above the blue horizontal line) and a sparse blue tail (below the line).}
\label{fig:CMASS_k0}
\end{center}
\end{figure}

The CMASS sample is mainly composed of massive, luminous, red galaxies, which are favorite subjects to study galaxy clustering. Among them, 
however, there is an intrinsic bluer, star-forming population of massive galaxies (\citeauthor{Ross2012} \citeyear{Ross2012}; \citeauthor{Guo2012} 
\citeyear{Guo2012}), of which little is known. In the attempt to explore this bluer component to 
understand its contribution in the clustering properties,
we split the CMASS sample into its blue and red components by applying the color cut 
\begin{equation}
^{0.55}(g-i)=2.35
\label{eq:colorcut}
\end{equation}
constant in redshift and
$K$-corrected to the $z=0.55$ rest-frame using the code by
 \cite{Blanton2007}. \cite{Masters2011} applied this same color cut, with no $K$-corrections, to the BOSS CMASS DR8 sample to study the morphology of the 
LRG population; \cite{Ross2014} used a similar selection, $^{0.55}(r-i)=0.95$, to measure galaxy clustering at the BAO scale in CMASS DR10.
Figure~\ref{fig:CMASS_k0} presents our CMASS color selection, splitting the full sample into a red denser population (above the blue horizontal line) and a 
sparse blue tail (below the line), whose completeness dramatically increases when we move towards
 high redshift values ($z>0.55$). For our analysis, we focus on the high-redshift tail of the CMASS sample, selecting only galaxies with redshift beyond $z>0.55$.


\subsection{Weights}
\label{sec:weights}
Due to its structural features, a survey inevitably introduces some kind of spatial variation in its measurements. 
To avoid these distortions, we weight our pair counts by defining a linear combination of four different weights \citep{Anderson2012, Sanchez2012, Ross2012}:
\begin{equation}
w_{tot}=w_{FKP}\hspace{0.1cm}w_{sys}(w_{fc}+w_{zf}-1),
\label{eq:totweight}
\end{equation}
 each one correcting for a different effect. In the expression above, $w_{zf}$ accounts for targets with missing or corrupted redshift ($z$ failure); 
 $w_{fc}$ corrects for fiber collision, compensating the fact that fibers cannot be placed closer than 62$"$ on the survey plates. 
 This limitation prevents obtaining spectra of all galaxies with neighbors closer than this angular distance in a single observation. 
 The default value of $w_{zf}$ and $w_{fc}$ is set to unity for all galaxies. When a fiber collision is detected, we increment by one the 
 value of $w_{fc}$ for the first neighbor closer than 62$"$. In the same way, for the nighbor we increase by one the value of 
 $w_{zf}$ of the nearest galaxy with a good redshift. To minimize the error in the measured clustering signal, we also require a correction
  based on the redshift distribution of our sample, namely the $w_{FKP}$ factor \cite[]{Feldman1994}, that weights galaxies according to their number density, $n(z)$. It is defined as
\begin{equation}
w_{FKP}=\frac{1}{1+n(z)P_{FKP}},
\end{equation}
where $P_{FKP}$ is a constant that roughly corresponds to the amplitude of the CMASS power spectrum 
$P(k)$, at $k=0.1\hspace{0.1cm}h$ Mpc$^{-1}$. We assume $P_{FKP}=2\times10^4\hspace{0.1cm}h^3$ Mpc$^{-3}$, in \cite{Anderson2012}. 
The last weight, $w_{sys}$, accounts for a number of further systematic effects that could cause spurious angular fluctuations
 in the galaxy target density. These effects are treated in detail in \cite{Ross2012}, but we do not include them in this analysis, since they are not relevant at the scales 
 considered in this work. Therefore we set in $w_{sys}=1$ in the following analysis.


\section{Modeling full CMASS Sample}
\label{sec:Modeling_full_CMASS_Sample}


\subsection{Full CMASS Clustering} 
\label{sec:Mock_Catalog_Creation}

\begin{figure}{}
  \begin{center}
    \includegraphics[width=\linewidth]{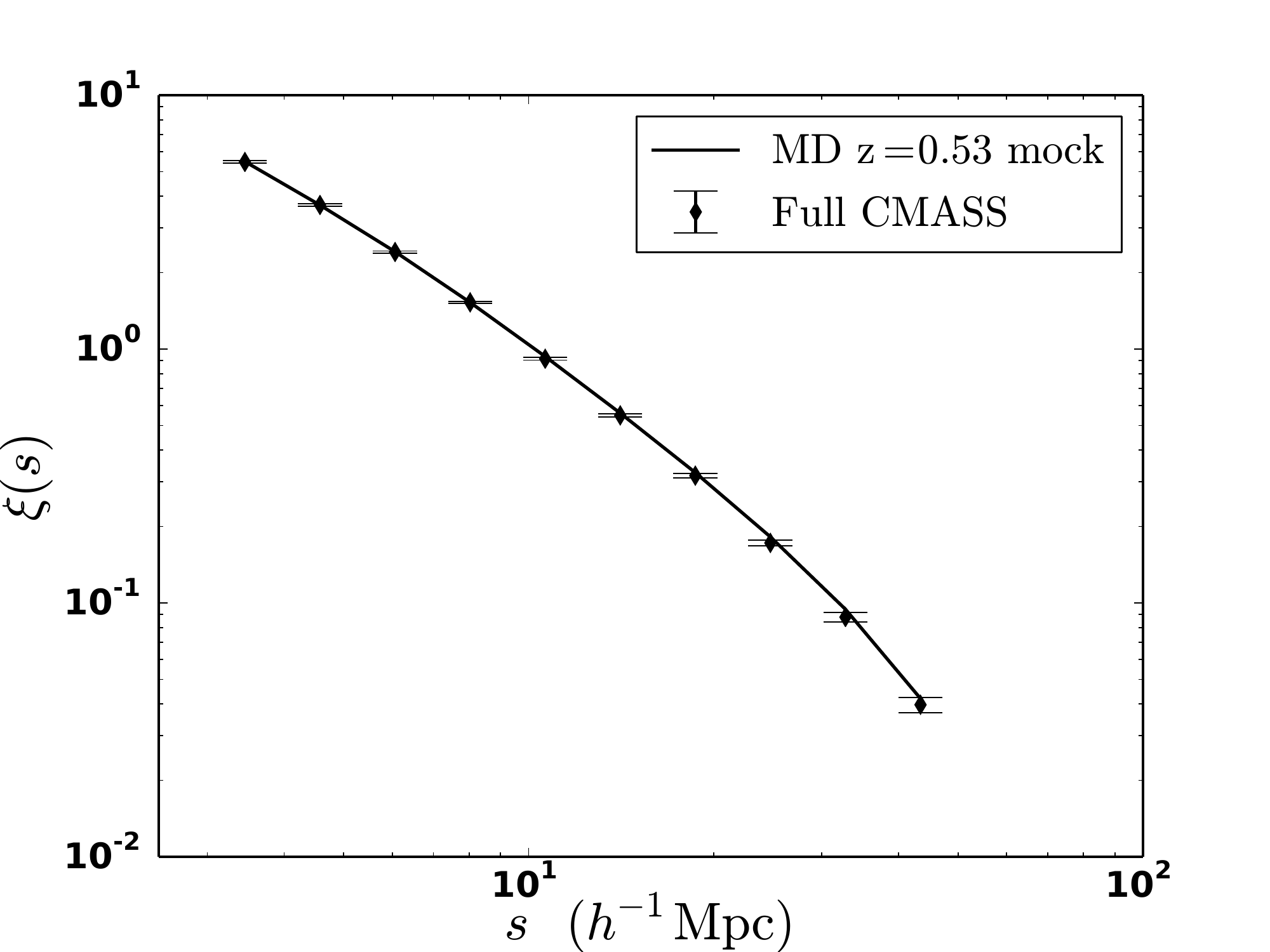}
  \end{center}
  \caption{Redshift-space monopole correlation functions of our $z=0.53$ MultiDark full mock 
  galaxy catalog (solid line) compared to BOSS CMASS DR11 measurements.  Error bars are estimated using 200 jackknife regions.}
  \label{fig:CF_full_mock_xi}
\end{figure}

We construct an HOD model using MultiDark halos and sub-halos (see model description in Section\;\ref{sec:HOD_general}), 
and produce a mock galaxy catalog which we compare to the full CMASS DR11 population.  
This mock is built by varying the HOD parameters to match $\xi(s)$, populating the MD simulation in each step, 
and using the peculiar velocities in the simulation to model redshift-space distortions. 
The intention is that changing the HOD will constrain the overall galaxy bias, hence we fit only one statistic. 
We then evaluate and further investigate these fits over the three clustering metrics: 
$\xi(s)$, $w_p(r_p)$ and $\Sigma(\pi)$.

 However, since implementing a formal fit to determine the optimal HOD parameters is beyond the scope of this work, 
 we improve the matching empirically, changing the input values until we find a suitable 
 ($\log M_{min}$, $M_0$, $M'_1$, $\alpha'$, $\sigma_{\log M}$) set that reproduces the observed $\xi(s)$ amplitude. 
 We fit only $ M_{min}$ (the minimum halo mass), $M'_1$ (the mass scale of the satellite cut-off profile) and $\alpha$ 
 (the satellite slope). 
 The remaining parameters are fixed to their default values given by \cite{White2011}: $\log M_0=12.8633$, $\sigma_{\log M}=0.5528$. 
 The specific choice of these three parameters arises from their connection to two physical quantities we want to 
 measure: (i) the satellite fraction, $f_{sat}$, that controls the slope of the $1$-halo term at small scales, 
 where sub-structures of the same halo dominate; (ii) the galaxy number density, $n(z)$, 
 affecting the $2$-halo term at larger scales, where correlations between sub-structures of different hosts become appreciable. 
 Figure~\ref{fig:HODvariation} in the Appendix illustrates how a change in $M_{min}$, $M'_1$ and $\alpha$ affects the projected correlation function.

\begin{figure}
  \begin{center}
    \includegraphics[width=\linewidth]{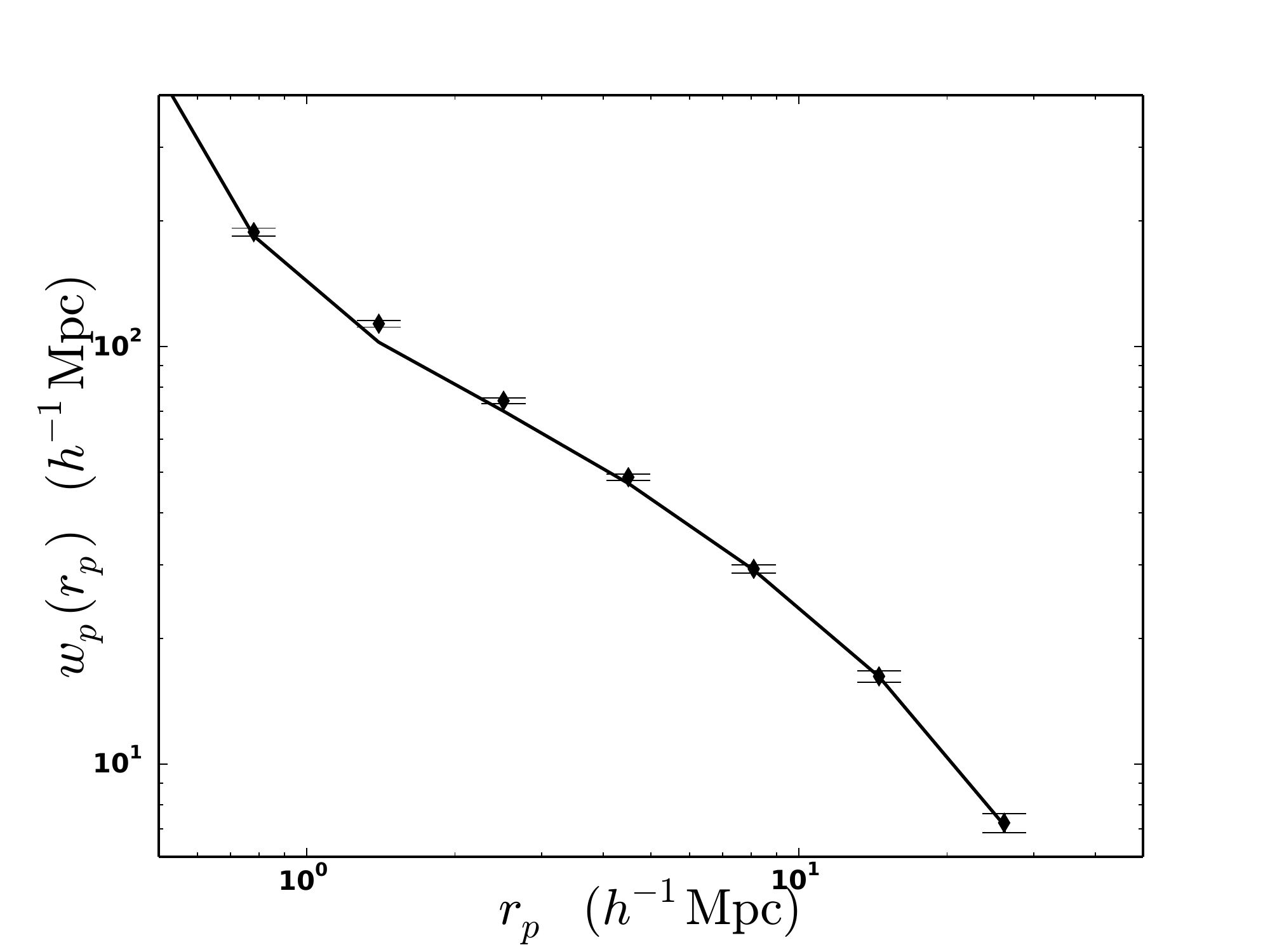}\hfill\\
    \includegraphics[width=\linewidth]{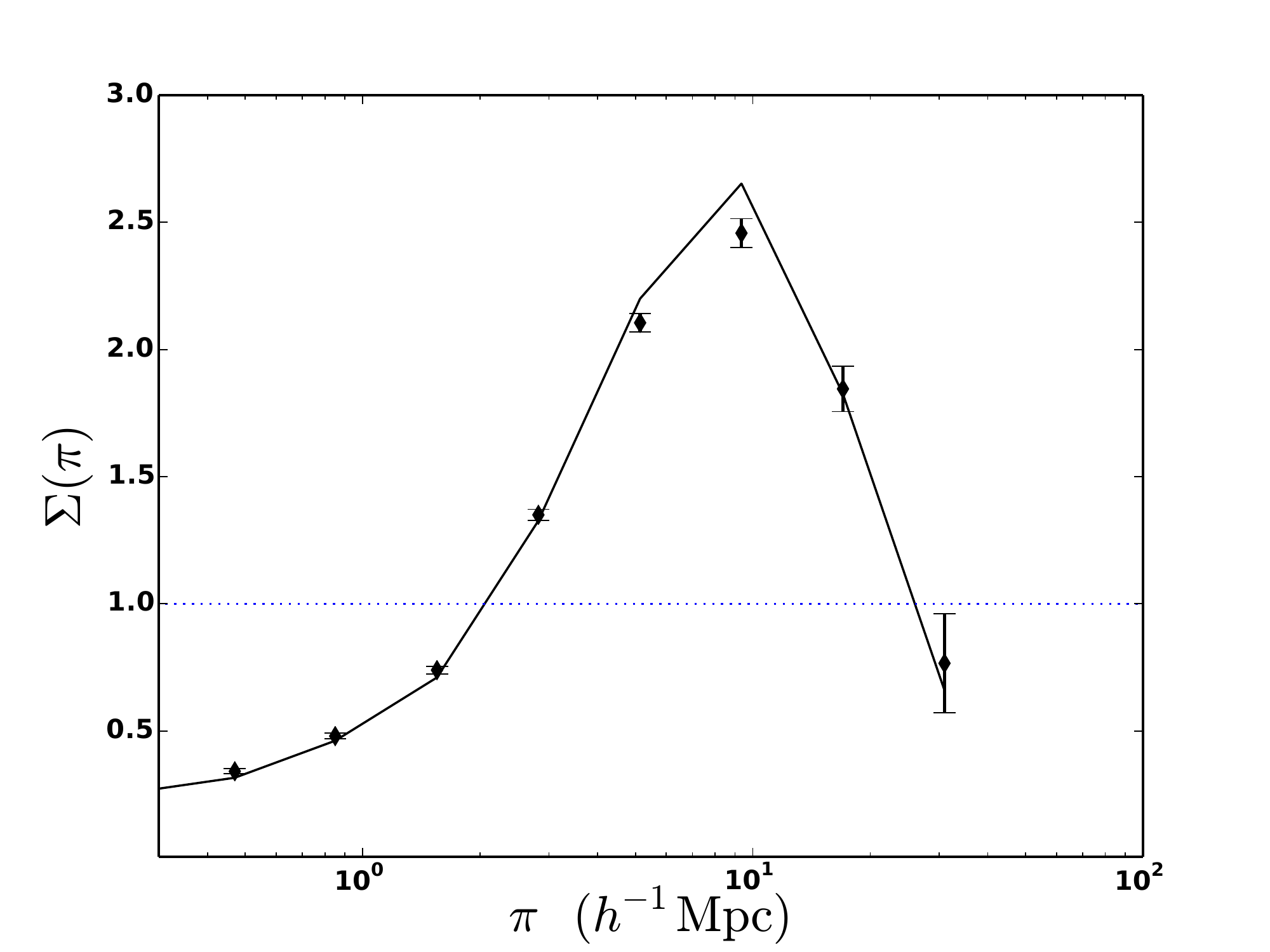}    
  \end{center}
  \caption{Projected correlation function (top) and $\Sigma(\pi)$ (bottom) for the $z=0.53$ MultiDark full mock galaxy 
  catalog (solid line), compared to BOSS CMASS DR11 measurements. Error bars are estimated using 200 
  jackknife regions containing the same number of randoms.}
  \label{fig:CF_full_mock_wpSigma}
\end{figure}

Figure\;\ref{fig:CF_full_mock_xi} displays the redshift-space monopole corresponding to our 
empirical best fit ($\chi^2=11.08/7\,dof$ including the full covariance matrix computed with jackknife; the HOD
parameters are given in Table \ref{tab:HODparam}) mock galaxy catalog from the MultiDark simulation.
 The projected correlation function, $w_p(r_p)$, and the line-of-sight statistic, $\Sigma(\pi)$, 
 corresponding to this model are shown in Figure\;\ref{fig:CF_full_mock_wpSigma}.
In agreement with many previous works \citep{Zehavi2004, Zehavi2005b, Guo2012}, we find that 
CMASS galaxies are more highly clustered at small scales ($1$-halo regime);  then, as the spatial 
separation between the pairs increases, the clustering strength drops  ($2$-halo term). Compared to \cite{White2011}, 
our best-fit mock has a much lower satellite slope, $\alpha$, and $M_1'$, resulting in a higher satellite fraction (about 
27$\%$); however, our mean satellite occupation function is compatible with results from \cite{Guo2015}.
Overall, the amplitude of our model galaxies is in good agreement with observations. Error bars are estimated using 200 jackknife 
regions gridded in right ascension and declination as follows:  10$\,$RA$\times$15$\,$DEC cells for the
 CMASS North Galactic Cap ($N_{res}=150$), plus 5$\,$RA$\times$10$\,$DEC regions for the South Galactic Cap,
  ($N_{res}=50$). This approach produces $200$ equal areas of about $100\deg^2$ each.

In the calculation of the full CMASS (MD mock) $\Sigma(\pi)$ through Eq.\;\ref{eq:sigma_new_metrics}, we use the best-fit 
power-law to the full CMASS (MD mock) $w_p(r_p)$. The relative $r_0$ and $\gamma$
 estimates are given in Figure\;\ref{fig:PL}. 
Beyond $8-10 \; \hmpc$, where the Kaiser squashing becomes 
predominant, the jackknife uncertainties on $\Sigma(\pi)$ are wider. This measurement reveals that 
the deviation of $\xi(\bar{r}_p,\pi)$ from the real-space behavior dramatically changes according to 
the scale of the problem: at very small redshift separations, $\pi\leq2 \; \hmpc$, where the 
finger-of-god dominate, the contribution of peculiar velocities pushes $\Sigma(\pi)$ below unity. 
Above $3 \; \hmpc$, $\Sigma(\pi)$ increases sharply and peaks 
around $8 \; \hmpc$. On larger scales, the correlation between pairs of galaxies is compressed 
along the line of sight since the Kaiser infall dominates and $\Sigma(\pi)$ drops.


\subsection{Modeling Redshift-Space Distortions and Galaxy Bias}
\label{sec:Modeling_z_distortions}

\begin{figure}
  \begin{center}
    \includegraphics[width=\linewidth]{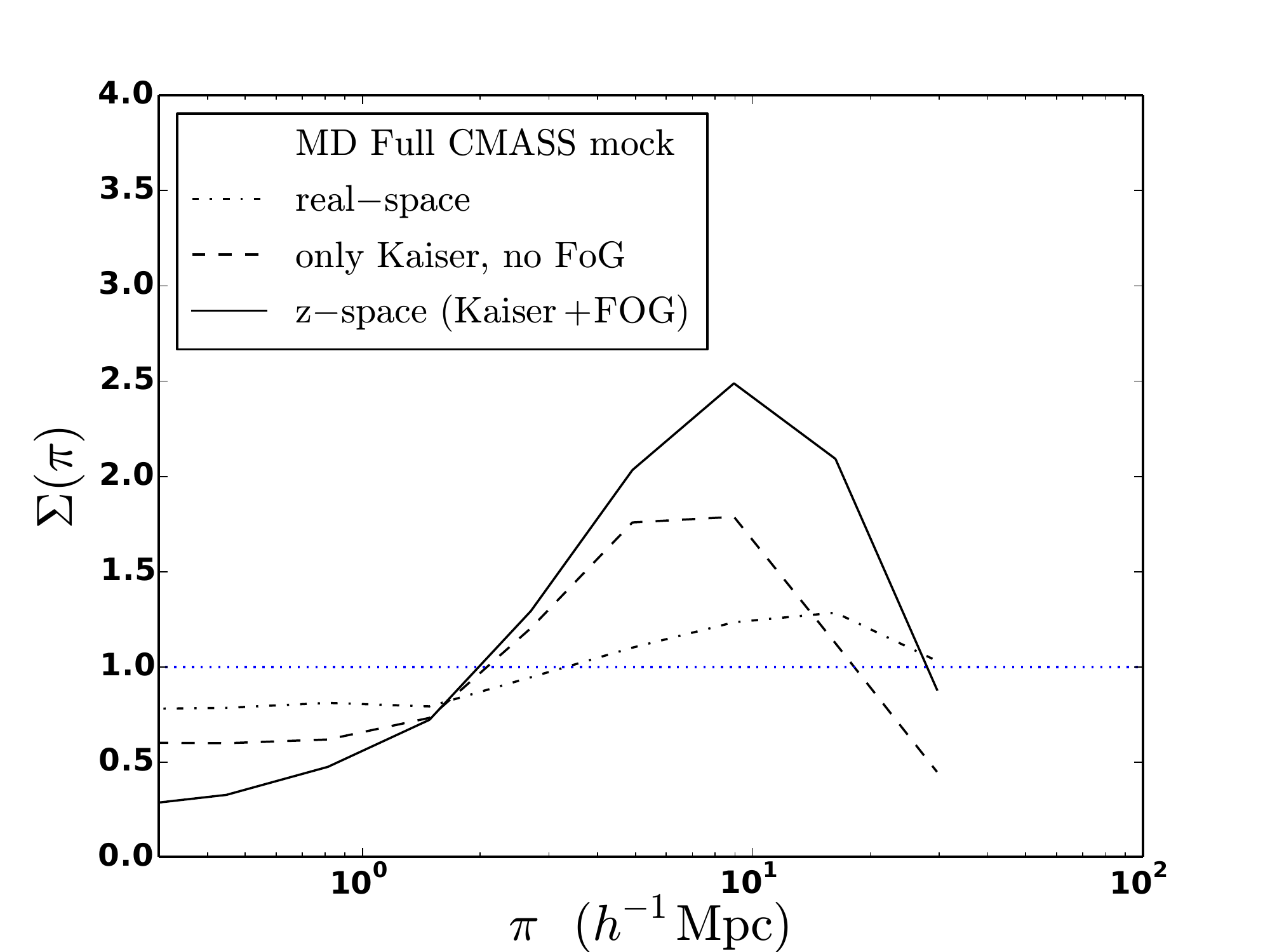}
    \caption{$\Sigma(\pi)$ in real-space (dot-dashed line), redshift-space with 
    only Kaiser contribution (dashed) and Kaiser plus finger-of-god (solid). 
    As expected, the real-space behavior is close to unity at all scales.}
    \label{fig:sigma_real_kaiser_zspace}   
  \end{center}
\end{figure}
In redshift-space, two different distortion features are observed: the finger-of-god effect which dominates 
below $2\; \hmpc$, and the Kaiser flattening, which becomes important beyond 10 \textemdash\; 15$ \; \hmpc$. 
These phenomena preferentially manifest themselves on different scales, but a certain degree of entanglement is 
unavoidable in both regimes. In order to better separate the two effects, we examine $\Sigma(\pi)$ in our 
MultiDark full mock catalog in three different configurations: real-space, redshift-space with only Kaiser 
effect and full redshift-space (FoG+Kaiser), as shown in Figure \ref{fig:sigma_real_kaiser_zspace}. 
The real-space $\Sigma(\pi)$ is defined in Eq.\;\ref{eq:sigma_new_metrics}, omitting the peculiar velocities 
both in the numerator and in $w_p(r_p)$ to which we fit the power law at the denominator. Since $\Sigma(\pi)$ is 
the ratio between two spherically averaged power laws, we expect it to be close to unity at all scales. 
Hence, the dot-dashed line in Figure \ref{fig:sigma_real_kaiser_zspace} is compatible with expectations.
The redshift-space case with only Kaiser contribution (dashed line) is computed by assigning satellite galaxies their 
parental $v_{pec}$ value. In this way, each satellite shares the coherent 
motion of its parent, but it does not show any random motion with respect to it. The last case considered is 
the full redshift-space $\Sigma(\pi)$ (solid line), in which satellite galaxies have their own peculiar velocity, which is 
independent from their parents. 
\begin{figure*}
  \begin{center}
    \includegraphics[width=0.5\linewidth]{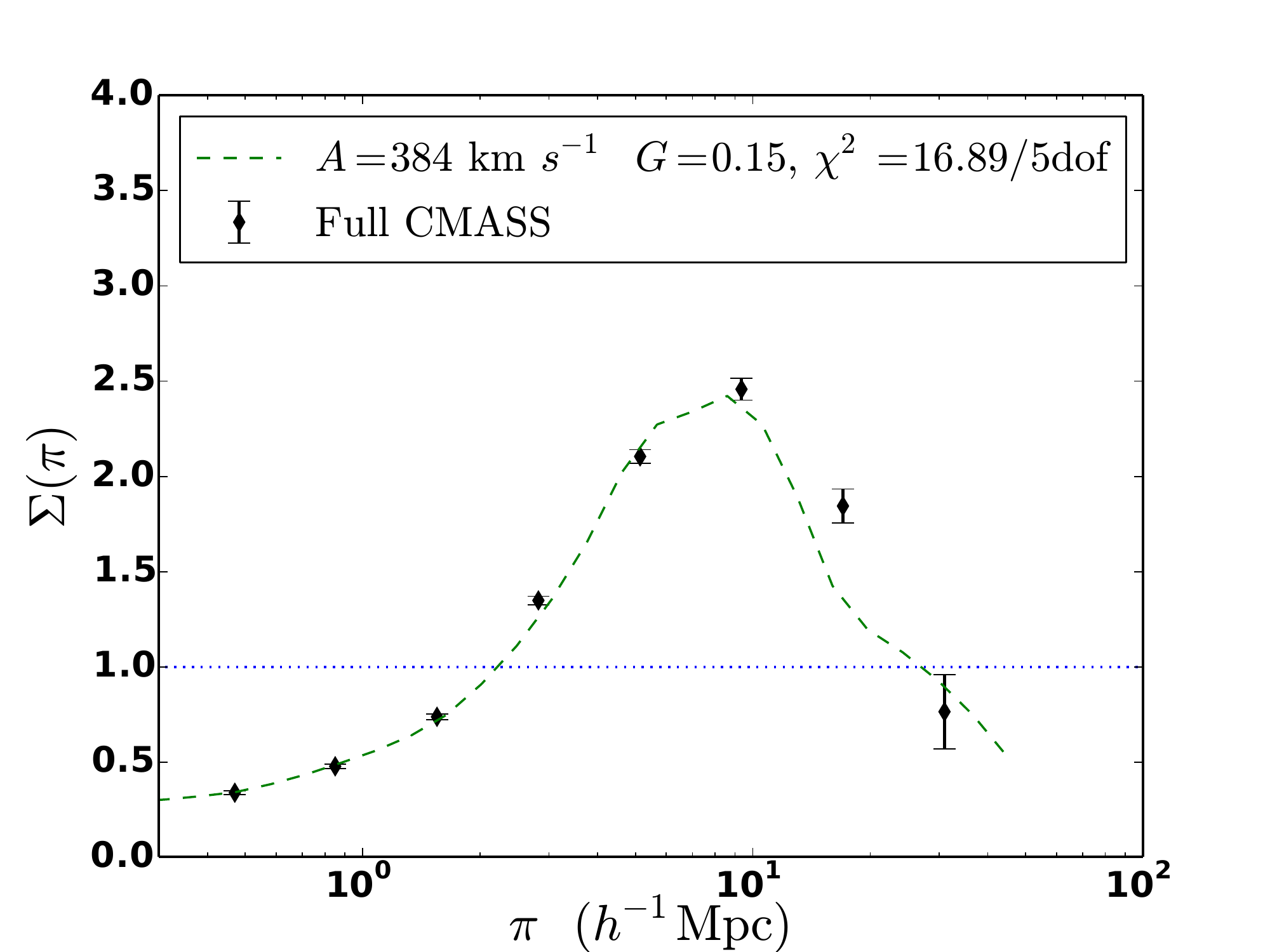}\hfill
    \includegraphics[width=0.5\linewidth]{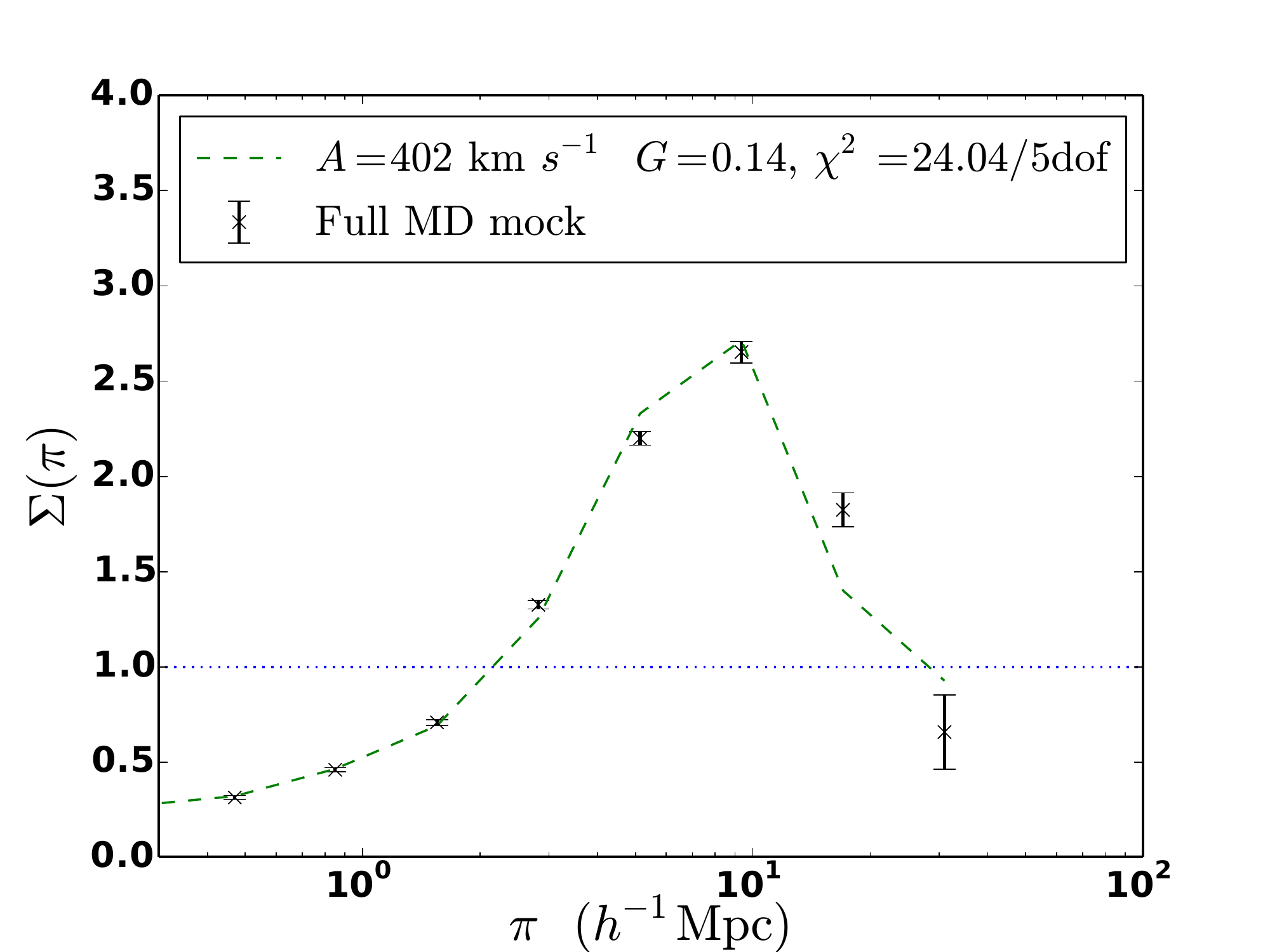}\hfill
  \end{center}
  \caption{$\Sigma(\pi)$ full CMASS DR11 measurement (left panel)  and our MultiDark $z=0.53$ mock (right panel), versus their $A,G$ analytic model
   (dashed lines). For both data and mock sets we assume the errors are given by our jackknife estimate, computed using 200 resamplings. The fits are performed by 
  using the full covariance matrix. These plots reveal that the full CMASS sample and the MultiDark model galaxies share almost the same large-scale bias value, while the peculiar velocity contribution is higher in the mocks.}
  \label{fig:sigma_model_mocksall}
\end{figure*}

We are now able to provide a full description of our $\Sigma(\pi)$ results by modeling them through Eq.\;\ref{eq:modello_analitico_sigma}, 
in terms of four parameters: the power-law correlation length, $r_0$, its slope $\gamma$, the pairwise velocity dispersion, 
$A$ and the $G$ parameter, which is inversely proportional to the linear galaxy bias, $b$, through Eq. \ref{eq:bias_approx_AG}.

The linear galaxy bias is scale dependent and has been computed \citep[e.g., ][]{Nuza2013}
as the ratio between the galaxy and matter correlation functions,
\begin{equation}
b(s)=\sqrt{\frac{\xi(s)}{\xi_m(s)}}.
\label{eq:bias_red_blue}
\end{equation}

Our goal is to provide an estimate of both the peculiar velocity field causing the distortions we observe in 
redshift-space in our clustering measurements and the large scale bias, using the $A,G$ values we find from our 
full, red and blue CMASS and MultiDark $\Sigma(\pi)$ modeling. To this purpose, we do not compute the bias as \cite{Nuza2013}, through Eq. \ref{eq:bias_red_blue}, 
but we estimate it from Eq. \ref{eq:bias_approx_AG}. 

Figure \ref{fig:sigma_model_mocksall} displays the $A,G$ models (dashed curves) for our CMASS measurements 
(left panel, squares) and full MultiDark mock catalog (right panel, crosses). 
All the model fits are performed including the full covariance matrix, estimated by using 200 jackknife re-samplings (Sec. \ref{sec:Mock_Catalog_Creation}). For
the MultiDark mock, we assume the same scatter of the CMASS data. 

Adopting a normal function (Eq. \ref{eq:vpec_norm}) to mimic  the contribution of peculiar velocities (see Table \ref{tab:err_AG}), results in the 
MD model galaxies that have slightly higher bias \textemdash\; which means a lower $G$ value \textemdash\; than the full CMASS population and higher peculiar velocity contribution \textemdash\; 
higher $A$ value. This result is in agreement with the bottom panel of Figure \ref{fig:CF_full_mock_wpSigma}: CMASS data points (diamonds) 
experience a stronger Kaiser squashing at $\sim10$ Mpc h$^{-1}$i.e., they have a smaller large-scale bias, compared to 
the MultiDark model galaxies (solid line). From these $A,G$ values, we conclude that our full MD mock catalog can be considered a 
reliable representation of the full CMASS sample.

The reduced $\chi^2$ values we derive from the full CMASS and MultiDark $\Sigma(\pi)$ model fits are relatively high, compared to the $\chi^2$ values 
we find for $\xi(s)$, which are reported in the caption of Table \ref{tab:HODparam}. The main reason for this result is the nature of $\Sigma(\pi)$, which is a ``derived" clustering measurement, 
in the sense that it is built from the ratio (Eq. \ref{eq:sigma_new_metrics}) of the 2PCF \textemdash\; spherically averaged in the range $0.5\leq r_p<2\,h^{-1}$Mpc \textemdash\; over a real-space term.
 In order to mimic this average in our model, we must perform a 
double integration in ($r_p,\pi$) of the convolution ($i.e.$ the inner integral in the numerator of Eq. \ref{eq:modello_analitico_sigma}) of the real-space term with the peculiar velocity contribution, $f(w)$. 
Such a double integration has to be computed numerically, in ($r_p,\pi$) bins. In this way, we eliminate the dependence on 
$r_p$ \textemdash\; we remain with a single ``mean" $r_p$ value, representing the 0.5 \textemdash\; 2$\,h^{-1}$Mpc bin \textemdash\; and maintain the $\pi$ dependence \textemdash\; we remain with a ´mean" $\pi$ value for each $\pi$ bin. 
Thus, the $A,G$ model reproduces the $\Sigma(\pi)$ measurement in bins of ($r_p$,$\pi$) and not analytically in each point. This is a first approximation. 

Also, we assume for the peculiar velocity term, $f(w)$, a Gaussian functional form (Eq. \ref{eq:vpec_norm}), but this is an arbitrary choice, which introduces another approximation. 
In addition, the denominator in Eq.\ref{eq:modello_analitico_sigma}, which is nothing but the best-fit power law to $w_p(r_p)$, spherically averaged in the way described above, 
presents the same numerical issues of the numerator. 

In conclusion, we are applying a 
series of approximations that are necessary in order to define our $\Sigma(\pi)$ model, but they unavoidably affect the accuracy of the fit. 

Since our goal is to give a qualitative estimate, in terms of linear bias and redshift-space distortions, of the full, red and blue CMASS samples, we 
do not heavily focus on the goodness of our $\Sigma(\pi)$ model fits, but instead we stress the importance of a cross-comparison, in terms of $A,G$ values, 
between the three CMASS populations and the MultiDark model galaxies. In particular, for the full CMASS case $b\sim3$, which is 
relatively high, compared to the value $b\sim 2$, reported in \cite{Nuza2013}. This disagreement can be justified by recalling that we are selecting only the 
high-redshift tail of the CMASS sample, beyond $z>0.55$, and for those galaxies the bias 
is expected to be higher as compared to the full CMASS bias in \cite{Nuza2013}.


\subsection{Full CMASS Covariance}
\label{sec:full_covariance}
We compute the full CMASS jackknife covariance matrix for the three metrics of interest using Eq.\;\ref{eq:covma}, in which 
$\xi$ is either $\xi(s)$, $w_p(r_p)$, or $\Sigma(\pi)$. We estimate the goodness of our model fits
to the CMASS measurements by computing the relative $\chi^2$ values as
\begin{equation}
\chi^2=A^TC_{\star}^{-1}A,
\label{eq:chi2}
\end{equation}
where $A=(\xi_{data}^i-\xi_{model}^i)$ is a vector with $i=1,..., n_{b}$ components and $C_{\star}^{-1}$ is an unbiased estimate 
of the inverse covariance matrix \citep{Hartlap2007,Percival2014},
\begin{equation}
C_{\star}^{-1}=(1-D)C^{-1},\,\,\,D=\frac{n_b+1}{N_{res}-1}.
\label{eq:psi_correction}
\end{equation}
In the equation above, $n_b$ is the number of observations and $N_{res}$ the number of jackknife re-samplings.
 For the full CMASS population, the correction factor $(1-D)$ represents a $8\%$ effect on the final $\chi^2$ value.
 
 In Appendix \ref{sec:Jackknife_Poisson}, we test our jackknife error estimates using a set of 100 Quick Particle Mesh \citep[QPM;][]{White2014} galaxy mock catalogs.


\section{Modeling Color Sub-samples}
\label{sec:Modeling_Color}

We repeat the same analysis described in Section~\ref{sec:Modeling_full_CMASS_Sample} on the red and blue color 
sub-samples.  We first use $\xi(s)$ to fit a HOD and match the overall clustering, then use our analytic 
model to obtain fits for $A$ and $G$.  There remains a question on how to model the sub-populations in the mocks; 
we explore two methods.

\subsection{Independent Red and Blue models}
\label{sec:Independent_Red_Blue_mocks}

\begin{figure*}
  \begin{center}
    \includegraphics[width=0.33\linewidth]{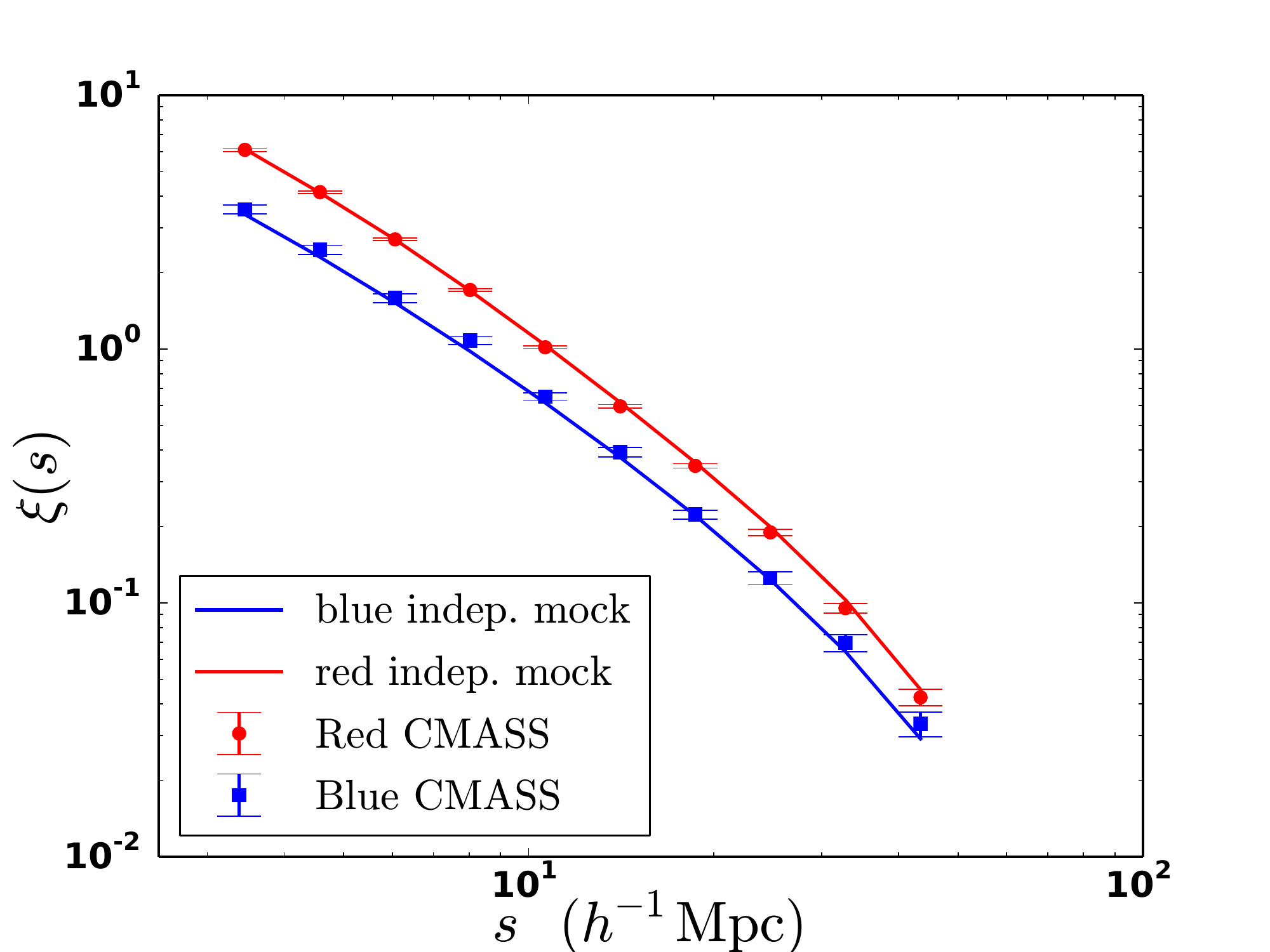}\hfill%
    \includegraphics[width=0.33\linewidth]{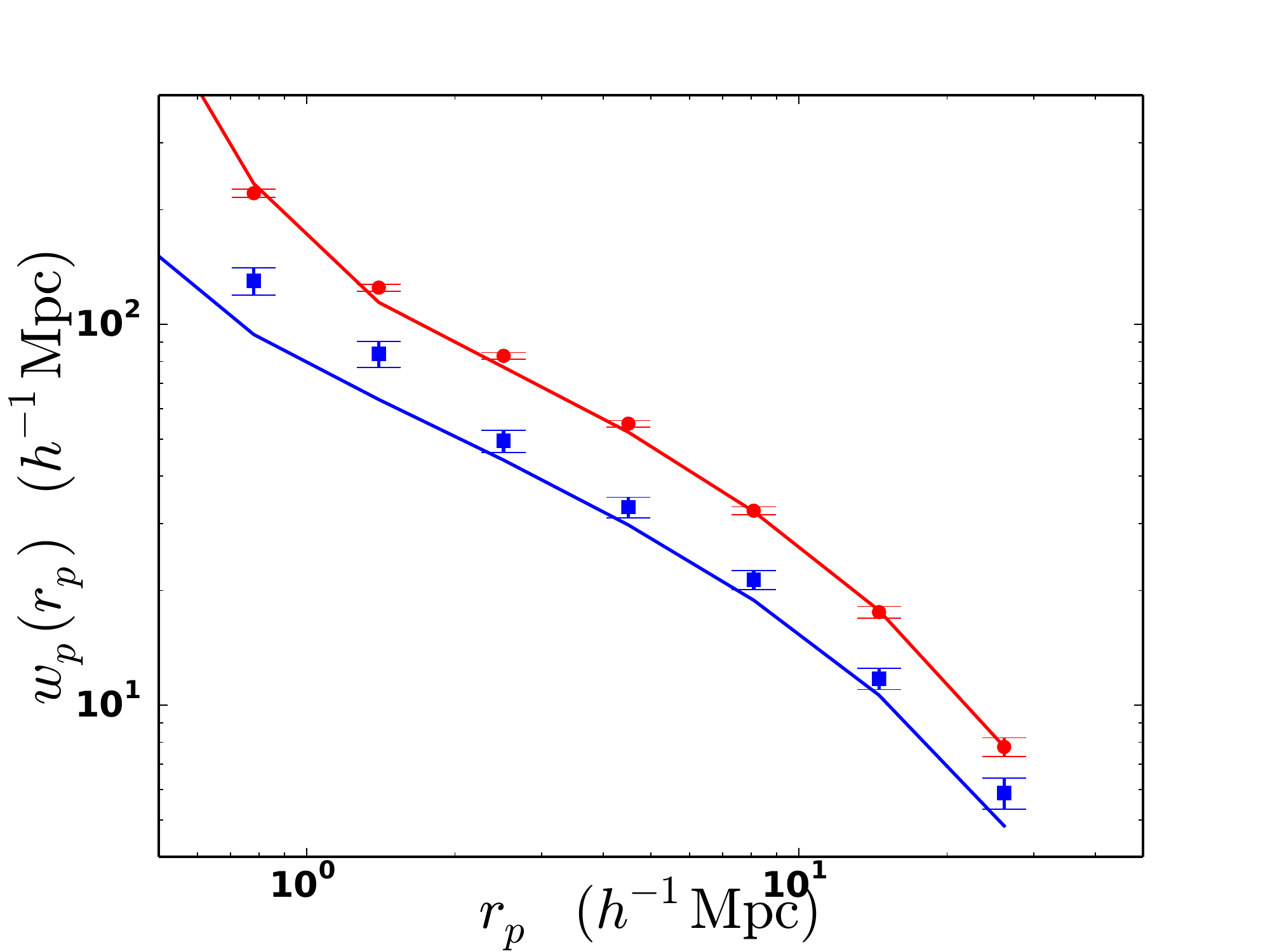}\hfill%
\includegraphics[width=0.33\linewidth]{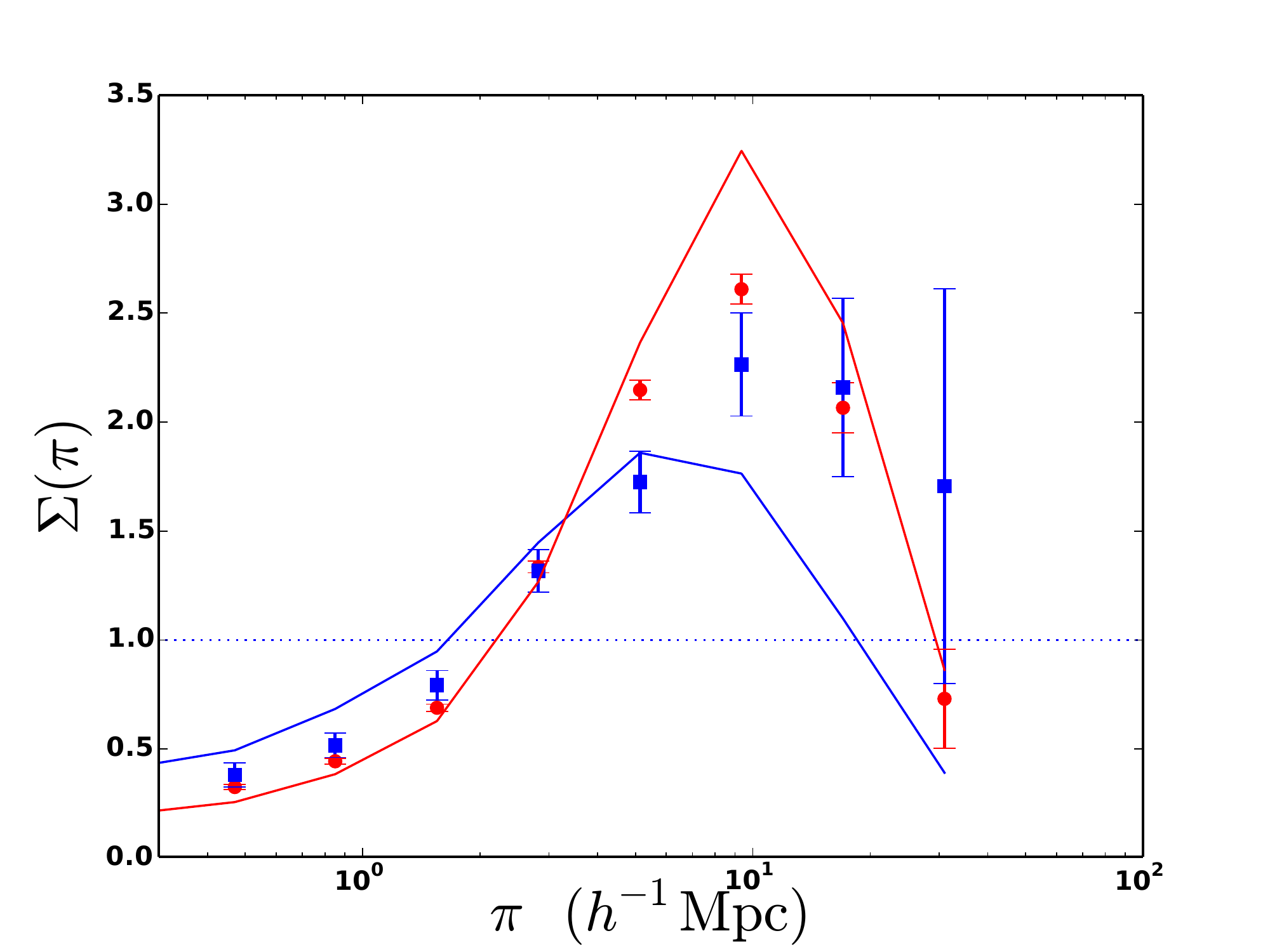}
  \end{center}
  \caption{Independent mock catalogs designed to model CMASS DR11 red 
  and blue $\xi(s)$, $w_p(r_p)$ and $\Sigma(\pi)$ measurements (points and squares).
  The error bars are the $1\sigma$ regions estimated using 200 jackknife re-samplings of the data. Despite we fit only $\xi(s)$, we find good agreement between data and mocks in all our three statistics. As expected, red galaxies show a higher clustering amplitude compared to the blue population.}
  \label{fig:CF_indep_models}
\end{figure*}

\begin{table}
  \centering
  \begin{tabular}{|l r r r |}
    \hline
    \hline
    &$\,\,\,$Total$\,\,\,$&$\,\,\,$Red$\,\,\,$&$\,\,\,$Blue\\
    \hline
    $\log M_{min}$$\,\,\,$&13.00 & 13.10&12.50 \\ 
    $\log M'_1$$\,\,\,$ &13.30  &13.02 & 13.85 \\
    $\alpha$$\,\,\,$ &0.20&0.22& 0.15\\
    $f_{sat}$\,\,\,$$&0.27  &0.33 & 0.11\\ 
    $\langle\log M_h\rangle$\,\,\,$$& 12.75& 13.00 &12.50 \\
    \hline
    \hline
  \end{tabular}
  \caption{Our best empirical estimates of the HOD parameters for the total, red and blue independent models of the CMASS populations. 
  We obtain these values only by fitting $\xi(s)$ with a three-dimensional grid in $\log M_{min}$, 
  $\log M'_1$ and $\alpha$. The resulting $\chi^2$ values are: 11.08/7dof (full CMASS), 13.54/7dof (red) and 14.91/7dof (blue).}
  \label{tab:HODparam}
\end{table}

For simplicity, our first attempt at the color sub-samples is to individually model the red and blue CMASS 
populations.  That is, we assume the clustering comes from a complete sample and we generate a HOD populating halos 
independently of whether a galaxy is red or blue.  By definition, there is no connection in the overlap and the 
same halo or sub-halo could host either a red and blue galaxy in the corresponding mocks.  This is an 
over-simplified view, as clearly a galaxy can be either red or blue and not both.  However, it is an 
assumption that is embedded within several related analyses \citep[]{Zehavi2004, 
Zehavi2005b, Guo2012, Guo2014}.

Figure~\ref{fig:CF_indep_models} shows the agreement between the CMASS monopole, projected 2PCF and 
$\Sigma(\pi)$ measurements and our independent red and blue model galaxies. Our empirical best-fit HOD parameter 
values are reported in Table \ref{tab:HODparam}, together with the satellite fraction; the fraction is higher 
for red than for blue galaxies, confirming that luminous red galaxies tend to live in a denser 
environment \citep[]{Wang2007, Zehavi2005b, Swanson2008}.
We conclude that we are able to fit correctly all our red and blue CMASS clustering results, by means of the 
same HOD technique, with small variations in its input parameters. However, these red and blue independent models 
are non-physical, because they allow the same galaxy to be either red or blue. In other words, they place both red 
and blue galaxies in the same hosting halos, which is not the case.\\
To overcome this problem, we propose an alternative halo occupation distribution approach (see next Section)
 in which the red and the blue models are obtained by splitting the full mock catalog into sub-populations that 
 match the observed red/blue CMASS galaxy fractions. In this way, the red and blue model galaxies 
 are no longer independent and, by construction, they cannot occupy the same positions in a given halo.


\subsection{Splitting Color Samples using Galaxy Fractions}
\label{sec:Splitting_Colors}

\begin{figure}
 \begin{center}
 \includegraphics[width=\linewidth]{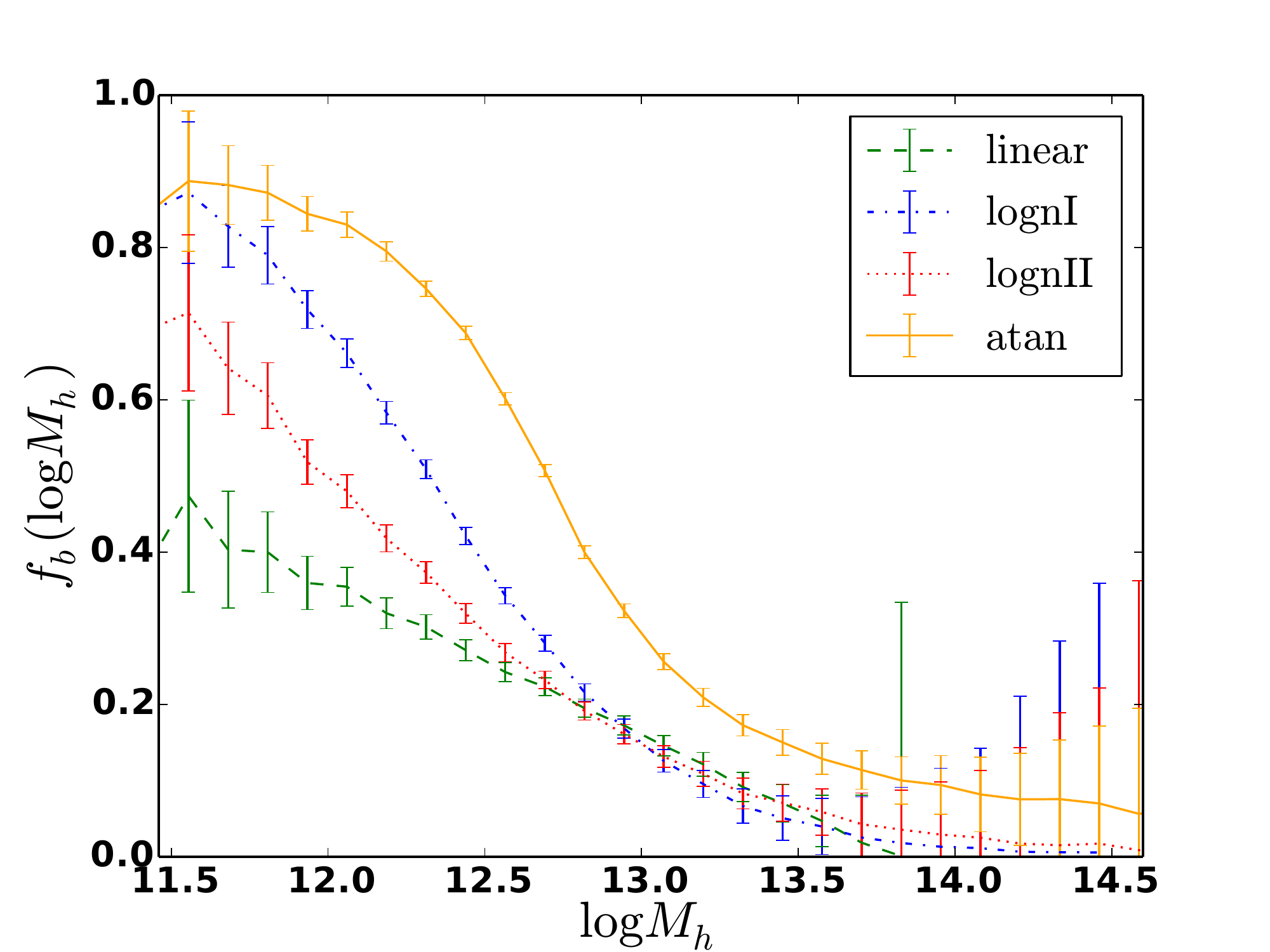}
 \caption{Blue galaxy fraction models, $f_b$, and the corresponding Poisson error, as 
 a function of the central halo mass: linear (dashed line), log-normal I (dot-dashed), log-normal II (dotted), 
 inverse tangent (solid). The red galaxy fractions are recovered by $f_r=1-f_b$.}
\label{fig:fbmodels}
\end{center}
\end{figure}

\begin{figure*}
  \begin{center}
      \includegraphics[width=0.33\linewidth]{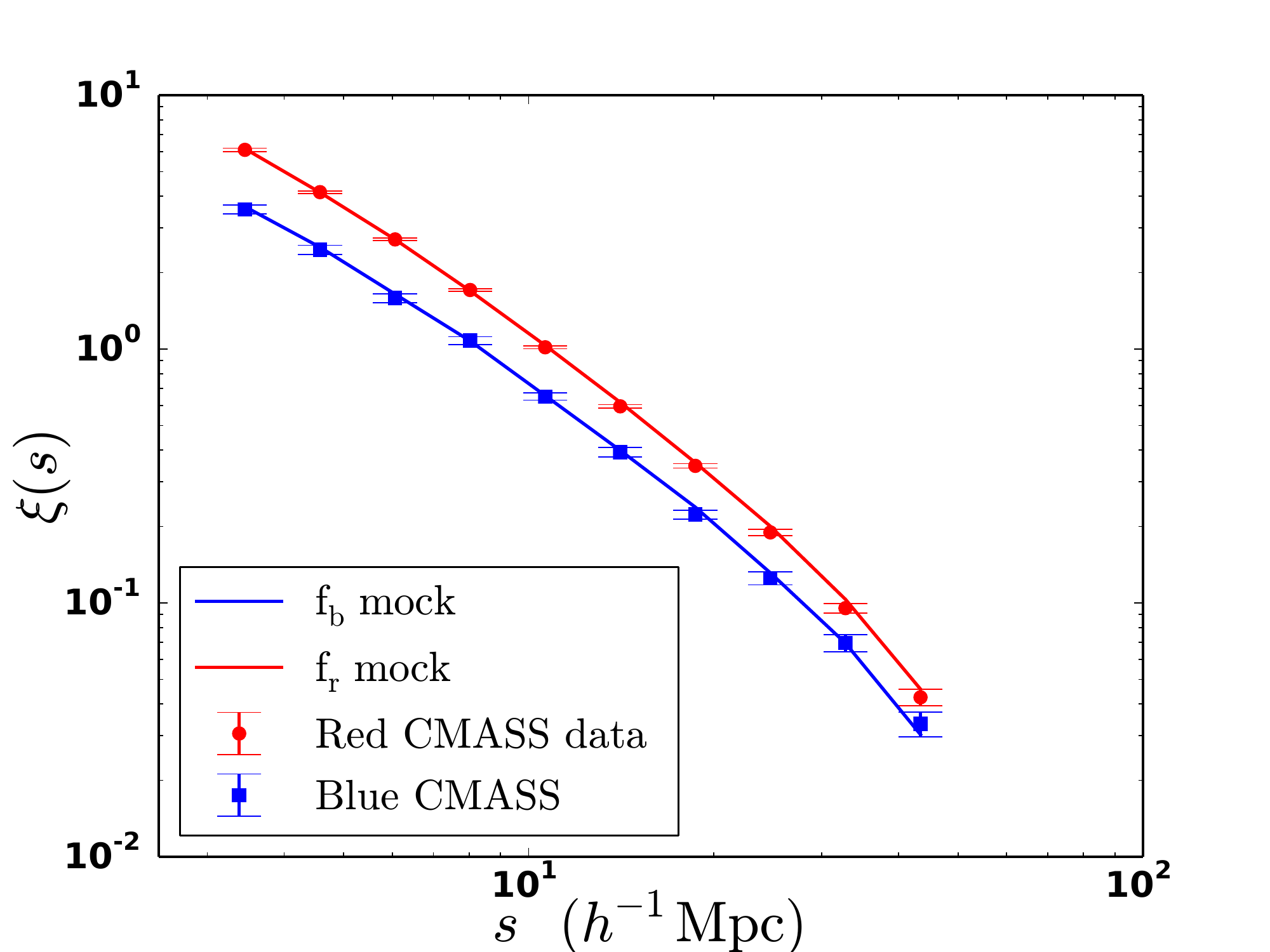}\hfill%
    \includegraphics[width=0.33\linewidth]{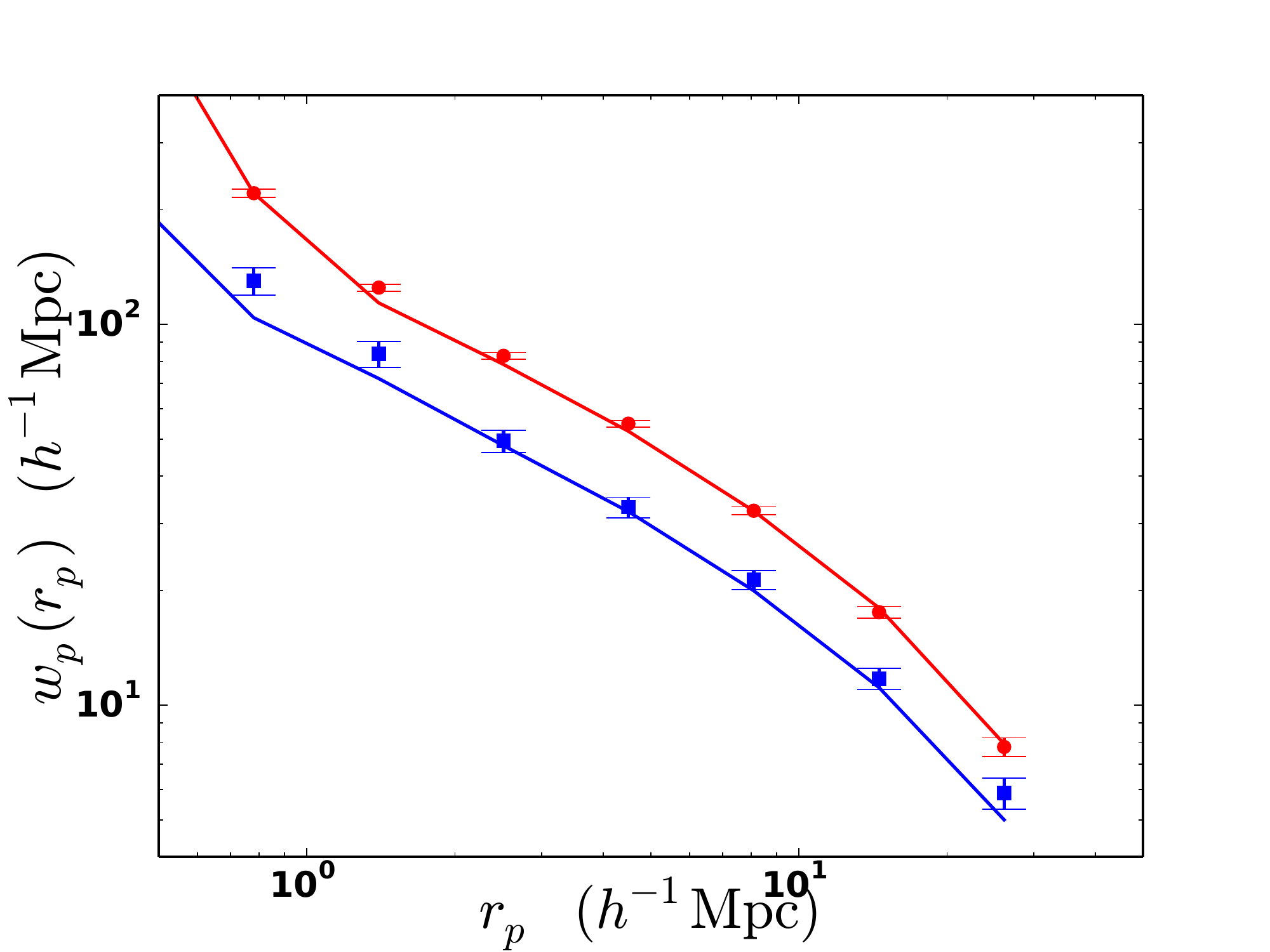}\hfill%
\includegraphics[width=0.33\linewidth]{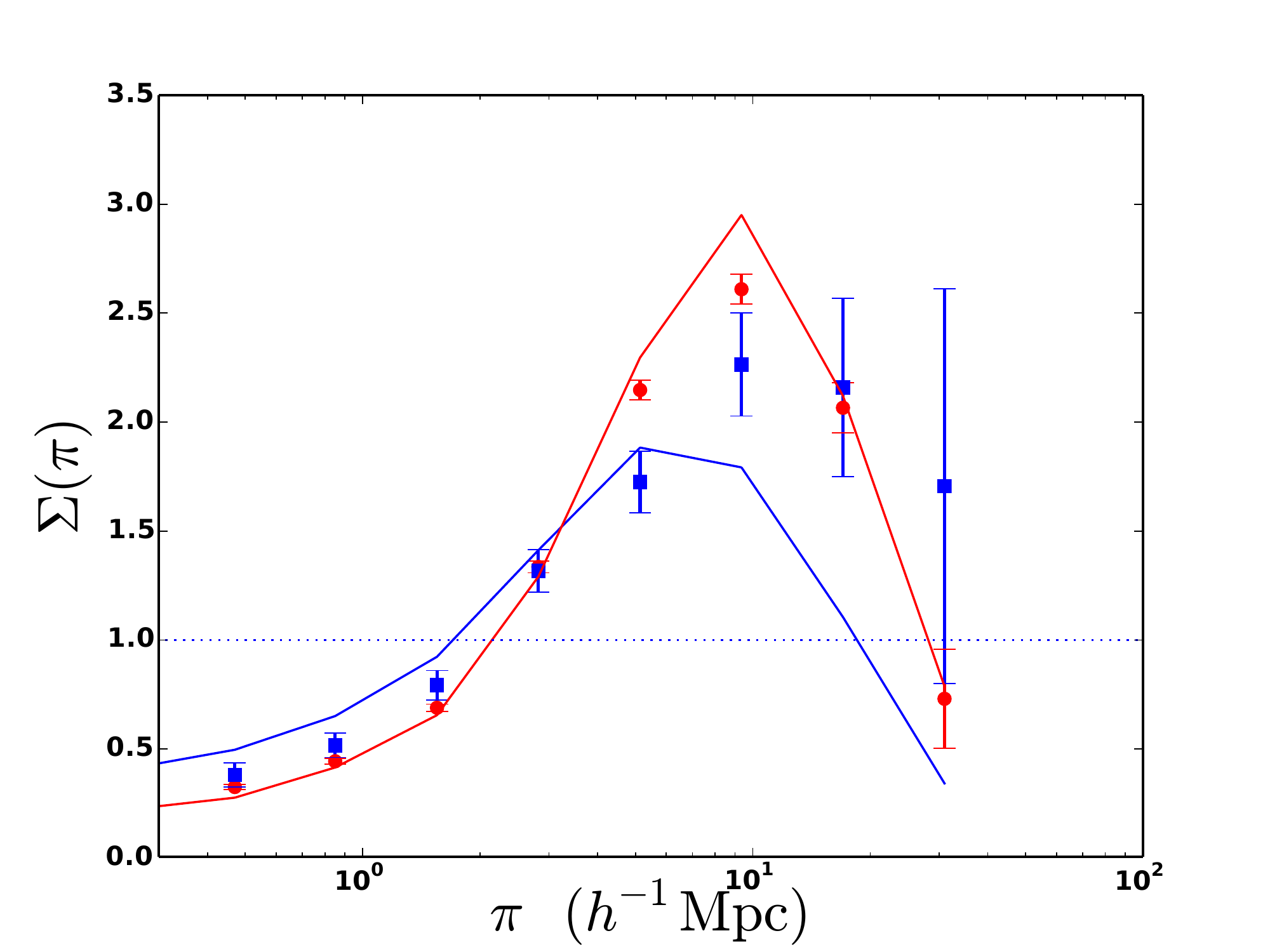}
  \end{center}
  \caption{CMASS DR11 red and blue clustering measurements versus mocks. The models are obtained by splitting the full MultiDark 
  mock into its red and blue components, matching the observed CMASS red/blue galaxy fraction, $f_{b,r}$. In this way, we prevent the same mock galaxy to be either red or blue, and guarantee the reliability of the model. We find good agreement between the CMASS measurements and our MultiDark mocks, and confirm that red galaxies leave in more dense environments compared to the blue population.}
  \label{fig:CF_indep_and_fbr_hod}
\end{figure*}

\begin{figure*}
  \begin{center}
   \includegraphics[width=0.48\linewidth]{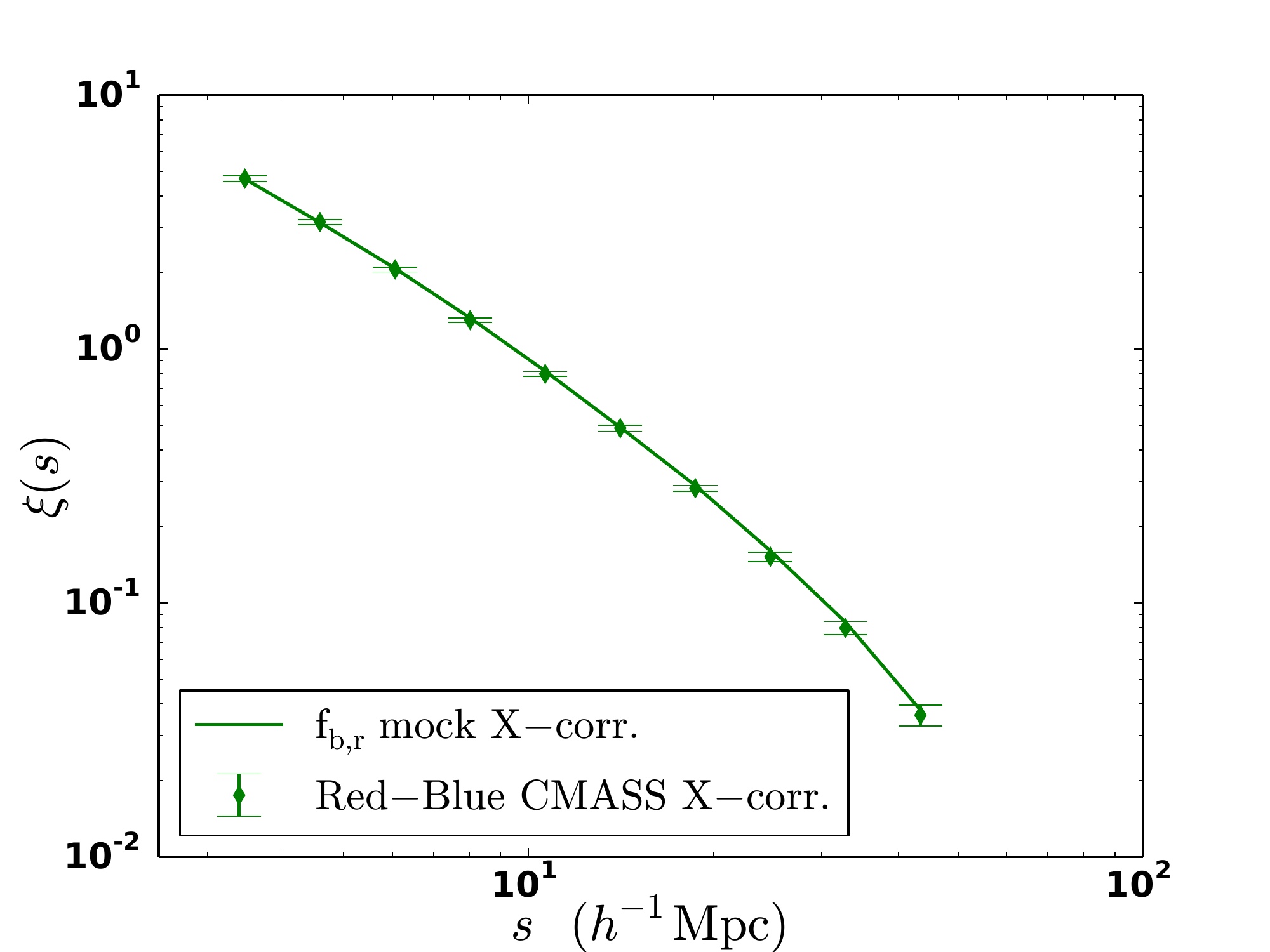}%
            \includegraphics[width=0.48\linewidth]{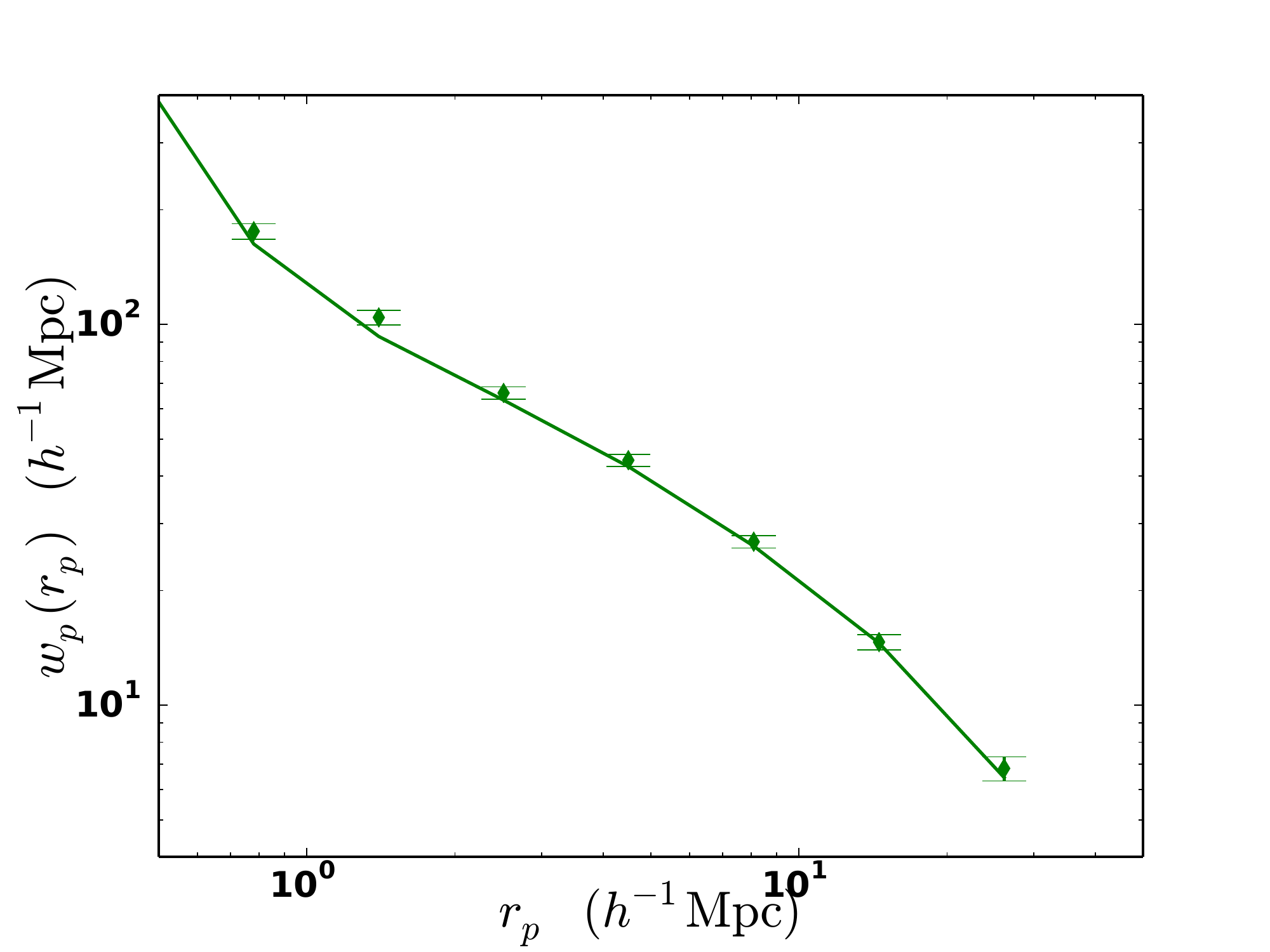}
  \end{center}
  \caption{Red-blue CMASS DR11 (diamonds) versus inverse tangent mock (lines) cross-correlation functions. These plots are useful to check the mutual behavior of the the red and the blue CMASS samples. In fact, as expected, we find that the cross-correlation of these galaxies lies in between their auto-correlation functions, and the size of the errorbars (computed with 200 jackknife resamplings) is consistent with the uncertainties on their individual clustering measurements.}
  \label{fig:cross_gal_frac}
\end{figure*}

Inspired by the result in the previous section, we developed a more physically motivated model of red/blue color separation.
In line with the standard halo model, we explore a splitting method based entirely on host halo mass, 
with each of them matching the corresponding observed CMASS galaxy fraction.  By modeling these red/blue fractions, 
$f_{b,r}$, as a function of the central halo mass, we are able to correlate the red and the blue mock catalogs to the full one, 
reducing the number of free parameters from 15 (5 for each independent HOD) to 5 (full HOD) plus 2 (constraint on galaxy 
fractions). Our galaxy fraction model must verify two conditions: (i) to obtain reliable results, the models must reproduce the overall 
$f_{b,r}$ values observed in our CMASS red/blue selection; this is done by requiring that
\begin{equation}
\begin{aligned}
&\Sigma_{i=1}^{N}f_b(\log M_h(i))/N=0.25,\\\\
&f_r(\log M_h)=1-f_b(\log M_h)=0.75
\end{aligned}
\label{eq:condition_mod}
\end{equation}
where we allow $20\%$ of scatter, and (ii) the red (blue) fraction must approach zero at low (high) mass scales.
 We build our theory as a function of the central halo mass only, omitting the dependence on satellite masses. Despite this 
 simplifying assumption, the resulting red and blue mocks match correctly the observed clustering amplitude.
 To mimic the red/blue split, we test different functional forms of $f_{b,r}$, starting with a basic linear one 
(Figure~\ref{fig:fbmodels}, dashed line) and two different log-normal models (dot-dashed and dotted curves) with three 
degrees of freedom each; 
they are treated in detail in Appendix \ref{sec:appendix4}.
 In order to produce a clear separation between the two populations, the best compromise is 
 an inverse tangent-like function (solid line), with only two free parameters. 
 The resulting functional form, as a function of the central halo mass, is
 \begin{equation}
\begin{aligned}
f_b(\log M_h)&=\frac{1}{2}-\frac{1}{\pi}\tan^{-1}\left[\frac{\log M_h-D}{10^C}\right],\\\\
f_r(\log M_h)&=1-f_b(\log M_h)
\end{aligned}
\label{eq:invtang}
\end{equation}
where the parameter $C$ determines how rapidly the blue fraction drops and $D$ establishes the half-width of the curve. 
Applying Eqs. \ref{eq:condition_mod}, \ref{eq:invtang} to the full CMASS mock catalog, we select the $(C,D)$ combination
 that globally best fits the observed red and blue redshift-space auto-correlation functions, $\xi(s)$. The best-fit values are $C=-0.50$, 
 $D=12.50$, with $\chi^2_{red}=15.43/5dof$, $\chi^2_{blue}=6.20/5dof$ and $\chi^2_{tot}=10.82/10dof$. 
 We use these red and blue inverse tangent mocks to match the other two statistics, $w_p(r_p)$ and 
 $\Sigma(\pi)$, which are shown in Figure \ref{fig:CF_indep_and_fbr_hod} and the cross-correlation 
 functions in Fig. \ref{fig:cross_gal_frac}. 
 The $\xi(s)$ fit is performed using the full covariance matrix and the uncertainties are estimated via 
 jackknife resampling (Sec.\;\ref{sec:Covariance_Estimation}). 

The cross-correlations between red and blue CMASS galaxies behave similarly to the auto-correlation
 functions: they are stronger on small scales and weaker when the pair separation increases. 

 These functions represent a consistency check of our red/blue fitting scheme and they provide robust information about red and 
 blue galaxy bias: the younger and more star-forming is the galaxy, the lower are its clustering amplitude and bias.

\begin{figure*}
  \begin{center}
 \includegraphics[width=0.5\linewidth]{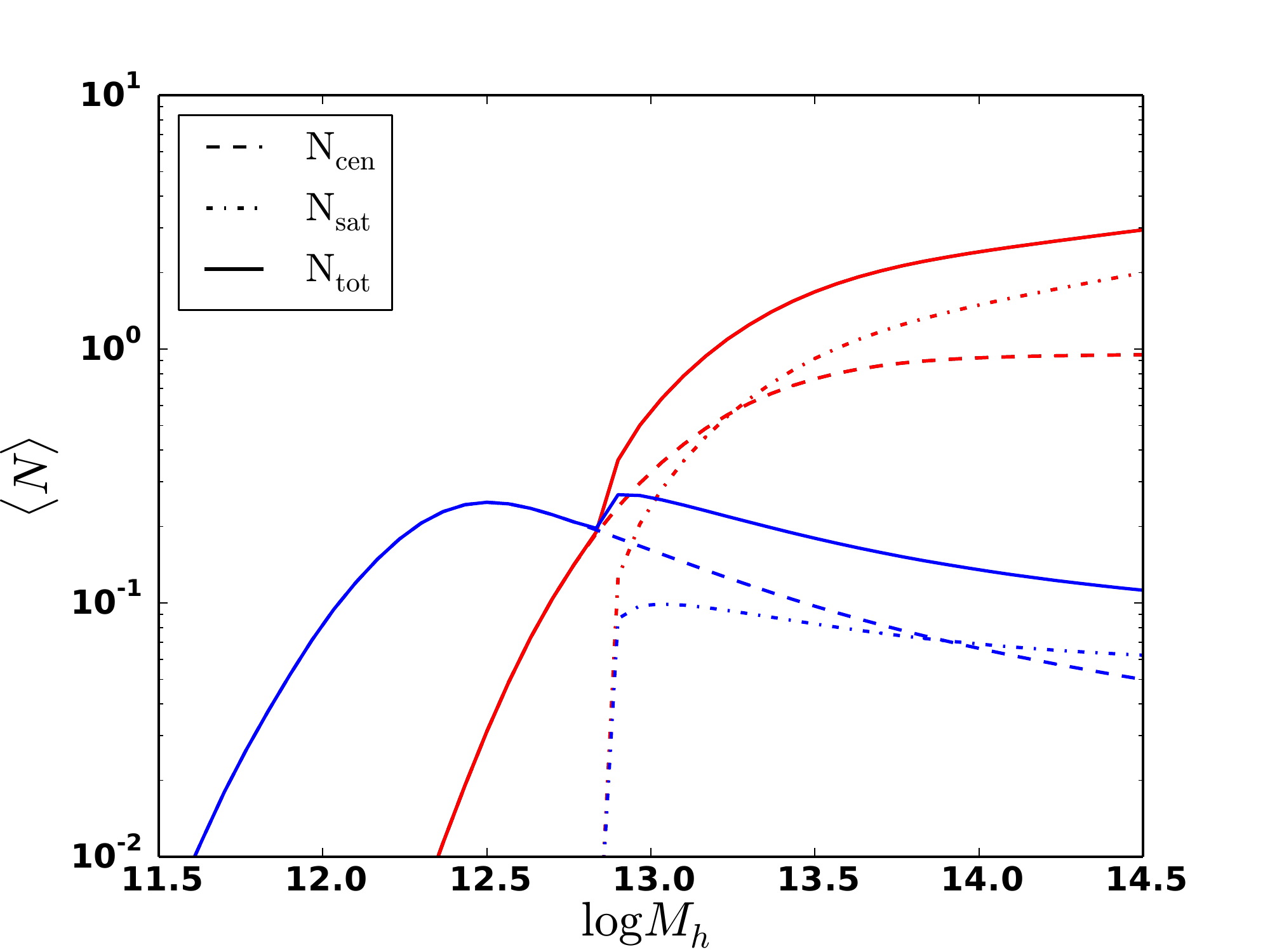}
    \caption{Red and blue HOD models obtained by applying the galaxy red/blue fraction condition to the 
    MultiDark mock catalog for the full CMASS population. The lines are the predictions computed by 
    normalizing $\langle N_{c}\rangle$, $\langle N_{s}\rangle$, $\langle N_{t}\rangle$ by $f_{b,r}$. For red galaxies, the HOD shape is consistent with Figure \ref{fig:HOD}, confirming that the red/blue galaxy separation we are imposing with the satellite fraction constraint is reliable for the red population. For blue mocks, the expected average number of galaxies per halo mass is about 10 times less than for red ones at $\log M_h=13.5$, and drops almost linearily as the halo mass increases. This reveals that blue star-forming galaxies preferentially populate low-mass halos.}
     \label{fig:red_blue_new_hod}
  \end{center}
\end{figure*}

Figure\;\ref{fig:red_blue_new_hod} displays the red and blue HOD models inferred by splitting the full MultiDark 
mock using the observed CMASS red/blue galaxy fraction. 
The lines are the predictions computed normalizing $\langle N_{c}\rangle$, $\langle N_{s}\rangle$, 
$\langle N_{t}\rangle$ by $f_{b,r}$. For red galaxies the HOD 
shape is compatible with the model shown in Figure\;\ref{fig:HOD}, confirming that the red/blue separation
 we imposed with the galaxy fraction constraint is reliable for the red population. For blue mocks, the average
  number of galaxies per halo mass is $\sim10$ times less compared to the red $\langle N_{cen}\rangle$, at $M_h=10^{13.5}\,h^{-1}$M$_{\odot}$ and drops 
  almost linearily ($3\%$ factor) as the halo mass increases. Such a trend reflecs the preference of blue star-forming galaxies to populate low-mass halos.

\begin{figure*}
\begin{center}
 \includegraphics[width=0.5\linewidth]{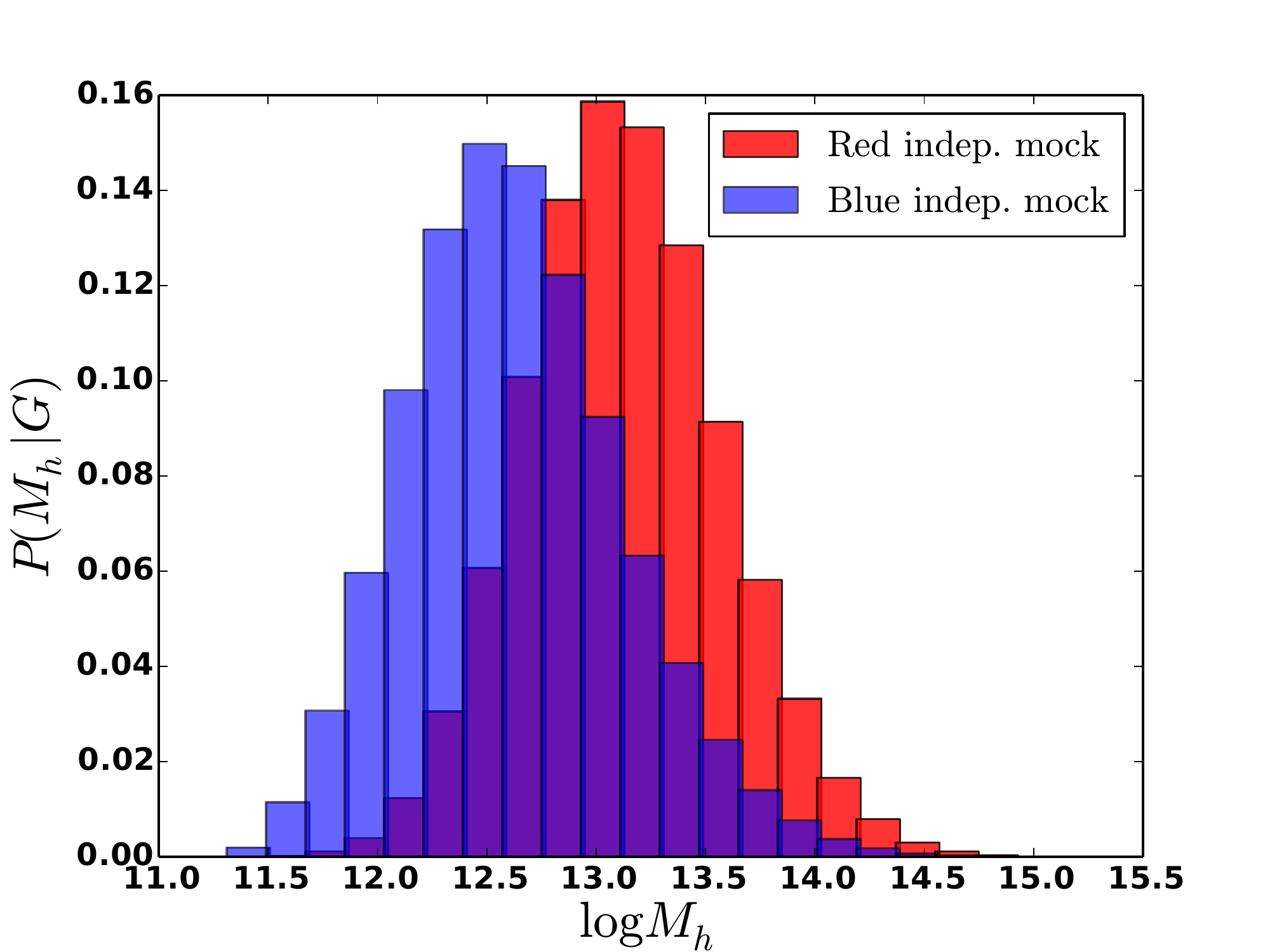}\hfill%
 \includegraphics[width=0.5\linewidth]{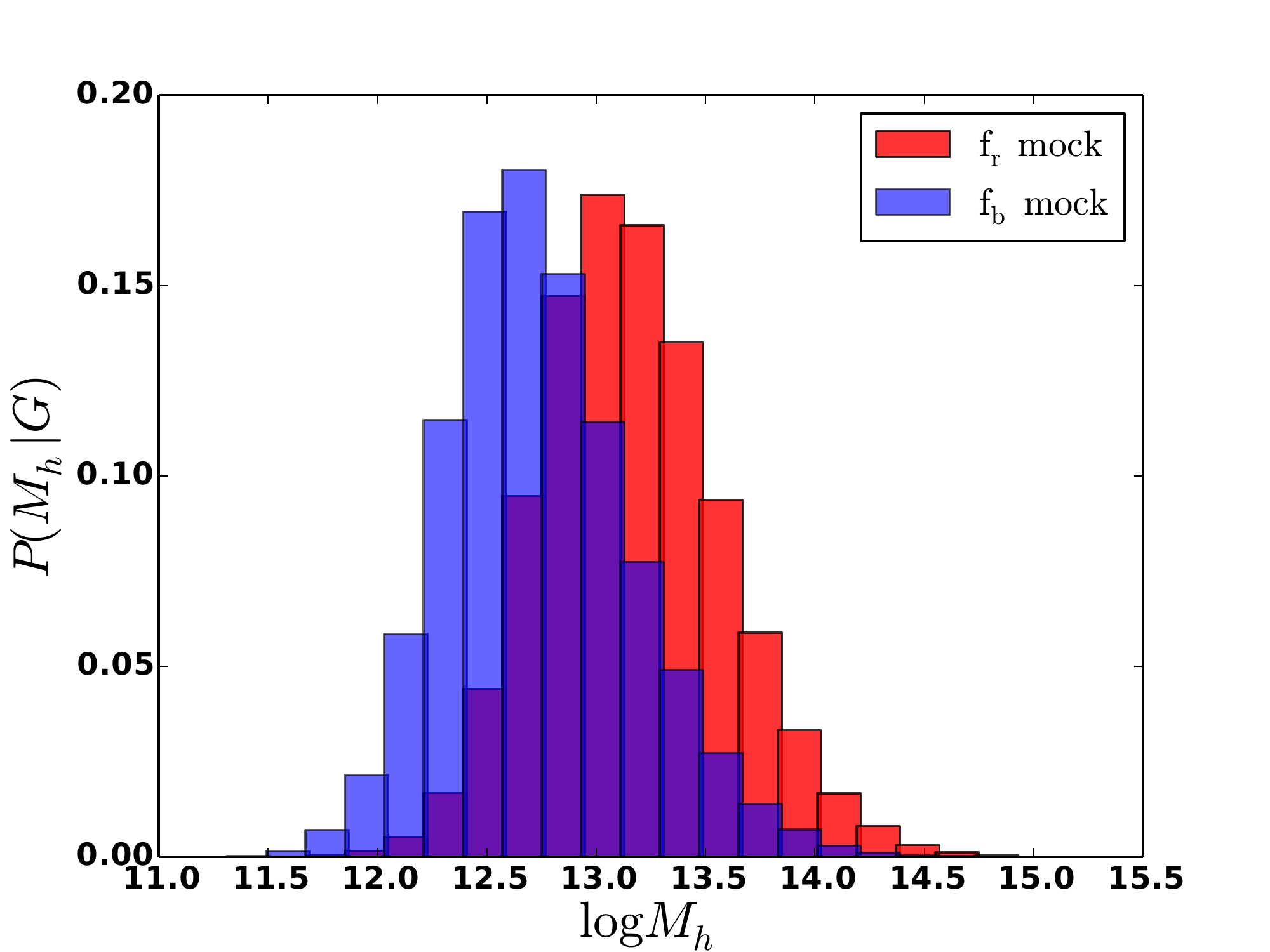}\hfill
 \caption{Conditional probability that a given galaxy $G$ with a specific color is hosted by a central halo 
 with mass $M_h$ obtained from our red and blue independent mock catalogs (left) and applying the galaxy fraction constraint (right). In both cases, as expected, we find that red galaxies live in more massive halos compared to the blue ones.}
 \label{fig:PMG}
 \end{center}
 \end{figure*}
From this analysis, we estimate the conditional probability, $P(M_h|G)$, that a galaxy $G$ with a specific color is
hosted by a central halo having mass $M_h$; see Figure~\ref{fig:PMG}. As expected, the result demonstratess that CMASS 
early-type redder galaxies are associated to more massive halos ($M_h\sim10^{13.1}\,h^{-1}$M$_{\odot}$), compared to the late-type bluer 
($M_h\sim10^{12.7}\,h^{-1}$M$_{\odot}$) companions.

 
\section{Results}
\label{sec:results}

\subsection{Red and Blue $A,G$ models}

\begin{figure*}
  \begin{center}
    \includegraphics[width=0.5\linewidth]{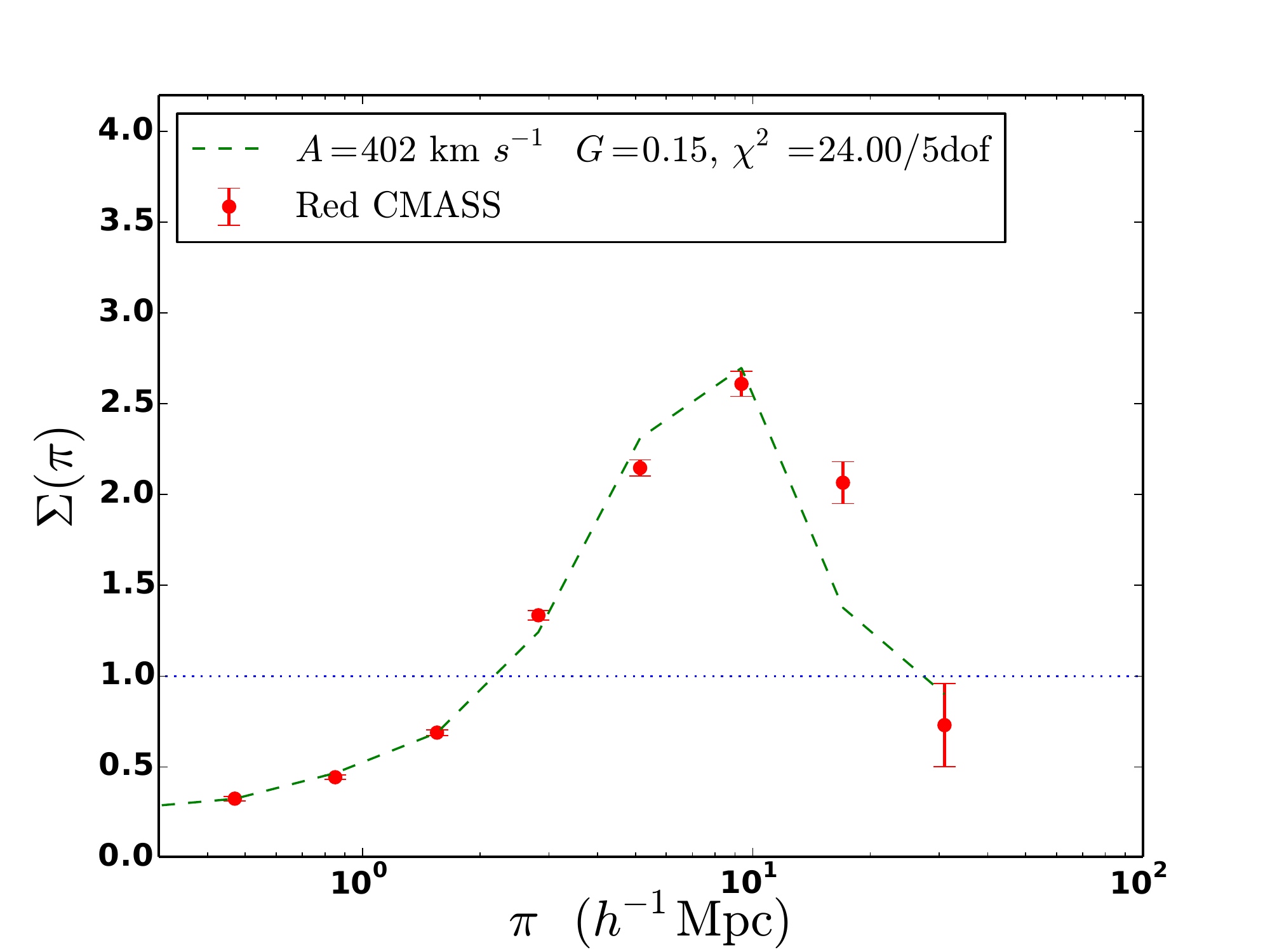}\hfill
    \includegraphics[width=0.5\linewidth]{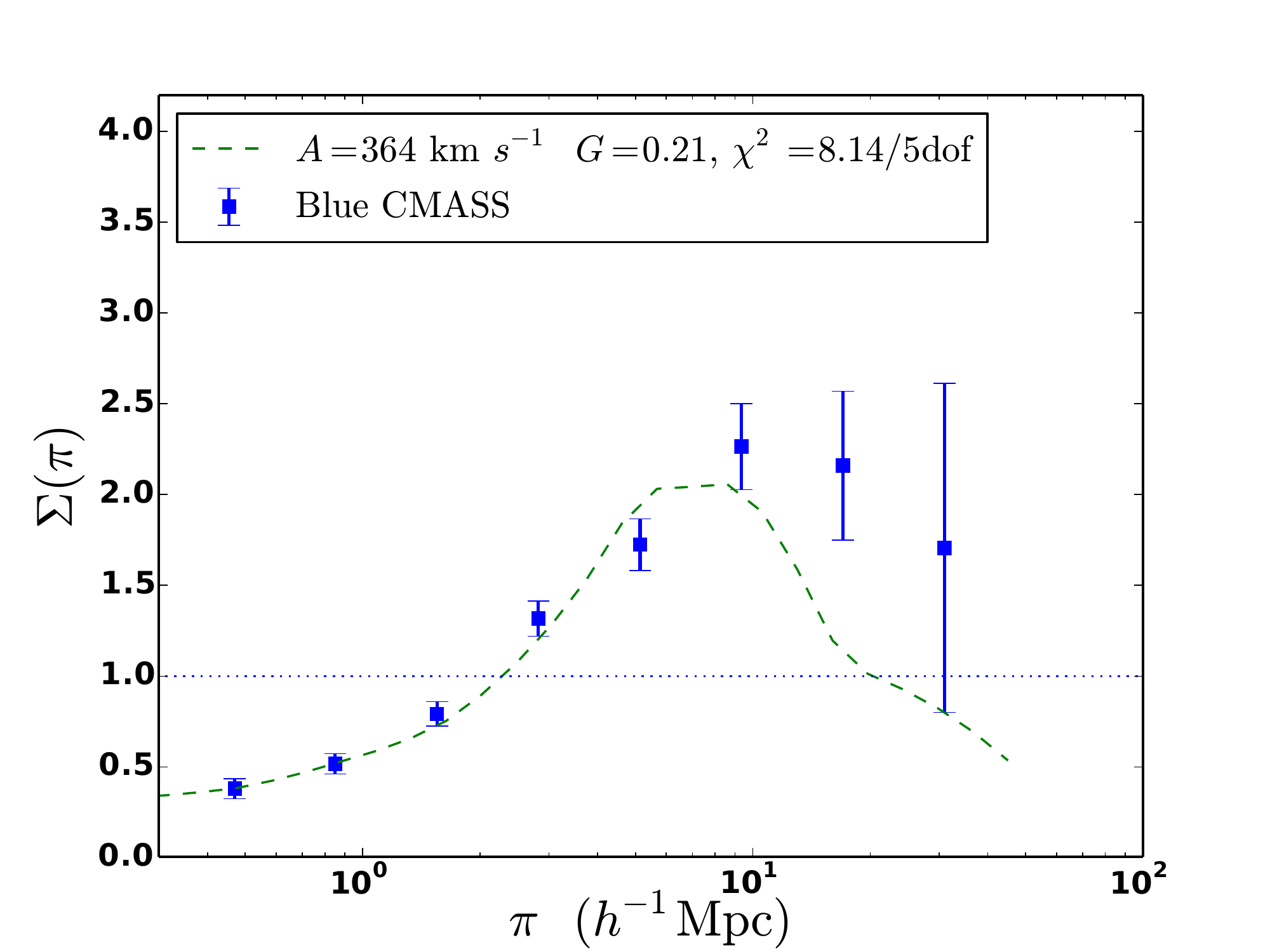}\hfill
    \includegraphics[width=0.5\linewidth]{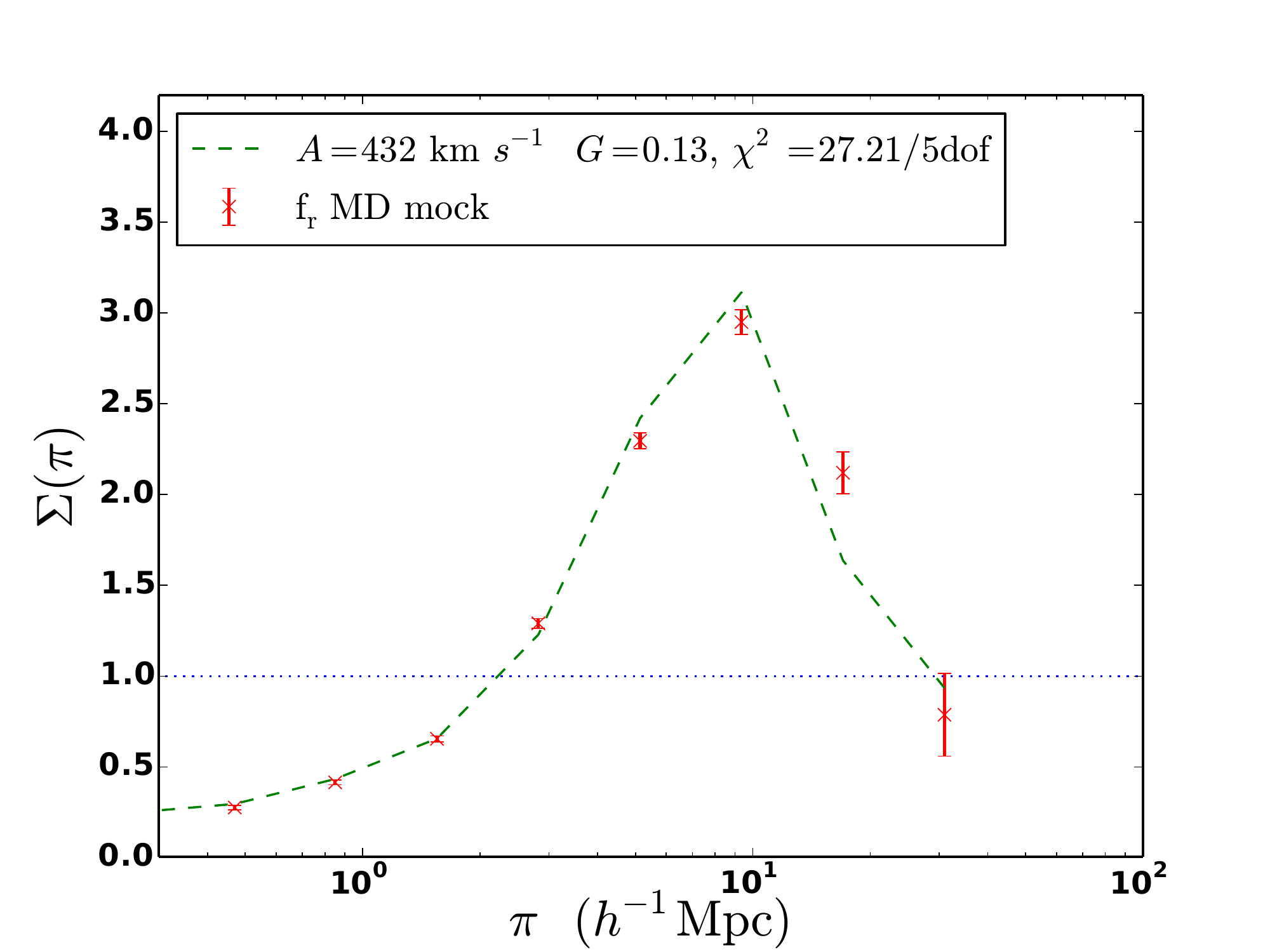}\hfill%
    \includegraphics[width=0.5\linewidth]{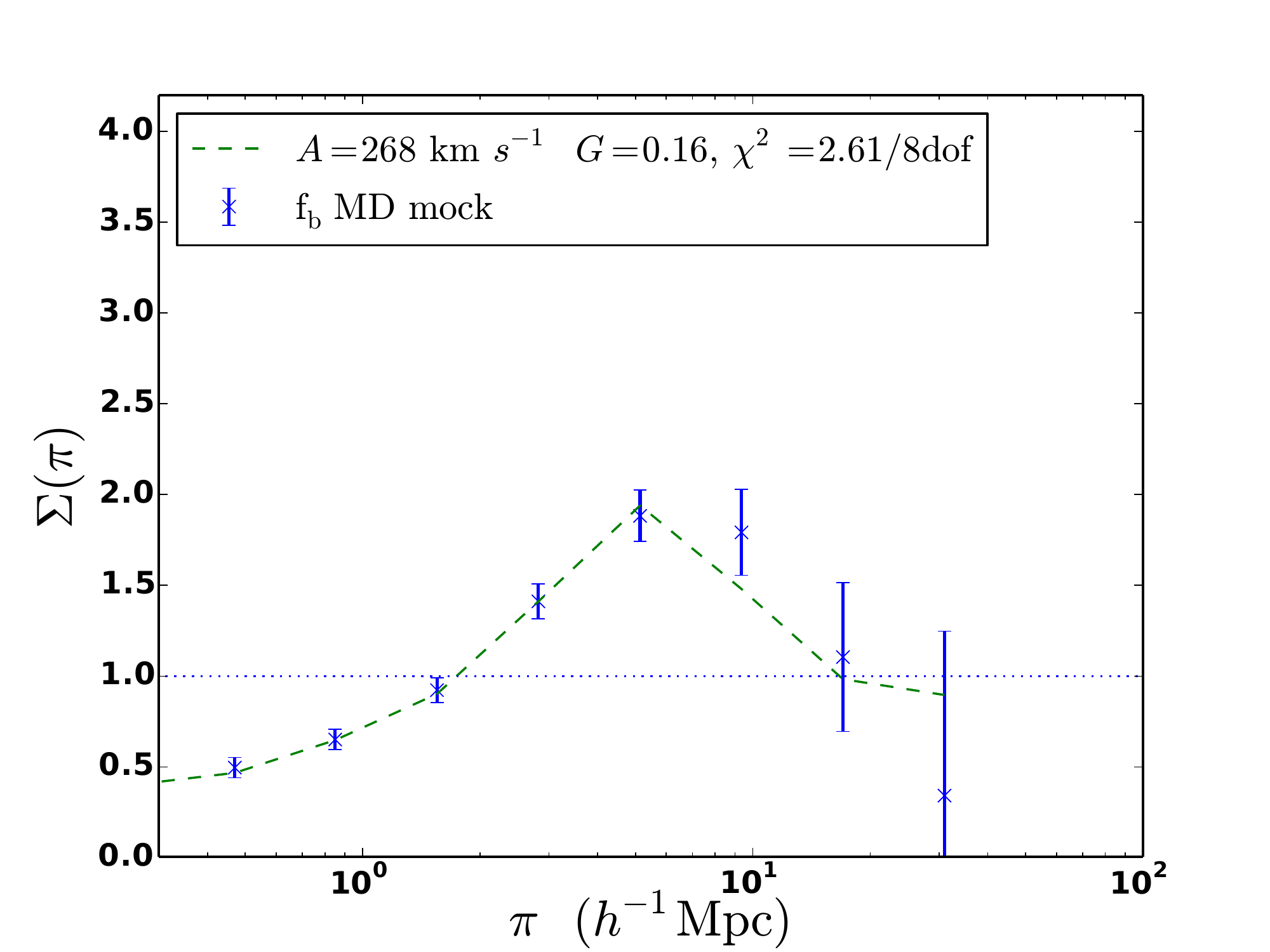}\hfill 
  \end{center}
  \caption{\textit{Top row:} CMASS DR11 $\Sigma(\pi)$ red (left) and blue (right) measurements and the $A,G$ analytic models (dashed lines). 
  \textit{Bottom row:} $f_{b,r}$ MultiDark mocks (solid curves) and their models (dashed lines). For the mocks we adopt the jackknife errors estimated for 
  the blue CMASS data doing jackknife. These fits are fully covariant. From these plots we conclude that blue CMASS galaxies are less biased and show a lower peculiar velocity contribution compared to the red population.}
  \label{fig:sigma_model_mocks_red_blueall}
\end{figure*}

We apply the same $A,G$ modeling performed in Section \ref{sec:Modeling_z_distortions} for the full CMASS sample and the MultiDark full 
mock galaxy catalog to the red and blue data samples and $f_{b,r}$ mocks, to quantify how significant their differences are 
at the level of large-scale bias and redshift-space distortions. Our main results are presented in Figure~\ref{fig:sigma_model_mocks_red_blueall}: 
the top row displays the red and blue $\Sigma(\pi)$ CMASS measurements (points and squares), versus the analytic models (dashed lines); 
in the bottom row are the results for the red and blue MD mocks (solid lines), versus their models (dashed curves). For both CMASS data and MD mocks we assume the errors are 
given by our jackknife estimate, done using 200 resamplings. All the model fits are fully covariant and our best estimate of the $A,G$ parameters are reported in Table \ref{tab:err_AG}.

As expected, the blue CMASS galaxies are less biased than the red population and have lower peculiar 
velocity contribution, which results in a lower clustering amplitude. A similar behavior is seen in a comparison of the red and the blue MultiDark model galaxies, we see a similar 
behavior, suggesting that we are correctly modeling our results in terms of redshift-space distortions and large-scale bias. As previously discussed in Section 
\ref{sec:Modeling_z_distortions}, our relatively high bias values are due to the fact that we are selecting the high-redshift tail ($z>0.55$) of the CMASS galaxies, for whom the bias is expected 
to be higher than the typical value reported by \cite{Nuza2013}, $b\sim2$. Also, the fact that our analysis produces high $\chi^2$ values is due to how the $\Sigma(\pi)$ 
measurement is built in terms of the 2PCF and to the numerical limitations of the $A,G$ model.
\begin{table*}
  \centering
  \begin{tabular}{||l l l l l||}
    \hline
    \hline
        &$\rm{\,A\,\,(km\,s^{-1})\,}$&$\rm{\,G\,}$&$\rm{\,b\,}$&$\rm{\,\chi^2}$\\
    \hline
     Full CMASS&384$\pm$6&0.15$\pm$0.01&$\sim3$&16.89/5dof \\ \\
    Full mock&$402^{+9}_{-6}$ & $0.14^{+0.01}_{-0.02}$&$\sim3$&36.20/6dof \\ 
        \hline
    Red CMASS&$402^{+8}_{-9}$ & $0.15^{+0.01}_{-0.02}$&$\sim3$&24.00/5dof \\\\
    Red mock & $432^{+10}_{-8}$&$0.13\pm0.01$&$\sim3.5$&27.21/5dof  \\ 
        \hline
    Blue CMASS &$364^{+47}_{-39}$&$0.21^{+0.05}_{-0.04}$&$\sim2$& 8.14/5dof\\\\
    Blue mock&$268\pm35$ &$0.16^{+0.07}_{-0.09}$&$\sim2.8$&2.61/8dof\ \\ 
    \hline
    \hline
  \end{tabular}
  \caption{Best-fit values of the $A,G$ parameters that model $\Sigma(\pi)$ in both full, red, blue CMASS measurements and MultiDark mocks. 
  All the fits are fully covariant. The bias is computed using the approximation given in Eq. \ref{eq:bias_approx}, where $\beta$ is our $G$ parameter, see Section \ref{sec:Analytic_Models}.}
  \label{tab:err_AG}
\end{table*}

\begin{figure}
 \begin{center}
 \includegraphics[width=\linewidth]{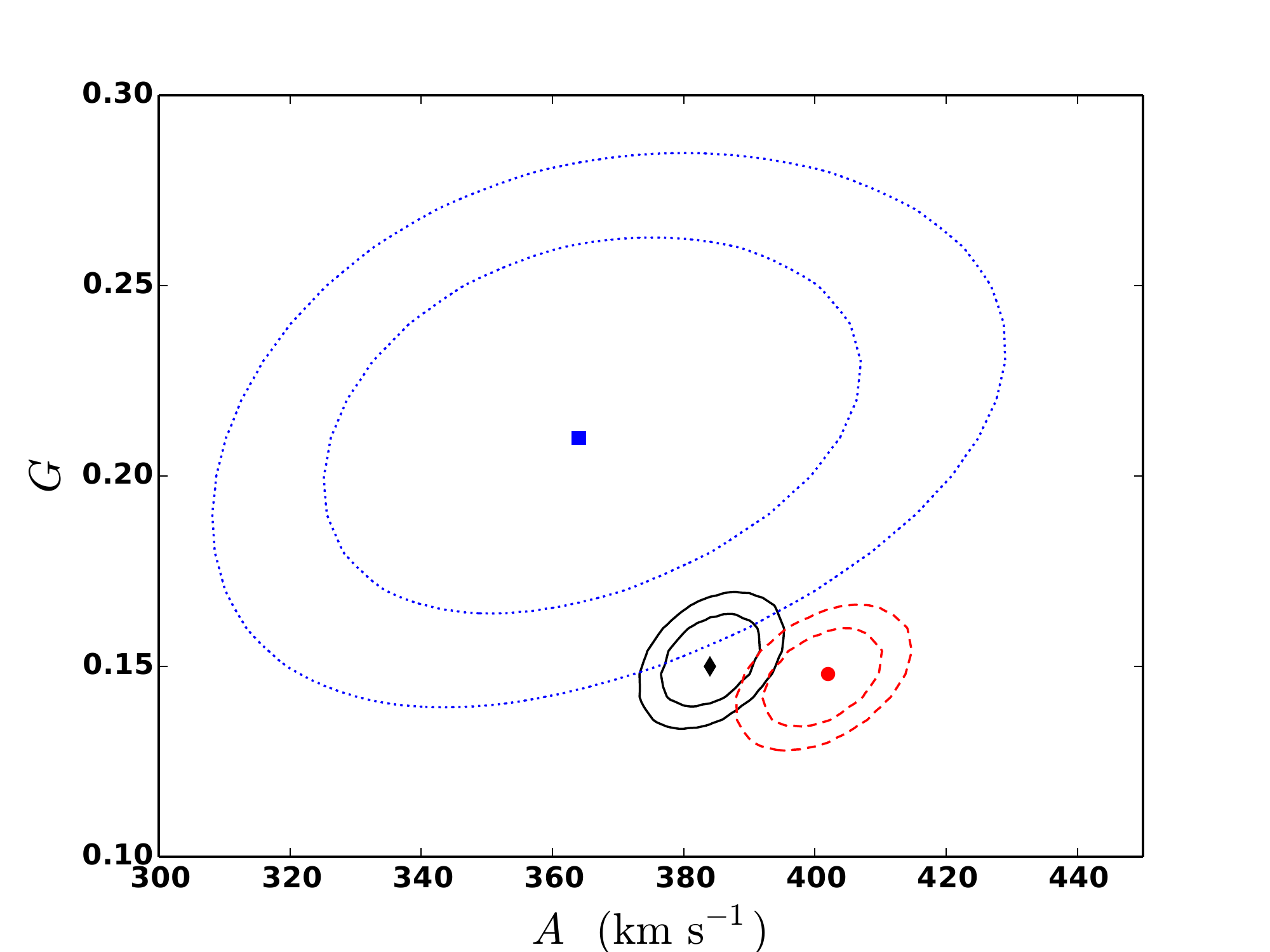} 
 \caption{68\% and 95\% confidence levels of the full (solid), red (dashed) and blue (dotted) $\Sigma(\pi)$ CMASS 
 measurements shown in Figs.\;\ref{fig:sigma_model_mocksall} (left panel) and \ref{fig:sigma_model_mocks_red_blueall} (top row). 
 All the contours include covariances. Consistently with the size of the error bars in Figure \ref{fig:sigma_model_mocks_red_blueall}, the blue contours are much less tight 
than the red and full ones. The blue CMASS galaxies are less biased and have lower velocity dispersion than the red and full populations.}
 \label{fig:chi2_surf}
\end{center}
\end{figure}

Figure~\ref{fig:chi2_surf} presents the $68\%$ and $95\%$ covariant confidence regions of the $A,G$
 models for the CMASS measurements. The $1\sigma$ blue region is spread out: 
 due to their larger uncertainties, blue galaxies have less power to constrain the $A,G$ values 
compared to the red and full CMASS populations. The dots indicate the position of the best-fit models for the three samples. 
As seen in Figure~\ref{fig:sigma_model_mocks_red_blueall}, red CMASS galaxies possess higher velocity dispersion 
and large-scale bias compared to the blue sample.

\subsection{large-scale bias}
\begin{figure}
 \begin{center}
  \includegraphics[width=\linewidth]{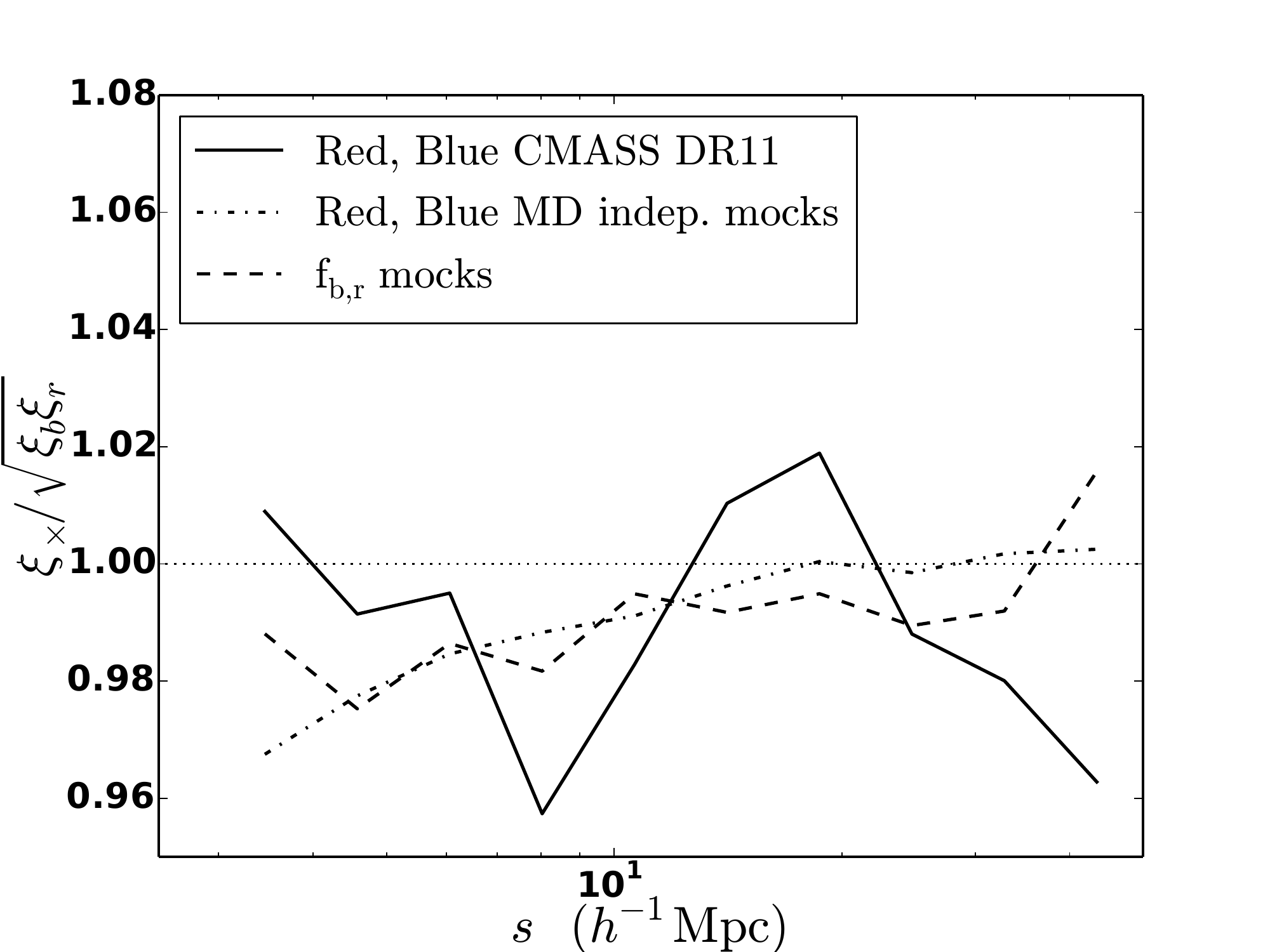}
 \caption{Ratio of the quantity $b_bb_r$ computed using the red-blue cross-correlation function, over the same quantity computed using the red and blue auto-correlation measurements. CMASS data (solid) versus independent (dot-dashed) and inverse tangent (dashed) mocks. Compatibly with expectations, the result is consistent with unity within $5\%$ and the fluctuations are Poisson noise.}
\label{fig:ratiobias}
\end{center}
\end{figure}
The linear bias factor $b$, defined in Eq. \ref{eq:bias_red_blue}, is related to the red-blue cross-correlation, $\xi_{\times}(s)$, by
\begin{equation}
b_r(s)b_b(s)=\frac{\xi_{\times}(s)}{\xi_m(s)}.
\label{eq:bias_red_blue_cross}
\end{equation}
where the subscripts $r,b$ indicate, respectively, red and blue galaxies, and $\xi_m(s)$ is the dark matter correlation function.
We then expect that the ratio $\xi_{\times}(s)/\sqrt{\xi_r(s)\xi_b(s)}$ \textemdash\; where each term in the denominator is given by Eq.\;\ref{eq:bias_red_blue} \textemdash\;
is close to unity. Figure\;\ref{fig:ratiobias} shows that our analysis produces a result that is consistent with expectations within $5\%$. 


\section{Discussion and Conclusions}
\label{sec:discussion}

We present a qualitative analysis of the galaxy clustering signal as a function of color in the BOSS CMASS DR11 sample. 
Applying the color cut defined in Eq.\;\ref{eq:colorcut}, we divide the full sample in a red and a blue component and 
compute the redshift-space and projected correlation functions, at small and intermediate scales ($0.1\leq r \leq 50 \;\hmpc$).  
Our measurements (see Section\;\ref{sec:Modeling_full_CMASS_Sample}) are consistent with previous results by 
\cite{Wang2007}, \cite{Zehavi2005b}, \cite{Swanson2008} and confirm that blue star-forming galaxies preferentially populate 
less dense environments, compared to the red ones. 

In addition, we describe a new metric, $\Sigma(\pi)$, defined in Eq. \ref{eq:sigma_new_metrics}, which provides robust information about 
nonlinear small-scale redshift-space distortions and large-scale bias. We map these results into the MultiDark cosmological 
simulation (Section \ref{sec:MD_Simulation}), using a five-parameter halo occupation distribution model (Section \ref{sec:HOD_general}), 
to generate reliable mock galaxy catalogs able to reproduce the observed clustering signal in the full, red and blue CMASS samples. \\

We separately model the full (Section \ref{sec:Modeling_full_CMASS_Sample}), red, and blue (Section \ref{sec:Independent_Red_Blue_mocks}) 
CMASS populations, building three independent mock galaxy catalogs (three different HOD models, with five dof each). 
We match our full, red and blue CMASS clustering measurements by empirically changing the HOD input parameters, 
until we find a set that reproduces the observed clustering amplitude. To simplify the task, we choose to vary only three parameters, 
specifically those values related to physical quantities we want to measure: $M_{min}$, the minimum host halo mass, which is connected 
to the galaxy number density, $M'_1$, governing the satellite fraction, and $\alpha$, the slope of the satellite contribution. 
Our best empirical estimates for the independent HODs are reported in Table~\ref{tab:HODparam} and confirm that red galaxies 
preferentially populate more clustered environments, where the satellite fraction is higher than for blue-star forming galaxies. 
This HOD model attempt suggests that we are able to individually match the clustering of full, red and blue CMASS samples, with small variations 
in the input parameters. 
Using these independent mocks, we calculate the probability, $P(M_h|G)$, that a specific galaxy $G$ is hosted by a halo with central mass $M_h$ 
(left panel of Figure~\ref{fig:PMG}), and estimate the mean central halo masses of our red and blue model galaxies. We find $M_h\sim10^{13.1},\,10^{12.5}\,M_{\odot}\,h^{-1}$, 
respectively for star-forming bluer and late-type redder galaxies, which again confirms that red galaxies live in more massive halos.

The traditional HOD formulation reproduces both red and blue CMASS clustering; however, it is based on a non-physical assumption: 
being independent, the red and blue models share a certain number of mock galaxies. This means the same galaxy 
can be either red or blue, whatever its mass is. 
In order to address this problem, we modify our HOD assignment to be able to infer both red and blue models from the full one, in such a way 
they are complementary and do not overlap. 
 To this purpose, we split the full mock catalog by using an appropriate model that reproduces the observed CMASS red/blue galaxy fraction, $f_{b,r}$ (Eq. \ref{eq:condition_mod}). 
 We test four different functional forms of $f_{b,r}$ (see Appendix \ref{sec:appendix4} for details), depending on a different number of parameters, and conclude that the best functional $f_{b,r}$ form 
 is an inverse-tangent-like function (Eq.\;\ref{eq:invtang}). The specific shape only has two free parameters, $C$ and $D$, that
 respectively govern how fast the blue (red) fraction drops (grows) as the halo mass increases and the position of the half-width point of the curve.  With this new HOD formulation, 
 we reduce to five the number of free parameters needed to build red and blue models from the full mock (i.e., five from the full mock, plus two from the $f_{b,r}$ condition).
 Our main results are presented in Figure\;\ref{fig:CF_indep_and_fbr_hod} and show good agreement between our model galaxies and the observations.
\\

We then quantify the differences in the blue and red populations from the point of view of the redshift-space distortions and large-scale bias (Section\;\ref{sec:results}). 
Two regimes are interesting to this purpose: 
on large scales, the gravitational infall of galaxies to density inhomogeneities compresses the two-point correlation function along the line-of-sight direction; on small scales, 
the 2PCF experiences an elongation effect due to the nonlinear peculiar velocities of galaxies, with respect to the Hubble flow (see Sec.
 \ref{sec:Clustering_Measurements}). In order to separate the two contributions and study the small scale stretching effect, we build the new metric $\Sigma(\pi)$, defined
  in Eq.\;\ref{eq:sigma_new_metrics} as the ratio between $\xi(r_p,\pi)$ \textemdash\; averaged in the range $0.5\leq r_p<2\,h^{-1}$Mpc to maximize the FoG effect \textemdash\; 
  and the best-fit spherical averaged power law to the projected correlation function, $w_p(r_p)$. Using this approach, we derive a robust prediction of the deviation of $\xi(r_p,\pi)$
   from the real space behavior. To estimate the contribution of both effects, we model $\Sigma(\pi)$ by convolving the real-space best-fit power 
   law to $w_p(r_p)$, with a peculiar velocity term, assumed to be a normal function (Eq.\;\ref{eq:vpec_norm}) and the Kaiser factor (Eq.~\ref{eq:kaiser_fac}). The resulting model 
   only depends on two parameters: $G$, that measures the Kaiser compression and is proportional to the inverse of the linear bias, $b$, and $A$, that is the pairwise velocity 
   dispersion, which quantifies the FoG elongation effect. Fitting this $A,G$ parametrization to our full, red, blue $\Sigma(\pi)$ CMASS and MD mock results demonstrates (see Table \ref{tab:err_AG}) that blue galaxies are less 
   biased than red ones and have a lower peculiar velocity contribution, which leads to a smaller clustering amplitude.    

\section*{Acknowledgments}

GF is supported by the Ministerio de Educaci\'{o}n y Ciencia of the Spanish Government through FPI grant AYA2010-2131-C02-01 and wishes to thank the Smithsonian 
Center for Astrophysics and the Harvard University, Astronomy Department, for the hospitality during the creation of this work. GF acknowledges support from the 
Spanish Government through EEBB-I-13-07167 and EEBB-I-12-05220 grants.

GF, FP, and coauthors acknowledge support from the Spanish MICINNs Consolider-Ingenio 2010 Programme under grant MultiDark CSD2009-00064, 
MINECO Centro de Excelencia Severo Ochoa Programme under grant SEV-2012-0249, and MINECO grant AYA2014-60641-C2-1-P.

Funding for SDSS-III has been provided by the Alfred P. Sloan Foundation, the Participating Institutions, the National Science Foundation and the U.S. Department of Energy.

SDSS-III is managed by the Astrophysical Research Consortium for the Participating Institutions of the SDSS-III Collaboration including the University of Arizona, the Brazilian Participation Group, Brookhaven National Laboratory, University of Cambridge, Carnegie Mellon University, University of Florida, the French Participation Group, the German Participation Group, Harvard University, the Instituto de Astrof\'{i}sica de Canarias, the Michigan State/Notre Dame/JINA Participation Group, Johns Hopkins University, Lawrence Berkeley National Laboratory, Max Planck Institute for Astrophysics, Max Planck Institute for Extraterrestrial Physics, New Mexico State University, New York University, Ohio State University, Pennsylvania State University, University of Portsmouth, Princeton University, the Spanish Participation Group, University of Tokyo, University of Utah, Vanderbilt University, University of Virginia, University of Washington and Yale University.


\bibliography{./bibliography}

\appendix

\section{Clustering Sensitivity on HOD Parameters}
\label{sec:hod_gradient}

\begin{figure*}
\begin{center}$
\begin{array}{c}
\includegraphics[width=0.4\linewidth]{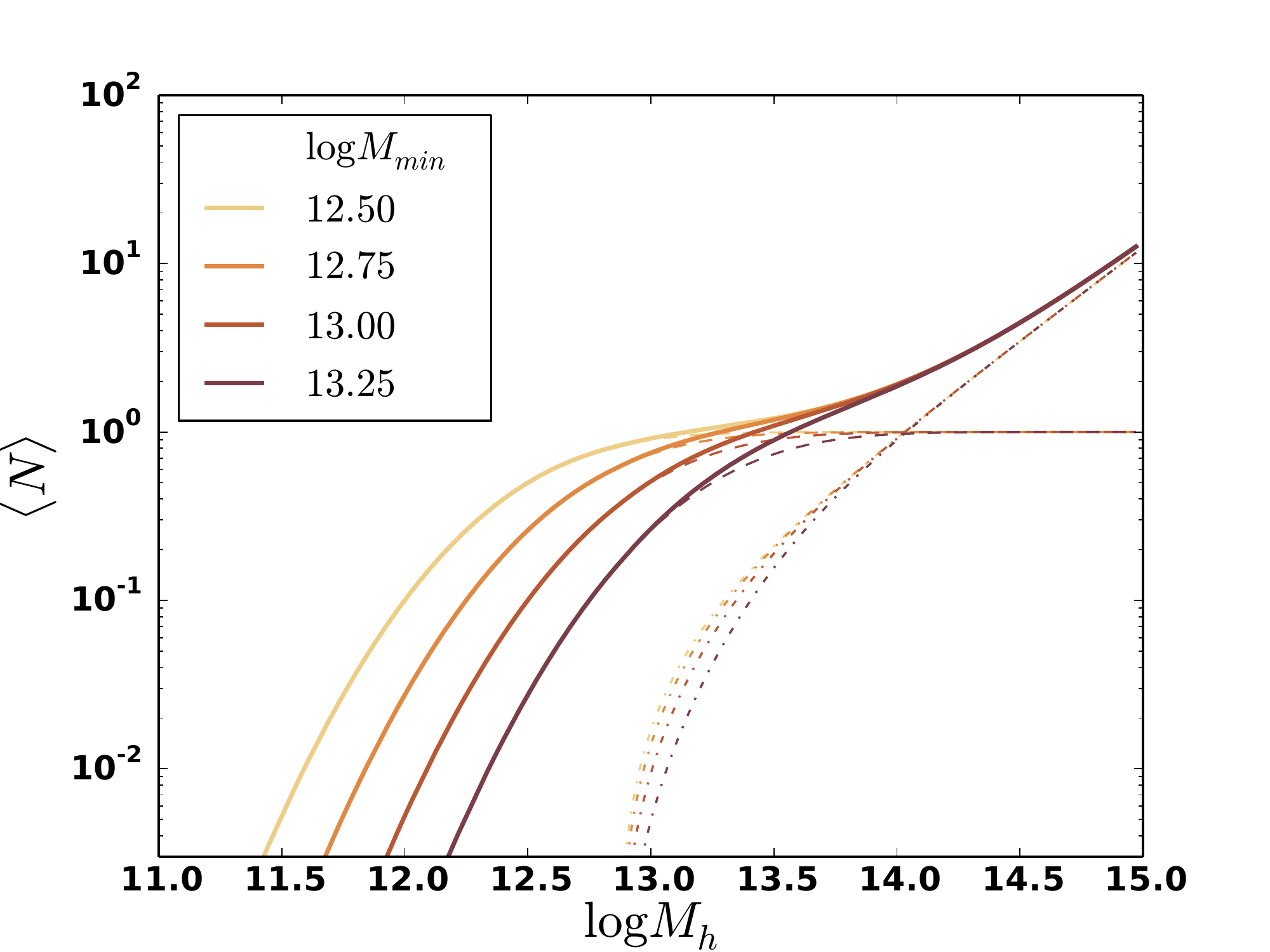} \hfill%
\includegraphics[width=0.4\linewidth]{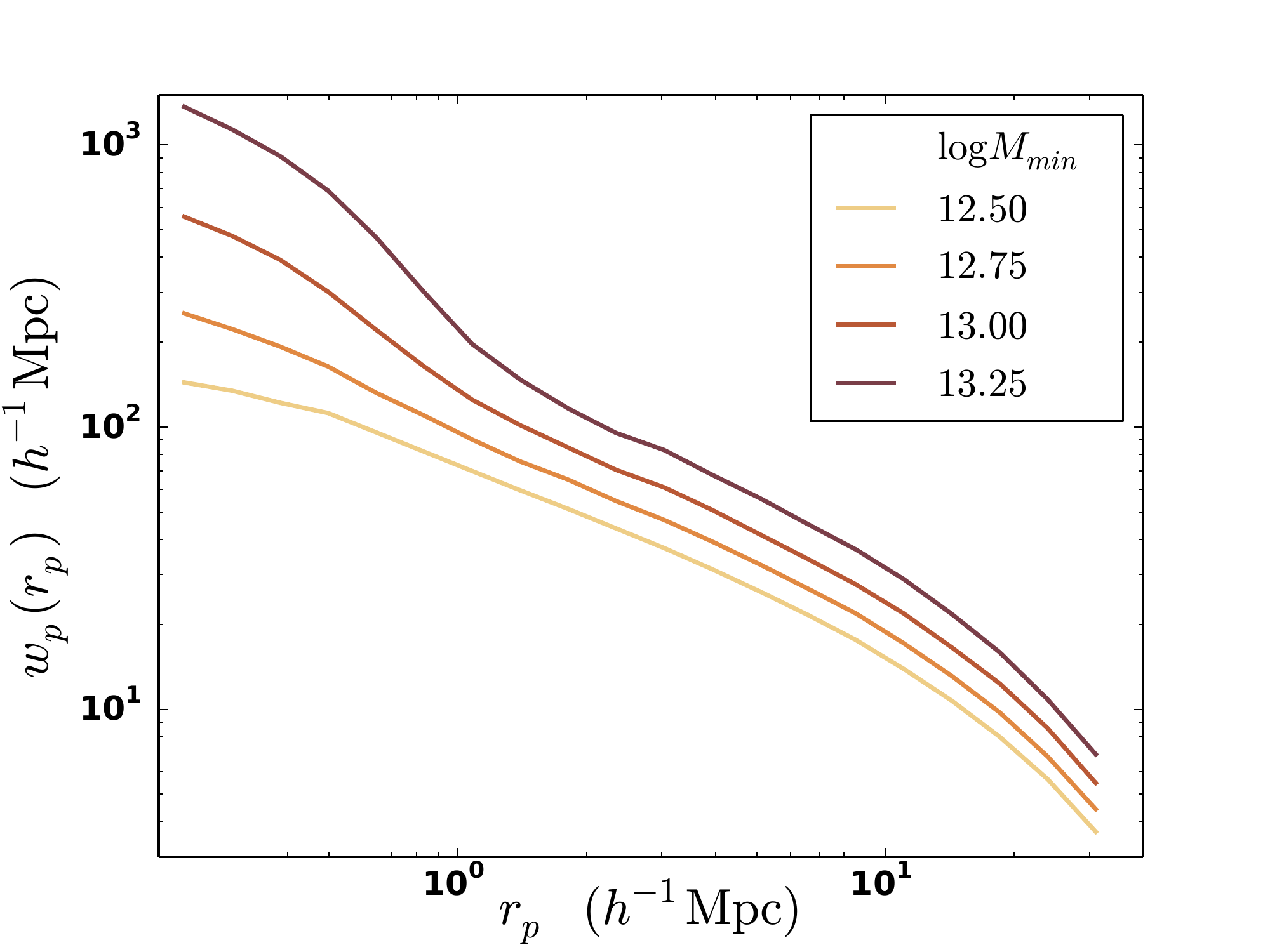}\hfill\\
\includegraphics[width=0.4\linewidth]{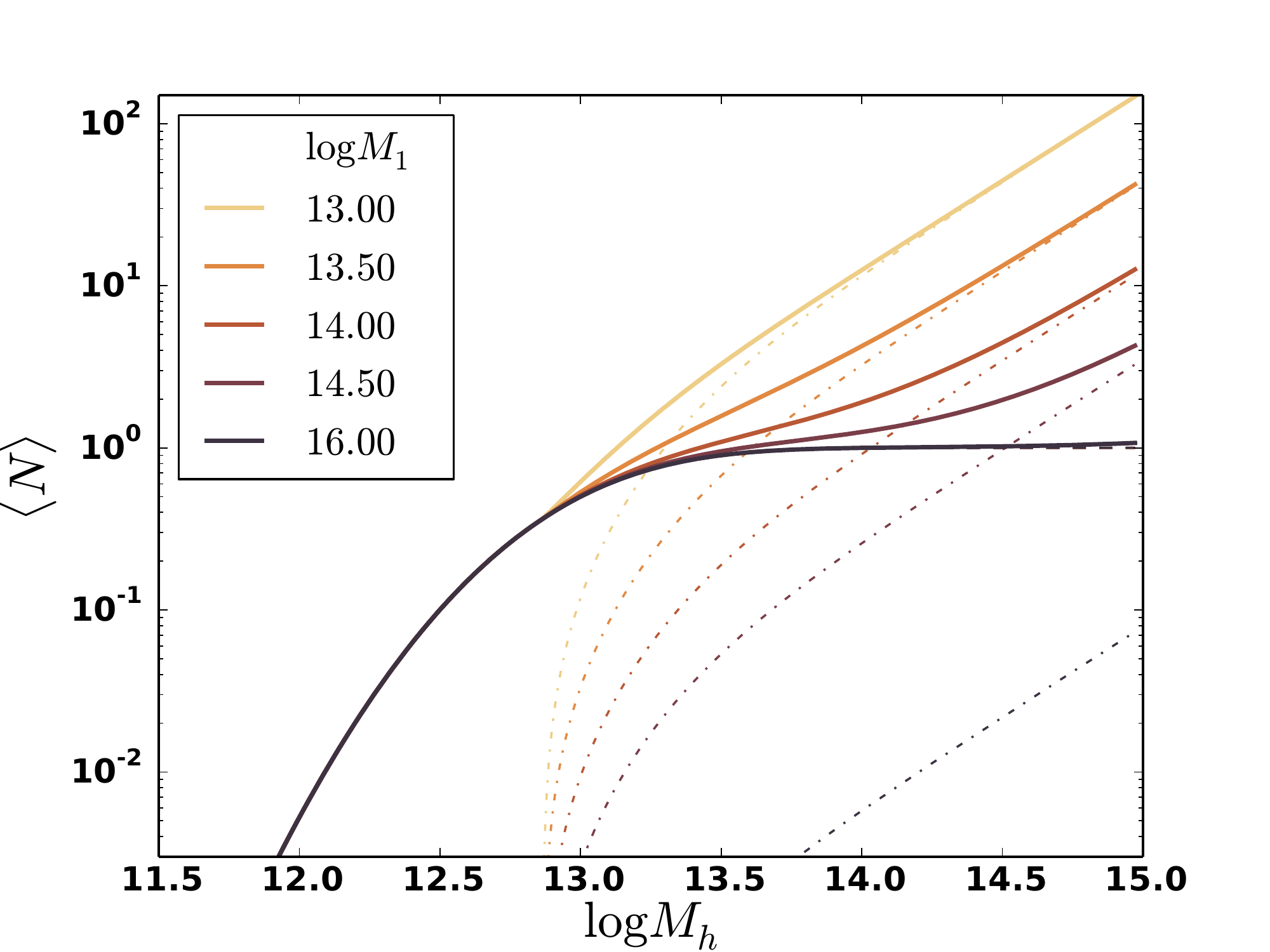} \hfill%
\includegraphics[width=0.4\linewidth]{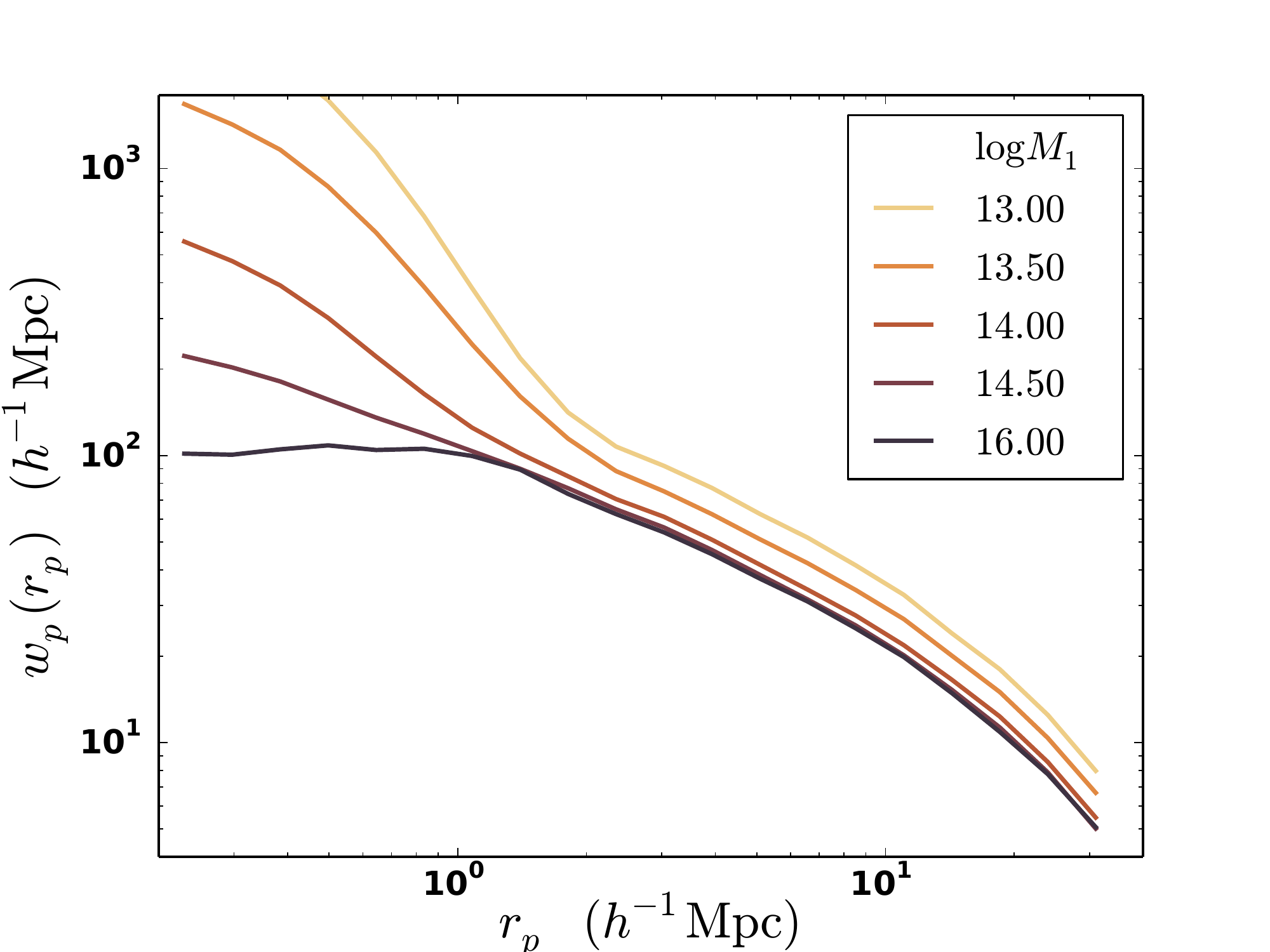}\\
\includegraphics[width=0.4\linewidth]{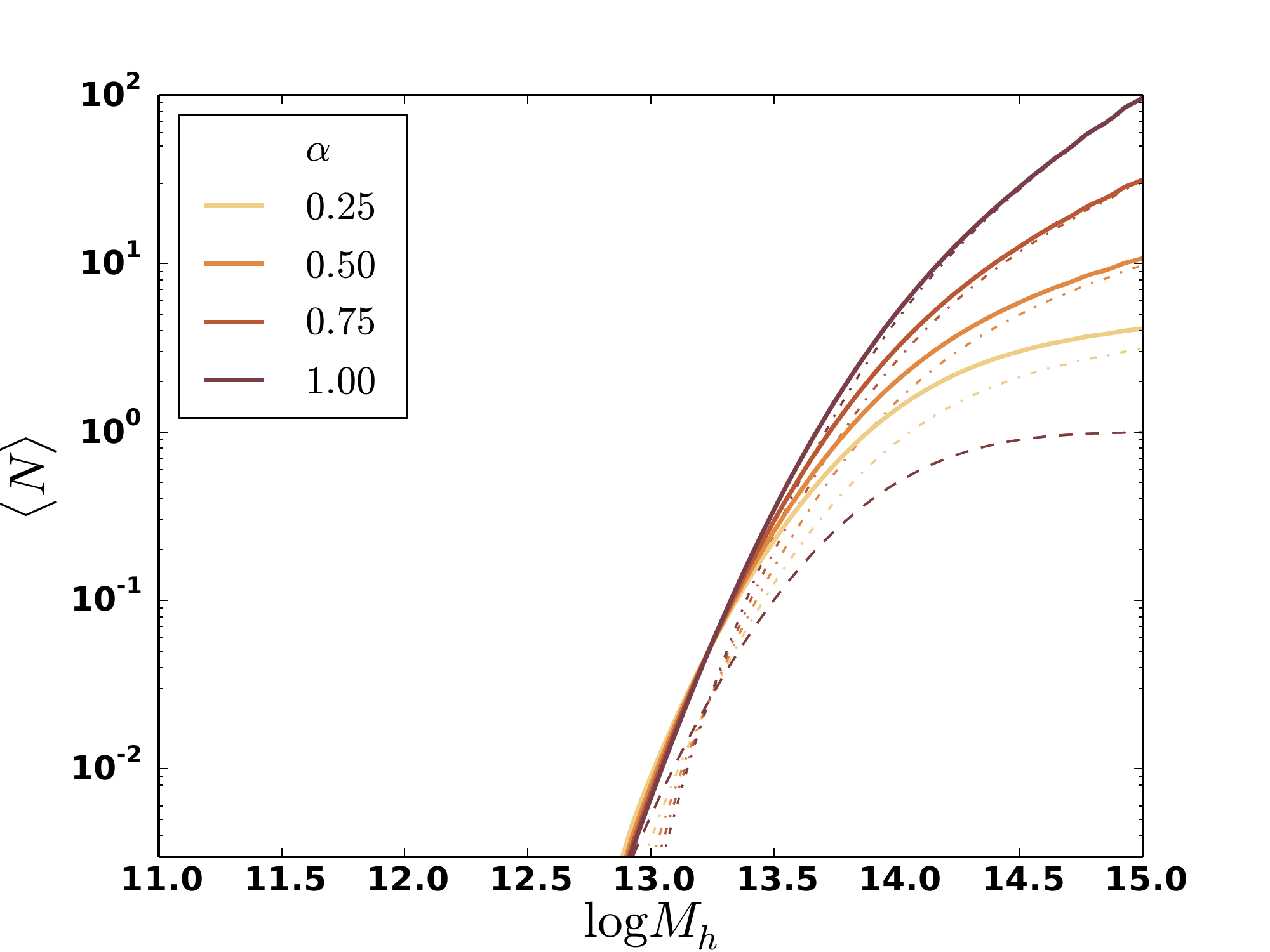} \hfill%
\includegraphics[width=0.4\linewidth]{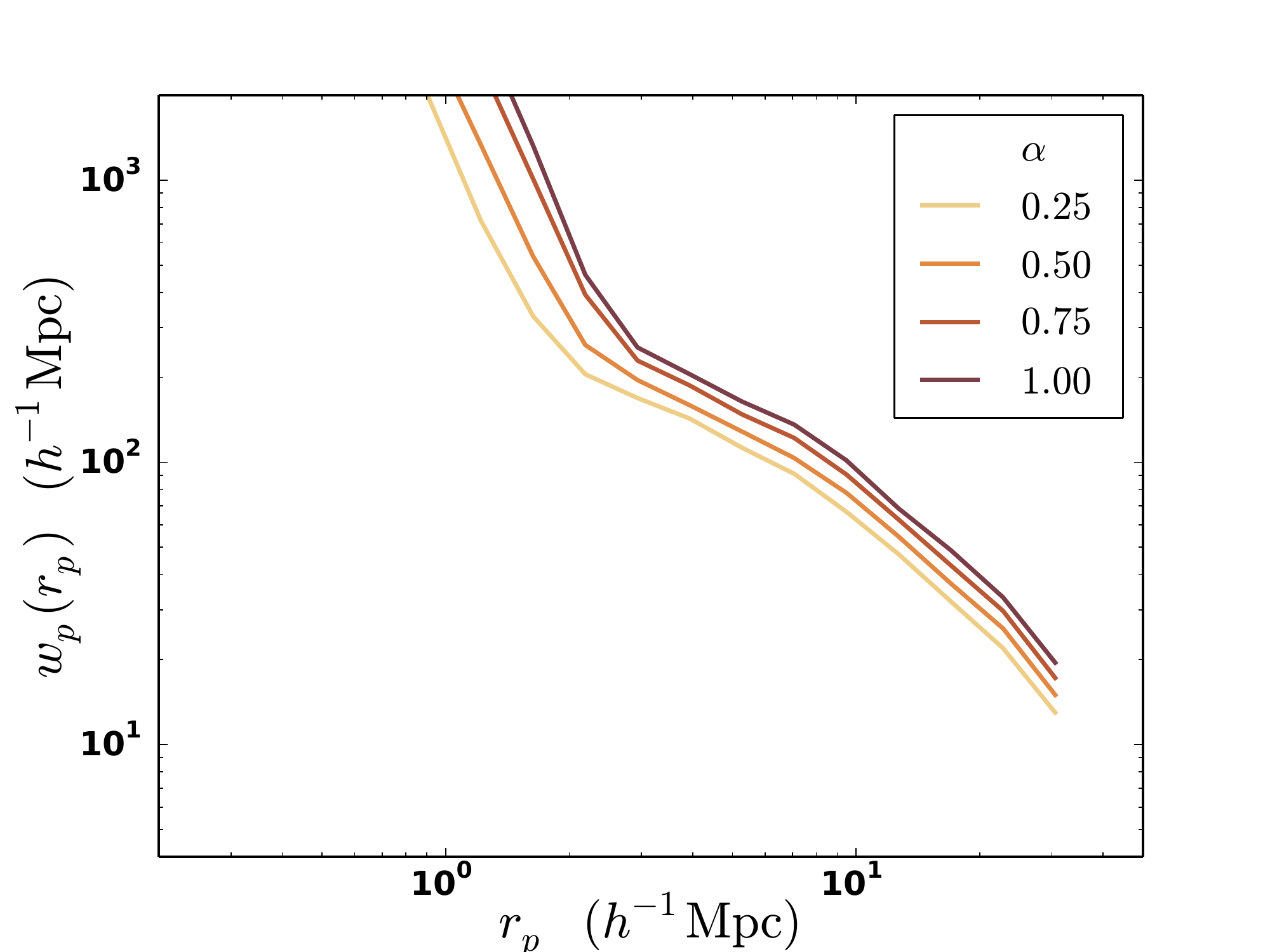}
\end{array}$
\end{center}
\caption{Implication of a change in the HOD input parameters (left column)  on the projected correlation function (right column). 
We allow only one parameter to vary at a time: $M_{min}$, in the top row, especially affects the 2-halo term; 
$M'_1$ ($\log M_{min}=13.00$) and $\alpha$, respectively in the middle and bottom row, have a strong effect on the 1-halo term. 
The resulting correlation functions are degenerate with respect to the variation of these three parameters. 
The remaining two parameters are fixed at the default values given by \citeauthor{White2011} (\citeyear{White2011}): $\log M_0=12.8633$, $\sigma_{\log M}=0.5528$. }
\label{fig:HODvariation}
\end{figure*}

The left column in Figure\;\ref{fig:HODvariation} presents our HOD model (see Section \ref{sec:HOD_general}) as a function 
of three parameters: $M_{min}$ (top row), $M'_1$ (middle), and $\alpha$ (bottom); the remaining two parameters are fixed to the default values 
given by \cite{White2011}: $\log M_0=12.8633$, $\sigma_{\log M}=0.5528$ . The projected correlation
 functions based on these mocks are shown in the right column. 
Increasing the value of  $M_{min}$ (top row, from lighter to darker solid lines) globally enhances the clustering amplitude,
 with a strong contribution from sub-structures belonging to different hosts ($2$-halo term). On the other side, the interaction 
 between satellites belonging to the same central halo ($1$-halo term) weakens as $M'_1$ increases (bottom row, from lighter 
 to darker solid lines), resulting in a smoother slope at scales $r_p\leq1 \;\hmpc$. The extreme case is achieved when 
 $\log M_1=$16.00, where the satellite contribution becomes almost negligible, and $f_{sat}=5.45\times10^{-4}\simeq0$.


\section{Red and Blue Galaxy Fraction models}
\label{sec:appendix4}

In addition to the inverse tangent fraction model defined in Eq.\;\ref{eq:invtang}, to mimic the red and blue galaxy fractions as a
 function of the central halo mass, we test also a linear model
\begin{equation}
f_b(\log M_h)=-M\log M_h+N,
\label{eq: linear}
\end{equation}
and two log-normal like functions, with three degrees of freedom each. The first one (Logn I) is given by
\begin{equation}
f_b(\log M_h)=\frac{P_b}{P_b+P_r},
\label{eq: lognI}
\end{equation}
where 
\begin{equation}
P_{b,r}=\exp\left(-\frac{(\log M_h-\mu_{b,r})^2}{2\sigma^2}\right)
\label{eq:density}
\end{equation}
is a density function. The parameters $\mu_{b,r}$ are the blue and red, mean galaxy masses, respectively, and $\sigma$ is the log-normal width.
The second version (Logn II) has fixed amplitude $\sigma$, and a new parameter $k$, that controls the mutual heights of the red and blue peaks. We have
\begin{equation}
f_b(\log M_h)=\frac{P_b}{P_b+kP_r},
\label{eq:lognII}
\end{equation}
where $P_{b,r}$ is given by Eq.\;\ref{eq:density}.
After applying these constraints to the full MultiDark mock catalog, we split it into its red and blue components. We then fit the clustering 
amplitudes of our model galaxies to the CMASS red and blue samples.


\section{Testing the errors \textemdash\; jackknife versus QPM mocks}
\label{sec:Jackknife_Poisson}

\begin{figure}
 \begin{center}

\includegraphics[width=\linewidth]{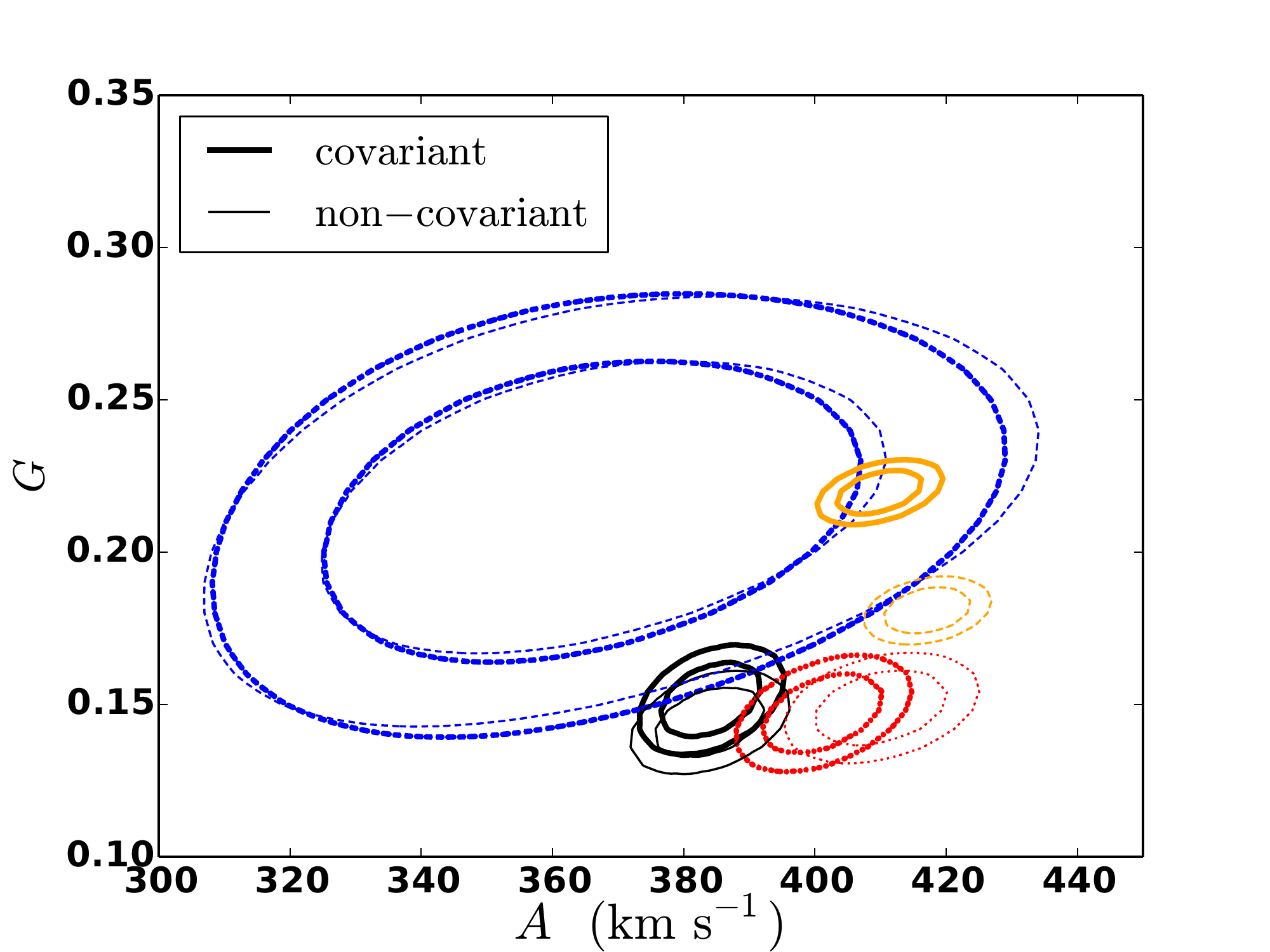}
 \caption{Covariant (thick contours) versus non-covariant (weak lines) $68\%$ and $95\%$ confidence levels of the $A,G$ 
 models for the $\Sigma(\pi)$ full (black solid), red (red dotted) and blue (blue dashed) CMASS measurements versus QPM mocks (orange dashed). QPMs have slightly different cosmology: $\Omega_m=0.29$.
The inclusion of covariances is almost negligible for the blue population, and weakly appreciable in the full case. Inversely, in the red population, covariances 
slightly move the fit towards higher velocity values; for QPMs, this shift is significant and drives the contours towards smaller bias values and slightly higher velocities.}
 \label{fig:chi2surf}
\end{center}
\end{figure}

We test our full CMASS jackknife error estimates by computing the $\xi(s)$, $w_p(r_p)$, and $\Sigma(\pi)$ covariance matrices from a set of 100 Quick Particle 
Mesh (QPM; \citeauthor{White2014} \citeyear{White2014}) mock catalogs, with slightly different cosmology: $\Omega_m=0.29$. 
Since these mocks are all independent of each other,  we can compute their covariance as 
\begin{equation}
C_{kl}^{QPM}=\frac{1}{n_{QPM}-1}\sum_{b=1}^{n_{QPM}} (\xi^b_{k}-\bar{\xi}_{k})(\xi^b_{l}-\bar{\xi}_{l}),
\label{eq:cov_matrix_qpm}
\end{equation} 
where $n_{QPM}=100$, and $\bar{\xi}_{k}$ is the mean QPM correlation function in the $k^{th}$ bin,
\begin{equation} 
  \bar{\xi}_k=\sum_{b=1}^{n_{QPM}} \xi^b_{k}/n_{QPM}.
\end{equation}

Figure \ref{fig:chi2surf} shows the covariant (thick lines) and the non-covariant (weak) $A,\, G$ contours of the full, red and blue CMASS $\Sigma(\pi)$ models versus QPM mocks (orange). 
The inclusion of covariances is almost negligible for the blue CMASS model, while it moves the full and red models toward smaller 
 bias values and higher velocity dispersion values, respectively. QPM contours are narrow, analogously to the full CMASS sample, and the inclusion of covariances in this case significantly moves the fit towards lower bias values and slightly higher velocities.

\begin{figure*}
 \begin{center}
 \includegraphics[width=\linewidth]{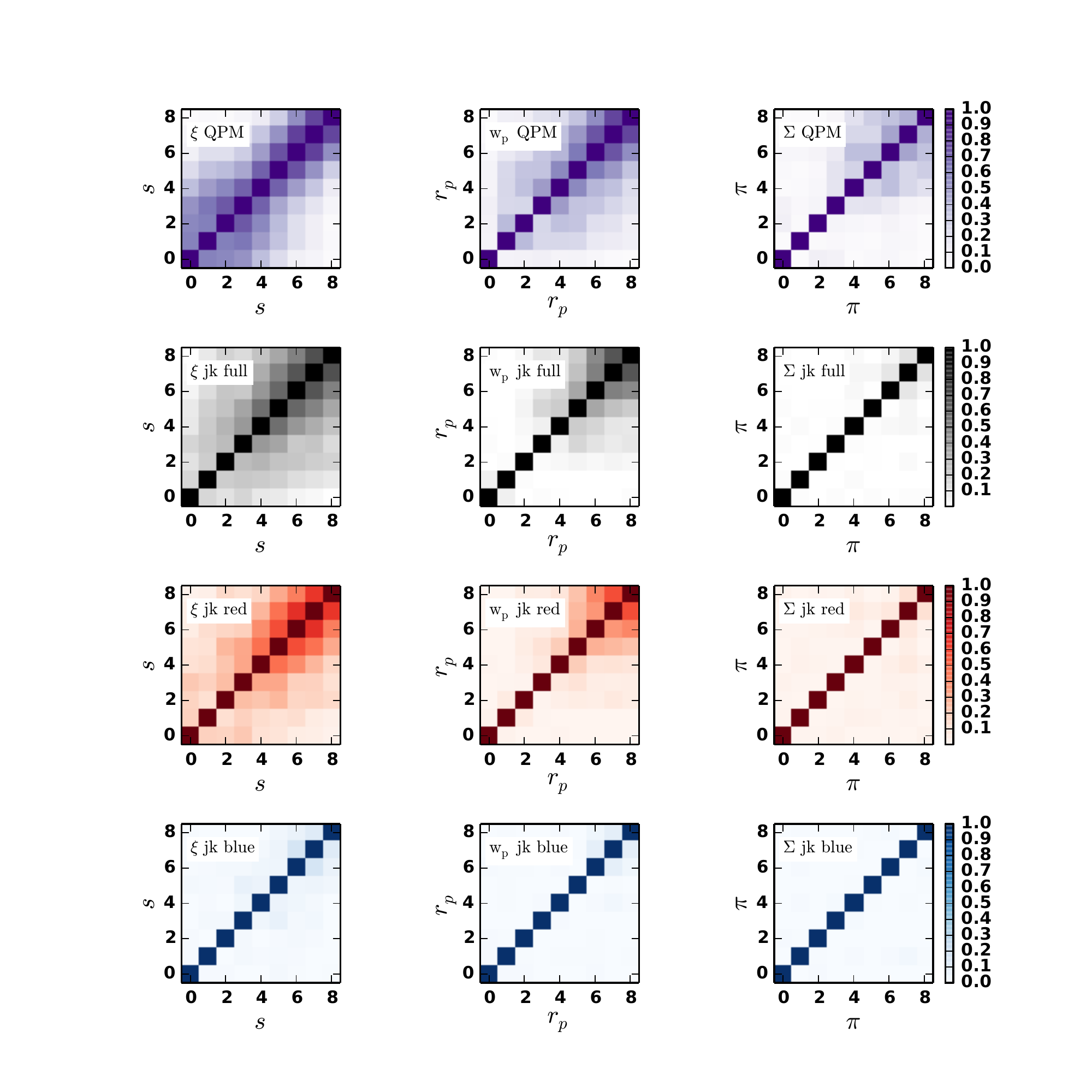}
 \caption{Normalized QPM (first row from the top) versus full (second row), red (third row) and blue (bottom row) CMASS jackknife covariance matrices for $\xi(s)$ (left column), $w_p(r_p)$ (central), and 
 $\Sigma(\pi)$ (right), as a function of the $s$, $r_p$ and $\pi$ bins, respectively. We adopt a ten-step logarithmic binning 
 scheme in the range $3-50\,h^{-1}$Mpc for $s$, $0.1-35\,h^{-1}$Mpc for $r_p$, and $0.1-40\,h^{-1}$Mpc for $\pi$. Overall, QPM mocks show higher covariances compared to the full, red, and blue CMASS samples, confirming the result shown in Figure \ref{fig:chi2surf}. The left column reveals that covariances become appreciable in the red and full redshift-space 2PCFs at intermediate scales (i.e., $s\geq 8\,h^{-1}$Mpc), while they are almost negligible in the blue population.  The red and full CMASS projected 2PCF (central column) are covariant at $r_p\geq2\,h^{-1}$Mpc, while the blue case is almost covariance-free at all scales. The $\Sigma(\pi)$ measurements (right column) are significantly less covariant than the other two clustering statistics: QPM mocks show appreciable covariances only above $\pi\sim3\,h^{-1}$Mpc, while the three CMASS samples are substantially covariance-free.}
 \label{fig:normalized_covma}
\end{center}
\end{figure*}

Figure \ref{fig:normalized_covma} compares the normalized $\xi(s)$, $w_p(r_p)$, and $\Sigma(\pi)$ (respectively from left to right column) 
 covariance matrices estimated using the QPM mocks (top row) and the jackknife 
re-samplings of the full, red and CMASS galaxy samples, to test the correlation between our observations at different scales.
Overall, the QPM mocks show stronger covariances than jackknife in all three metrics. $\Sigma(\pi)$ is less correlated than the redshift-space and projected correlation functions; 
this is due to its definition, see Eq. \ref{eq:sigma_new_metrics}.  Since $\Sigma(\pi)$ is the ratio of two clustering measurements, both errors propagate in it, resulting in a smoother correlation at all scales. The red CMASS sample includes the majority of the CMASS galaxies, thus it is reasonable that its covariance matrices behave similarly to the ones of the full sample. 
The blue case is slightly different: errors are larger and covariances are almost negligible in all the three measurements, especially in $\Sigma(\pi)$.

\end{document}